\documentclass[journal=acsnano,manuscript=article]{achemso}
\usepackage[utf8]{inputenc}
\usepackage{graphicx}
\usepackage{amsmath}
\usepackage{amsfonts}
\usepackage{mdframed}
\usepackage{lipsum}
\usepackage[T1]{fontenc}

\usepackage[%  
    colorlinks=true,
    pdfborder={0 0 0},
    linkcolor=red
]{hyperref}

\title{Opportunities and Challenges of Solid-State Quantum Nonlinear Optics}
\author{Abhinav Kala}
\affiliation{Department of Electrical and Computer Engineering, University of Washington, Seattle, WA 98195, USA}
\author{David Sharp}
\affiliation{Department of Physics, University of Washington, Seattle, WA 98195, USA}
\author{Minho Choi}
\affiliation{Department of Electrical and Computer Engineering, University of Washington, Seattle, WA 98195, USA}
\author{Arnab Manna}
\affiliation{Department of Physics, University of Washington, Seattle, WA 98195, USA}

\author{Prathmesh Deshmukh}
\affiliation{Department of Physics, The Graduate Center, City University of New York, NY 10016, USA}
\alsoaffiliation{Department of Physics, City College of New York, City University of New York, NY 10031, USA}

\author{Vijin Kizhake Veetil}
\affiliation{Department of Physics, UMBC (University of Maryland, Baltimore County), Baltimore, Maryland 21250, United States }

\author{Vinod Menon}
\affiliation{Department of Physics, The Graduate Center, City University of New York, NY 10016, USA}
\alsoaffiliation{Department of Physics, City College of New York, City University of New York, NY 10031, USA}

\author{Matthew Pelton}
\affiliation{Department of Physics, UMBC (University of Maryland, Baltimore County), Baltimore, Maryland 21250, United States }

\author{Edo Waks}
\affiliation{Institute for Research in Electronics and Applied Physics and Joint Quantum Institute, University of Maryland, College Park, Maryland 20742, United States}
\alsoaffiliation{Department of Electrical and Computer Engineering, University of Maryland, College Park, Maryland 20740, United States}

\author{Arka Majumdar}
\affiliation{Department of Electrical and Computer Engineering, University of Washington, Seattle, WA 98195, USA}
\alsoaffiliation{Department of Physics, University of Washington, Seattle, WA 98195, USA}
\email{arka@uw.edu}

\date{February 2024}

\usepackage{subfiles}

\begin{document}

\section{Abstract}

Nonlinear interactions between single quantum particles are at the heart of any quantum information system, including analog quantum simulation and fault-tolerant quantum computing. This remains a particularly difficult problem for photonic qubits, as photons do not interact with each other. While engineering light-matter interaction can effectively create photon-photon interaction, the required photon number to observe any nonlinearity is very high, where any quantum mechanical signature disappears. However, with emerging low-dimensional materials, and engineered photonic resonators, the photon number can be potentially reduced to reach the quantum nonlinear optical regime. In this review paper, we discuss different mechanisms exploited in solid-state platforms to attain quantum nonlinear optics. We review emerging materials and optical resonator architecture with different dimensionalities. We also present new research directions and open problems in this field.

\section{Introduction}

Understanding strongly correlated quantum materials has immense potential to not only explain novel physical effects such as high-temperature superconductivity and the fractional quantum Hall effect, but also to help design these materials to transform quantum information technologies\cite{schafer2020tools, daley2022practical}. Unfortunately, \textit{ab-initio} simulations of many of these correlated quantum materials are difficult on a classical computer, which prevents us from understanding the physical mechanism and hinders their systematic design. To that end, analog quantum simulators present a unique opportunity: instead of simulating the system on a computer, we create an equivalent physical system that is scalable, controllable, and measurable. We can program this model quantum system to mimic the behavior of another quantum system and design new synthetic quantum metamaterials with properties not realizable in naturally occurring materials. Furthermore, there is theoretical evidence that many quantum simulation problems can be solved without fault-tolerance\cite{trivedi2022quantum}. An archetypal quantum simulator consists of an array of coupled quantum nodes each with an intranode repulsive interaction and an internode hopping potential. While there is no clear consensus on which physical system is the most suitable for quantum simulations, superconducting circuits, trapped ions, and cold atoms are leading candidates. 

Instead of analog quantum simulation, a more universal platform will be a fault-tolerant quantum computing system, which can perform digital quantum simulation. Despite recent impressive proof-of-concept computing devices, such as Google’s 53-qubits Sycamore\cite{arute2019quantum} and IBM Q 1,121-qubits Condor\cite{choi2023ibm} in a superconducting platform and Harvard University/QuEra's 280-qubits in neutral atoms\cite{bluvstein2023logical}, the road to a practical, fault-tolerant quantum computer remains full of challenges due to the inherent fragility of quantum states. In fact, protecting these fragile quantum states from detrimental external noise remains an ongoing struggle. Moreover, a useful quantum computer will need millions of qubits, which can be realized in a distributed quantum computing system where $\sim 50$  qubit quantum computers are connected to each other in a network. This necessitates interfacing with a quantum communication system because the quantum states require transduction to optical photons without adding substantial noise, a difficult feat on its own. For both analog and digital quantum simulation, photonic qubit systems have two unique advantages\cite{o2009photonic}: quantum states of photons are easily maintained due to their high energy and extremely weak interaction with the external environment, and photonic quantum simulators are completely compatible with quantum communication. Additionally, optical photons allow direct measurement of multi-particle correlations and preserve the quantum states even at room temperature thanks to their high energy. Thus, despite theoretical similarities between photonic systems and superconducting circuits, a photonic quantum simulator is fundamentally more friendly to scalability. However, quantum optical systems (as used in atomic physics) have historically been bulky, and prone to misalignment. This has changed over the past decades, due to the advent of solid-state nanophotonics and meta-optics. Thanks to the ubiquitous nanofabrication and sophisticated electromagnetic simulators, it is possible to create large-scale photonic integrated circuits akin to their electronic counterpart. In fact, a nano-photonic system is arguably superior to any other quantum system in terms of small size and high-speed operation, both of which are critical to scale to practical quantum advantage \cite{beverland2022assessing}.

Unfortunately, photonic systems lack a critical component for quantum simulation: strong photon-photon interaction. The nonlinear optical interaction between single photons, i.e., photon blockade, is critical to mimic correlated electrons. In the photon blockade regime, the coupling of one external photon to the quantum system hinders the subsequent coupling of other photons. This effect is measured in the second-order autocorrelation $g^{(2)} (\tau)$ of the photons, where $\tau$ is the time delay between photons. A coherent state (e.g. laser) has $g^{(2)}(0)=1$, but the transmitted light exhibits sub-Poissonian statistics: $g^{(2)} (0)<1$, with $g^{(2)} (0)\approx0$ indicating perfect photon blockade and creation of hard-core bosons. The key to achieving single photon or quantum nonlinear optics (qNLO) and building an optical quantum simulator is to increase the light-matter interaction strength, which in turn mediates an effective photon-photon coupling. While the light-matter interaction strength is normally small, optical resonators can significantly enhance this interaction thanks to the spatial and temporal confinement of light in these resonators\cite{vahala2003optical}. To that end, the first demonstration of qNLO in terms of photon blockade was reported in atom-based cavity quantum electrodynamic (cQED) systems\cite{birnbaum2005photon}. However, the scalability of a quantum simulator necessitates the use of a solid-state platform, which has led to the extensive study of hybrid light-matter system polaritons\cite{byrnes2014exciton} formed by integrating quantum-confined structures to cavities. 

In this paper, we review the current state of qNLO in solid-state systems. We discuss different resonators and material systems that are necessary to reach the qNLO regime. Specifically, we explore the effect of dimensionality in both materials and optical resonators to understand the effect of confinement and modified density of states. We review the existing material systems where qNLO has been reported. We also discuss emerging low-dimensional material systems and novel resonator architectures, where qNLO is expected to improve.

\begin{figure}[H]
\centering
\includegraphics[scale=.5]{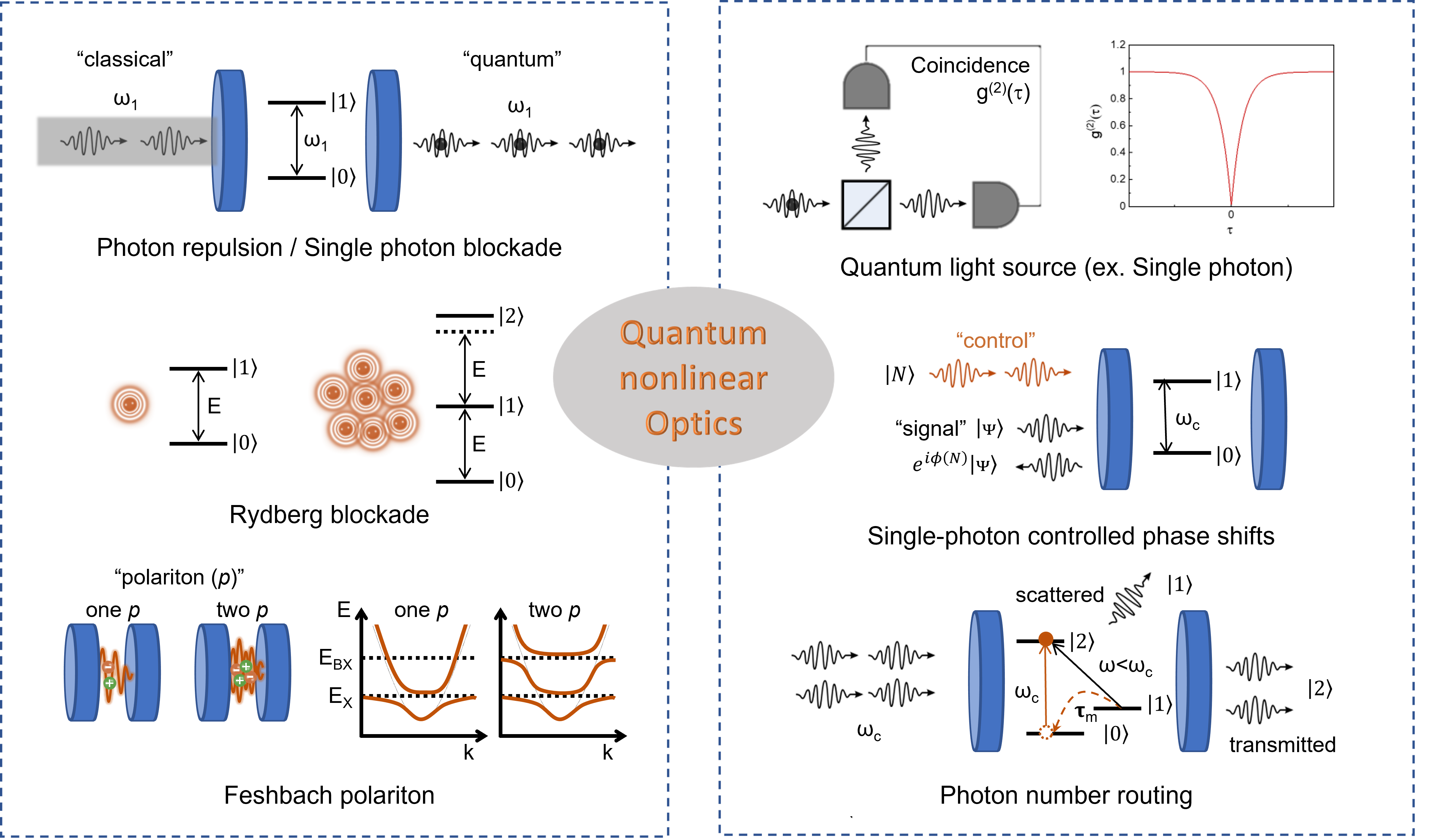}
\caption{Single-Photon Nonlinear Optics - The box on the left summarizes various quantum nonlinear optics (qNLO) phenomena. In the regime of qNLO, the interaction between single photons is facilitated by matter. For example, a single-photon blockade exploits the anharmonic energy structure of a strongly coupled two-level emitter-cavity system to create an effective repulsive interaction between single photons. Single photon blockade converts a classical beam of light into an antibunched quantum light beam. Another way to create such repulsive interaction and blockade in solid-state is by using the interaction between giant Rydberg excitons. The Rydberg blockade is achieved when the Rydberg excitons interact strongly enough to generate large shifts in the Rydberg excitonic energy level. Feshbach blockade is one more way to use excitons to achieve qNLO. The interaction among polaritons in a strongly coupled quantum well exciton-cavity system gets enhanced near the biexciton which can lead to the blockade at the level of individual photons. In the box shown on the right, some of the applications of qNLO are shown. Quantum light sources such as a single photon source, could be created by using repulsion between single photons. A single photon source is recognized by a vanishing second-order time correlation function at zero delay, i.e. - $g^{(2)}(0)$. One of the major applications of qNLO is in quantum information processing where single photon gates controlled phase shifts are required. In a cavity with a two-level quanutm dot, the reflection phase shift for a signal photon can be controlled by the photon number of a control beam single photon level\cite{fushman2008controlled}. Another application is photon number routing which has been achieved with a silicon vacancy inside a single mode cavity \cite{sipahigil2016integrated}. In the presence of a metastable ground state $|1\rangle$ in addition to the ground state $|0\rangle$, the preferential transmission and the scattering of a photon resonant with the cavity mode.}
\label{Introduction}
\end{figure}

\section{Physics of Quantum Nonlinear Optics Phenomena}
\begin{figure}[p]
\centering
\includegraphics[scale=0.52]{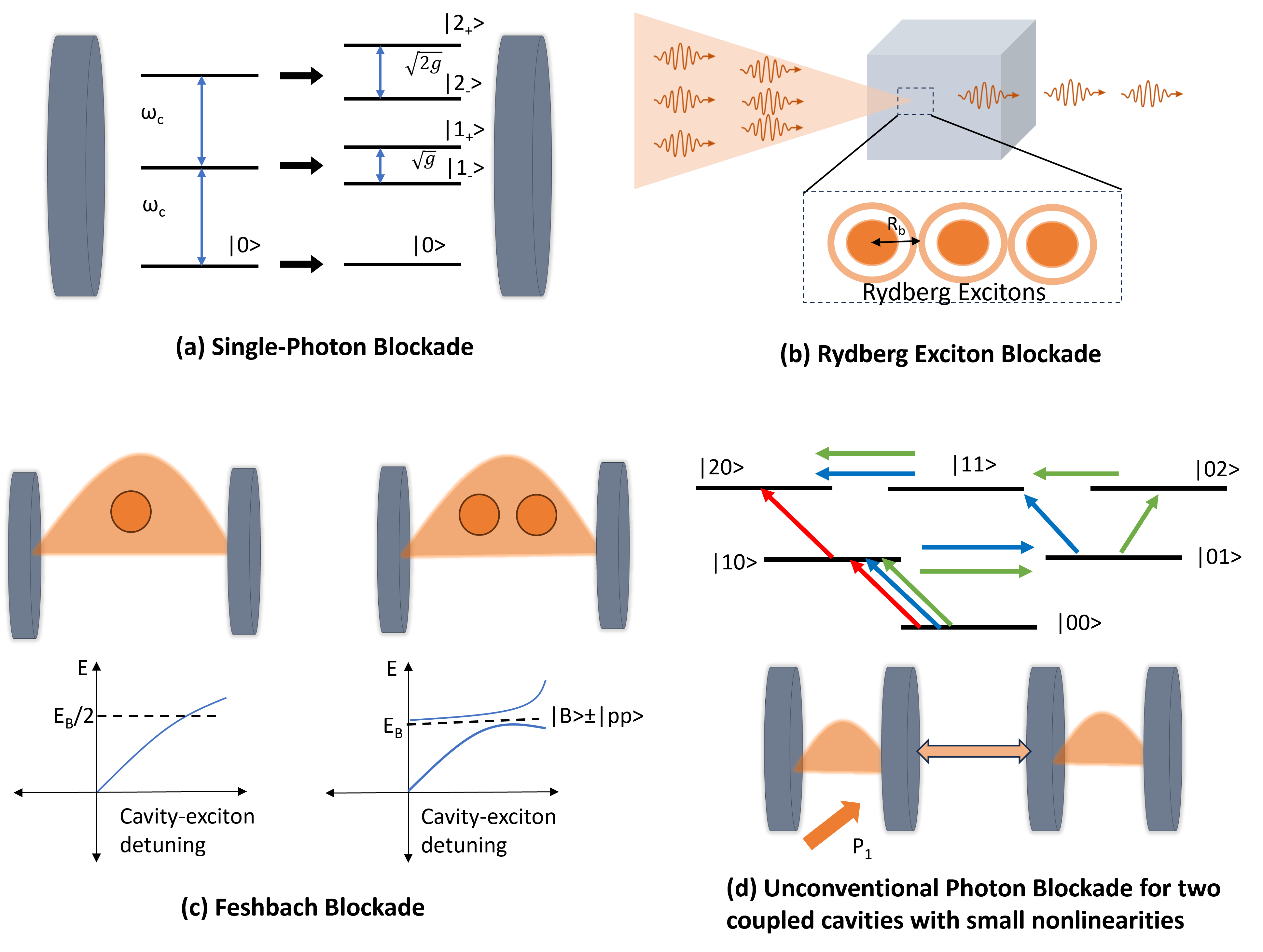}
\caption{Physical Origin of Single-Photon nonlinearity - (a) Single-photon blockade in a cavity coupled to a two-level emitter - (left) Bare cavity has a harmonic level structure with equal spacings of $\omega_c$. The energy level structure for the cavity with a two-level emitter is shown on the right. All the bare cavity levels above the vacuum state split in two with splitting dependent on the photon number. This anharmonic level structure prevents the absorption of two photons from the ground state and photon blockade is achieved. Other nonlinearities, such as those of quantum well polaritons and bulk material optical nonlinearities could also be used to achieve single photon blockade. (b) Rydberg exciton blockade - When in a solid the transition to Rydberg excitons is using a classical light source, the interactions among Rydberg excitons prevent the creation of more than one exciton within the blockade radius $R_b$. Therefore, the transmitted photons are antibunched. (c) Feshbach blockade - The dispersion of the lower polariton is plotted as a function of cavity-exciton detuning and is assumed to cross the energy E$_B$/2 at some detuning. In the presence of two polaritons, the interaction with biexciton leads to hybrid 2 polaritons-biexciton states - $|B\rangle \pm |pp\rangle$. The 2-polariton energy level gets split in two and this anharmonicity leads to the Feshbach blockade.\cite{carusotto2010feshbach} (d) Unconventional photon blockade with two identical coupled cavities with arbitrarily small nonlinearity - In the case of only one cavity being externally pumped by pump $P_1$, The blockade is achieved when the amplitudes shown in the figure for creation of two photons in cavity 1 cancel each other. Different pathways are shown with different colored arrows. The figure is adapted from a reference.\cite{flayac2017unconventional} }
\label{Physics}
\end{figure}

First, we will elaborate on the physics behind various qNLO phenomena that have been reported in solid-state systems. 

\noindent \textbf{Photon blockade with a single two-level emitter}

The most prevalent system to achieve qNLO is a two-level emitter coupled to an optical cavity. A two-level emitter is inherently nonlinear since the absorption of a single photon prevents the immediate absorption of another photon. However, such a nonlinear response is not readily achievable for a single emitter and a single photon due to very weak light-matter coupling. A cavity could trap a photon for a long time and spatially confine it to a small mode volume, increasing the probability of interaction. To quantify the effect of cavity confinement, we use the example of a Fabry-Perot cavity of length $L$, resonance frequency $\omega$, and corresponding resonance wavelength $\lambda$. Cavity finesse $\mathcal{F}$ is defined as the ratio of the free spectral range ($\delta\omega$) to the cavity linewidth ($\Delta \omega$). The Q-factor, that is the ratio of the linewidth to the resonance frequency $(Q=\Delta\omega/\omega)$, is related to the cavity finesse as $Q=2\mathcal{F}L/\lambda$\cite{hunger2010fiber}. The Q-factor, therefore, is proportional to the finesse for a fixed cavity length. If $\delta_E$ is the fraction of energy lost from the cavity mode in each round-trip then, finesse could also be written as\cite{hunger2010fiber} $\mathcal{F} = 2\pi/\delta_E$. Multiple reflections inside a cavity therefore lead to an increase in the effective cross-section of the emitter due to the cavity by a factor of\cite{reiserer2015cavity} $\mathcal{F}/\pi = Q\frac{\lambda}{2\pi L}$. We note that for most solid-state cavities, the Q-factor is a more relevant quantity than the finesse, as solid-state cavities generally support a much smaller number of modes due to their small size and material dispersion. However, for macroscopic cavities, such as the ones used in atomic physics, finesse is used as the primary parameter to quantify light-matter interaction.  In solid-state cavities, the stronger photon-emitter interaction requires higher Q-factor and lower mode volumes. If the coherent interaction strength under dipole approximation between an emitter resonant to a cavity mode an that cavity mode is $g$. And, the emitter is located at the cavity field maxima with dipole moment aligned along the cavity electric field direction. Then, $g = \sqrt{\mu^2\omega_c/{2\hbar V_m}}$. Where, $\mu$ is the dipole transition matrix element between the ground state ($|\mathtt{g}\rangle$) and the excited state ($| e\rangle$) of the emitter, i.e. - $\mu = \langle e|\hat{d}|\mathtt{g}\rangle$ where $\hat{d}$ is the dipole moment operator, $\omega_c$ is the cavity resonance frequency, and $V_m$ is the cavity mode volume given by $\frac{\int \epsilon(\vec{r})|E(\vec{r})|^2 d^3r}{max(\epsilon(\vec{r})|E(\vec{r})|^2)}$. 
Here, $E(\vec{r})$ is the cavity electric field and $\epsilon(\vec{r})$ is the material permittivity. With an emitter decay rate of $\gamma$ and cavity field decay rate of $\kappa$, we can define cooperativity ($C$) as $C = \frac{g^2}{2\gamma\kappa}= \frac{\mu^2 \omega}{2\hbar\gamma V_m \Delta \omega} = \frac{\mu^2 }{\hbar\gamma}\frac{Q}{V_m}$. Here we have used the relation $\kappa = \Delta\omega/2$, as the field decays twice as slow as the intensity. For deterministic interactions between a single emitter and a single photon, we want the cooperativity to be high $C>>1$\cite{reiserer2015cavity}.  

\noindent In the absence of losses, the dynamics of a cavity coupled with a two-level emitter is described by the Jaynes-Cummings (JC) Hamiltonian, which under rotating-wave approximation is given by  \cite{majumdar2012nonlinear}

{\centering$\hat{H} = \hbar\omega_c\hat{a}^{\dagger}\hat{a} + \hbar\omega_a\hat{\sigma}^{\dagger}\hat{\sigma}  + i\hbar g [\hat{a} \hat{\sigma}^{\dagger} - \hat{a}^{\dagger} \hat{\sigma}]$\par} 

\noindent Here, $\omega_c$ is the cavity resonance frequency, $\omega_a$ is the emitter resonance frequency, and $g$ is the coherent coupling strength. $\hat{a}$ is the annihilation operator for the cavity electromagnetic field and  $\hat{\sigma}$ is the lowering operator for emitter, $\hat{\sigma} = |\mathtt{g}\rangle\langle e|$. The third term in the Hamiltonian is the dipole light-matter interaction term. In what follows, we denote the cavity-emitter detuning ($\omega_a-\omega_c$) by $\Delta$. $N$-photon degenerate eigenstates of this Hamiltonian in the absence of interactions, i.e.- when $g=0$, are $|\mathtt{g},N\rangle$ and $|e,N-1\rangle$. These combined light-matter eigenstates are tensor products of emitter($\mathfrak{e}$) and cavity field (${c}$) quantum states, i.e. - $|\mathfrak{e},c\rangle = |\mathfrak{e}\rangle\otimes|c\rangle$. The total Hamiltonian ($\hat{H}$), in the basis of these bare states is the direct sum of N-photon Hamiltonians ($\hat{H}_N$), which only operate in the two-dimensional subspaces formed by $|g,N\rangle$ and $|e,N-1\rangle$ states for $N\ge1$. $N=0$ is the ground state of the system. In the basis of bare eigenstates, $\hat{H}_N$ has the following form

{\centering$\hat{H}_N = \begin{pmatrix}
 N\omega_c & - i\sqrt{N}g \\
 i\sqrt{N}g & N\omega_c + \Delta 
\end{pmatrix}$\par}

\noindent The energy eigenvalues of $\hat{H}_N$ are $E_N{\pm} = N\omega_c + \frac{1}{2}(\Delta \pm \sqrt{\Delta^2 + 4Ng^2})$
and the corresponding eigenstates are $|N_{\pm}\rangle = cos(\theta_N) |N-1,\mathit{e}\rangle \pm sin(\theta_N) |N,\mathtt{g}\rangle$. Where, $2\theta_N = tan^{-1}{[\frac{2\sqrt{N}g}{\Delta}]}$

\noindent Therefore, the coupling splits the degenerate energy levels, and the splitting strength depends on the number of photons present in the cavity. The splitting between modes for zero detuning ($\Delta=0$) in the $N^{th}$ manifold is $\sqrt{N}\hbar g$. This anharmonic nature of the Jaynes-Cummings model energy levels leads to single-photon nonlinearity. To understand the photon blockade, we consider the one and two-photon manifolds of the excited states. For an incoming laser beam with its frequency tuned to the transition $|0,g\rangle \longrightarrow |1_+\rangle$, if one photon is coupled to the cavity, the second photon cannot be resonantly coupled since the levels of the second manifold will be detuned from the laser due to $\sqrt{2}\hbar g$ splitting in the $2^{nd}$ manifold(Figure~\ref{Physics}(a)). This leads to what is known as the single photon blockade where the presence of the first photon in the cavity blocks transmission of any more photons. This nonlinear interaction strength thus is given by $(\sqrt{2}-1)\hbar g$. Thus for a single emitter, the nonlinear interaction strength is directly dependent on the light-matter interaction strength. Due to the presence of loss in real systems, the cavity mode and exciton have finite linewidths. If $\gamma$ is the emitter linewidth and $\kappa$ is the cavity linewidth, the eigenmode $|1\pm\rangle$ will have the linewidth of $\frac{\gamma+\kappa}{2}$\cite{dovzhenko2018light}. The level splitting ($2g$) must be larger than twice of this linewidth to so that $|1\pm\rangle$ states appear as two separate resonances. This light-matter coupling regime where $2g>\gamma +\kappa$, is also called the strong coupling regime. Furthermore, for single photon blockade the nonlinear interaction strength $(\sqrt{2}-1)\hbar g$ has to be larger than twice the linewidth of the eigenmodes, i.e. - $\gamma +\kappa$.

\noindent Single photon blockade is an example of a repulsive photon-photon interaction that generates antibunched photons, i.e. photons with zero delay second order correlation function $g^{(2)}(0) < 1$\cite{birnbaum2005photon}. For perfect blockade we can achieve $g^{(2)}(0)=0$. Attractive interactions between photons can also be implemented in a cavity-emitter system. A way to implement that is using incident photon beam that is detuned from the transition between the ground state and the states in the first manifold but at the same time is tuned for the two photon absorption transition in the second manifold. In that case, coupling two photons to the cavity will have higher probability than coupling one photon and one would effectively implement an attractive interaction between two single photons. In that case, the transmission of two photons becomes more probable than the transmission of single photons\cite{faraon2008coherent, majumdar2012probing}.

\noindent While Jaynes-Cummings nonlinearity is the most studied mechanism for qNLO, single photon blockade can potentially be achieved in a two-dimensional quantum well or using bulk nonlinearity as we show below. 

\noindent \textbf{Polariton blockade}

In two-dimensional quantum-confined materials, such as quantum wells, photon blockade is often termed as polariton blockade. The polaritons in quantum wells are hybrid light-matter states, similar to the eigenstates of Jaynes-Cummings model, formed under the strong coupling of excitons and photons. Two polariton modes are formed when an exciton is strongly coupled to a cavity mode. The higher energy mode is called the upper polariton and lower energy mode is called the lower polariton. Unlike photons, polaritons can interact with each other due to their excitonic components, ultimately leading to an effective photon-photon interaction. We consider quantum well excitons strongly coupled to a confined photonic mode and focus on the lower polariton branch. The interaction Hamiltonian for the polariton-polariton interactions has the form \cite{delteil2019towards} $\hat{H}_{pol} = \frac{U_{pol}}{2}\hat{p}^{\dagger}\hat{p}^{\dagger}\hat{p}\hat{p}$ where, $\hat{p}$ is the annihilation operator for the lower polariton.  The operator $\hat{p}$ can be expanded as the cavity mode annihilation operator ($\hat{c}$) and exciton annihilation operator ($\hat{x}$) such that $\hat{p} = c_{x}\hat{x} + c_c\hat{c}$  where $c_{c}$ and $c_{x}$ are the Hopfield coefficients representing the photonic fraction ($|c_c|^2$) and excitonic fraction ($|c_x|^2$) of the polariton. The polariton-polariton interaction strength ($U_{pol}$) could be expressed as a function of exciton-exciton interaction strength ($U_{ex}$) as $U_{pol} = U_{ex}|c_{x}|^4 $. $U_{ex}$ depends on the excitonic wavefunction $\alpha(\vec{r})$ as, $U_{ex} = K\int d^2\vec{r}|\alpha(\vec{r})|^4$.\cite{verger2006polariton} Where, $K$ is the strength of exciton-exciton interaction potential. The integral $\int d^2\vec{r}|\alpha(\vec{r})|^4$ is inversely proportional to the lateral photonic mode area ($S$) and $U_{ex}$ scales as $\sim\frac{\Delta_B a_B^2} {S}$.\cite{verger2006polariton, tassone1999exciton} Where, $\Delta_B$ is the exciton binding energy and $a_B$ is the exciton Bohr radius. Hence, a larger $U_{ex}$ requires a stronger field confinement.

\noindent Polariton blockade could be achieved by pumping the system with a laser resonant to or slightly red-detuned from the lower polariton.\cite{verger2006polariton} 
The cavity field under this drive becomes antibunched with an amplitude ($1-g^{(2)}(0)$) approximately equal to the ratio $U_{pol}/\gamma_p$ where, $\gamma_p$ is the polariton linewidth. We know from the previous section that close to zero detuning, $\gamma_p \approx \frac{\gamma + \kappa}{2}$ where, $\gamma$ and $\kappa$ are the exciton and cavity linewidths, respectively. Therefore, strong photon confinement and small exciton and cavity linewidths are essential for the realization of perfect polariton blockade i.e. $g^{(2)}(0) \approx 0$.

\noindent \textbf{Photon blockade using bulk optical nonlinearity}

Bulk material nonlinearity of cavity material can also give an anharmonic level structure. Various schemes have been proposed to create an anharmonic level structure for cavity materials with non-zero nonlinear susceptibilities. Here we discuss how a third-order ($\chi^{(3)}$) and a second-order ($\chi^{(2)}$) bulk material optical nonlinearity can be exploited to reach single photon nonlinearity.\cite{ferretti2012single, majumdar2013single}

\noindent For centrosymmetric materials, $\chi^{(2)}=0$ and therefore $\chi^{(3)}$ is the most significant nonlinearity. If the cavity material has non-zero $\chi^{(3)}$ then, the nonlinear interaction term of the Hamiltonian can be written as $\hat{H}_{I3} = U_{\chi^{(3)}}\hat{a}^{{\dagger}2}\hat{a}^{2}$\cite{ferretti2012single}. This interaction creates an anharmonic energy level structure where a single cavity photon has energy $\omega_c$ but, the energy for two photons is $2\omega_c + 2 U_{\chi^{(3)}} $\cite{ferretti2012single}.
The nonlinear shift $U_{\chi^{(3)}}$ could be expressed as $U_{\chi^{(3)}} \simeq \frac{3(\hbar\omega_c)^2}{4\epsilon_0 V_{eff}}\frac{\bar{\chi}^{(3)}}{\bar{\epsilon}_r ^2}$. Where $V_{eff}$ is the effective mode volume defined being equal to $\frac{1}{\int |\alpha(\vec{r})|^4d^3\vec{r}}$ with $\alpha(\vec{r})$ being the cavity electric field profile. Furthermore, $\bar{\chi}^{(3)}$ and $\bar{\epsilon}_r$ are the average values of the real part of nonlinear susceptibility and material's relative permittivity, respectively.

\noindent To observe the photon blockade, the cavity is driven by a weak continuous wave laser with frequency $\omega_L$. If the cavity decay rate is $\kappa$, then for the photons emitted from cavity, $g^{(2)}(0) = \frac{1 + 4(\Delta\omega/\kappa)^2}{1 + 4(\hbar\Delta\omega + U_{\chi^{(3)}})^2/\kappa^2}$ where $\Delta = \omega_L-\omega_c$ is the laser detuning from the cavity resonance. When $U_{\chi^{(3)}}/\kappa >>1$, $g^{(2)}(0)$ approaches $0$ and perfect single photon blockade is achieved. Since, $U_{\chi^{(3)}} \propto 1/V_{eff}$ and $\kappa$ is proportional to the cavity quality factor $Q$, the requirement for perfect blockade could also be rephrased as the requirement of large $Q/V_{eff}$.

\noindent Non-centrosymmetric materials have non-zero $\chi^{(2)}$. This type of nonlinearity tends to have a larger magnitude compared to $\chi^{(3)}$ nonlinearity. One way to harness $\chi^{(2)}$ nonlinearity for photon blockade is by using a doubly resonant cavity with modes at frequencies $\omega_1$ and $\omega_2$ such that $\omega_2 = 2\omega_1$.\cite{majumdar2013single} The interaction Hamiltonian in the case of $\chi^{(2)}$ nonlinearity can be expressed as\cite{majumdar2013single}, $\hat{H}_{I2} = U_{\chi^{(2)}}[\hat{b}\hat{a}^{{\dagger}2} + \hat{b}^{\dagger}\hat{a}^{2}]$. The eigenstates of the cavity without nonlinearity are the eigenstates of operator $\hat{N} = \hat{N_1}+\hat{N_2}$ where, $\hat{N_1}$ and $\hat{N_2}$ are number operators for mode $1$ and mode $2$, respectively.  These eigenstates possess degeneracy for $N_1>1$. In the presence of nonlinearity, the degenerate levels split and an anharmonic level structure is formed which could be used to implement single photon blockade. The magnitude of splitting and the number of split levels is dependent on $N$. For example, the two lowest degenerate levels with $N=2$ and $N=3$ levels split in two with splittings of $2\sqrt{2}U_{\chi^{(2)}}$ and $2\sqrt{6}U_{\chi^{(2)}}$ respectively. $N=4$ state splits in 3 levels with splittings of $\pm 4U_{\chi^{(2)}}$ about level with the energy of bare state\cite{majumdar2013single}.

\noindent Furthermore, an order of magnitude estimation of the nonlinear strength ($U_{\chi^{(2)}}$) could be made by assuming the same electric field profile $\alpha(\vec{r})$ for both modes. This gives $U_{\chi^{(2)}} = \epsilon_0(\frac{\hbar\omega_1}{\epsilon_0\epsilon_r})^{3/2} \frac{\chi^{(2)}}{\sqrt{V_{eff}}}$\cite{majumdar2013single}. The effective mode volume in this system is defined as $V_{eff}^{-1/2} = \int \alpha(r)^3 d^3\vec{r}$. Similar to the $\chi^{(3)}$ materials, the nonlinear interaction strength has an inverse dependence on the mode volume as well.

\noindent While the anharmonicity coming from the light-matter interaction is perhaps the most studied mechanism of qNLO, there are two other mechanisms, namely Rydberg blockade and Feshbach blockade, by which photon-photon interaction can be achieved, potentially leading to qNLO.

\noindent \textbf{Rydberg exciton blockade}

Rydberg atoms are atoms with a single valence electron with a large principal quantum number ($n$). 
Similar to atoms, excitons in semiconductors can also have a series of energy levels above the ground state. These excited state excitons are called Rydberg exciton due to similarity with the single electron atoms. Rydberg excitons have been observed in semiconductors such as cuprous oxide ($Cu_2O$), transition metal dichalcogenide (TMDC) monolayers, and perovskites\cite{kazimierczuk2014giant, hill2015observation, bao2019observation}.

\noindent Rydberg excitons interact with each other via van der Waals and dipole-dipole interactions. Using a hydrogenic atom to model the exciton, van der Waals interactions between neutral excitons scale as $-A/r^6$, where $r$ is the interexciton distance. However, $A$ scales as $n^{11}$ which is why for large `n' these interactions become quite significant and can lead to the phenomenon of Rydberg blockade at the level of individual photons\cite{assmann2020semiconductor, comparat2010dipole}. The dipole-dipole interaction term in the potential scales as $B/r^3$, where, $B$ scales as $n^4$. The interaction potential is much more complex when excitons have non-zero orbital angular momentum.\cite{assmann2020semiconductor}  \par

\noindent When a Rydberg exciton is created due to the absorption of a single photon, the van der Waals interaction shifts the Rydberg levels of the neighboring excitons. A large enough shift can prohibit the creation of another Rydberg exciton with the same energy in a neighboring volume within radius $R_b$. This phenomenon is called the  Rydberg blockade and $R_b$ is called the blockade radius (Fig.~\ref{Physics}(b)). The Rydberg exciton hence acts as an effective two-level system within the blockade volume. A monochromatic light source resonant with the Rydberg transition will therefore produce antibunched light similar to the Jaynes-Cummings-type blockade described above. 

\noindent \textbf{Feshbach Polariton Blockade}

In atomic physics, Feshbach resonance occurs when the energy of two colliding atoms matches with that of a temporary molecular state \cite{chin2010feshbach}. While creating a transient state, a significant enhancement of the scattering cross-section is observed. In solid-state systems, the interaction between a biexciton, a bound state of two excitons, and polaritons can create a similar phenomenon when the biexciton energy is twice the energy of exciton-polariton\cite{takemura2014polaritonic}.\par

\noindent Under strong coupling of the exciton (for example, in a quantum well) with a localized cavity mode, the system has two resonant exciton-polariton modes - upper polariton and lower polariton. Energies of both polaritons vary as a function of detuning between the exciton and cavity resonance, but do not cross each other. The minimum energy gap between upper and lower polaritons is equal to the Rabi splitting energy ($2g$) at zero detuning. Scattering of polaritons could also create a bound state of two excitons which is called the biexciton. The biexciton state has an energy of $E_B = 2 E_X - \Delta_B$, where $E_X$ is exciton energy and $\Delta_B$ is biexciton binding energy. If the Rabi splitting energy is much larger than the biexciton binding energy ($2g >> \Delta_B$) and lower polariton energy is matched with $E_B/2$ at the same time, the upper polariton cannot participate in creating a biexciton because it has much higher energy. In that case, the nonlinear interaction of lower polaritons leading to biexciton formation can be modeled using an interaction term in the Hamiltonian of the form $\hat{H}_{int} = g[\hat{b}^{\dagger}\hat{a}\hat{a} + \hat{a}^{\dagger}\hat{a}^{\dagger}\hat{b} ]$.\cite{carusotto2010feshbach} 
$\hat{a}$ and $\hat{b}$ are annihilation operators for polaritons and biexcitons respectively.\\
\noindent This interaction conserves the number $N = N_p + 2N_b$ where, $N_p$ is the number of lower polaritons and $N_b$  is the number of biexcitons. For $N = 1$, the energy eigenstate is a single LP. For $N = 2$, the eigenstates are superpositions of a biexciton and two lower polaritons with energy splitting of $\pm\sqrt{2}g$ from the energy of two non-interacting lower polaritons. This is an anharmonic level structure similar to the Jaynes-Cummings model and therefore leads to single-photon blockade referred to as the Feshbach blockade (Fig.~\ref{Physics}(c)).

\noindent \textbf{Unconventional Photon Blockade}

Unconventional photon blockade is a set of schemes that do not require strongly coupled cavity-emitter systems or interactions between polaritons. These schemes instead utilize the destructive interference between different pathways that generate two-photon states. There are multiple systems that have been proposed to achieve unconventional photon blockade such as coupled cavities and a two-level system weakly coupled to a cavity.

\noindent One of the simplest schemes involves two coupled cavities (with annihilation operators $\hat{a}_1$ for cavity 1 and $\hat{a}_2$ for cavity 2) with material nonlinearity, $U_i = u_i \hat{a_i}^{\dagger2} \hat{a_i}^{2}$, $i =1, 2$ as shown in Fig.~\ref{Physics}(d). There are multiple pathways to create two-photon state from the ground state in the left cavity, $|00\rangle \rightarrow|20\rangle$. Unconventional photon blockade occurs if all of these contributions add up to zero which could be achieved with appropriate coupling, detuning, and pumping parameters\cite{flayac2017unconventional}.

\noindent Notably, unconventional photon blockade can be achieved for arbitrarily small optical nonlinearity when both cavities are pumped with independent control over pump intensity and phase. However, the photon states generated by these mechanisms are similar to the amplitude-squeezed Gaussian states rather than the pure single-photon states of the conventional blockade\cite{flayac2017unconventional}. Therefore, it requires one to work in the limit of a very small cavity photon number.
\\

\begin{mdframed}
        \textbf{\begin{center}Box 1: Examples of applications of qNLO in solid-state systems
        \end{center}} 

       Quantum information processing is among the most sought-after application where qNLO is required. Here we describe two examples in solid-state which could serve as the building blocks of such applications. 
       The schematic of the first example is shown in the sub-figure `single-photon controlled phase shifts' of Fig.~\ref{Introduction}. This was demonstrated using a single two-level system, that is a quantum dot (QD), inside a single mode cavity.  \cite{fushman2008controlled}
       A signal beam and a control beam with $\sim 1$ photon per cavity lifetime and frequency $\omega$ near the cavity resonance ($\omega_c$) are reflected from the cavity. The reflected signal beam acquires a phase at reflection which could be measured by interfering it with a reference beam. The phase-shift ($\phi_N$) for the signal beam depends on the control beam intensity/photon number (N). The origin of this intensity dependence is in the saturation of the quantum dot interaction with the cavity field at the few photons per cavity lifetime intensities. The cavity-QD system therefore acts as an effective nonlinear medium in this regime and an effective nonlinear index ($n_2$) could be assigned to it using expression $\phi_N \approx \omega_c n_2(P_{in} n/\lambda^2 2\kappa).$\cite{fushman2008controlled} Where, $P_{in}$ is input power, $\lambda$ is the probe photon wavelength, and $\kappa$ is the cavity mode decay rate. The conditional phase shift could work for either same or different wavelengths of control and the probe. 

        One of the limitations of using a two-level emitter for qNLO is the relatively short lifetime of the excited state which requires the interacting photons to arrive almost instantaneously\cite{chang2014quantum}. To overcome this challenge, emitters with three or more relevant levels could be used. Several implementations in atomic physics have used the electromagnetic induced transparency in multilevel atoms\cite{fleischhauer2005electromagnetically}. The following example of a photon number router in solid-state takes advantage of a third level of the emitter. The system in consideration is a silicon vacancy center (SiV) defect in diamond crystal inside a cavity studied for photon number routing among other things\cite{sipahigil2016integrated}. The schematic of the system is shown in the `photon number routing' sub-figure in Fig.~\ref{Introduction}. For switching, three levels of the SiV were considered - two metastable ground states $|0\rangle$ and $|1\rangle$ and one excited state $|2\rangle$. The state $|1\rangle$ is a metastable state with a transition lifetime of $\tau_m$ much larger that the lifetime of $|0\rangle \rightarrow |2\rangle$ transition. The transition $|0\rangle \rightarrow |2\rangle$ is resonant with the cavity mode but not strongly coupled.\\
        The short timescale (of the order of $\tau_m$) and low photon number dynamics of this system shows nonlinear behavior which is observed as the photon bunching ($g^{(2)}(0) > 0$) in the transmitted field and anti-bunching ($g^{(2)}(0) < 0$) in the scattered field. At incident laser power where not more than two excitations are present in the cavity at one instant, this system effectively acts as the photon number router which preferentially scatters single photons but transmits two photons. Intuitively, this routing could be understood as preferential projection of the SiV in one of the ground states depending upon whether the scattered photon or the transmitted photon is detected. If the transmitted photon is detected then the SiV is in the state $|1\rangle$ and the cavity is transparent for a photon resonant with the $|0\rangle \rightarrow |2\rangle$ transition for a timescale of the order of $\tau_m$. It increases the probability of transmission of the next photon and bunching is observed. On the other hand, when a scattered photon is detected, the SiV is more likely to be in the state $|0\rangle$, and the two-level dynamics gives rise to antibunching.  
\end{mdframed}

\section{Current State of Quantum Nonlinear Optics}

The first set of qNLO experiments involved cold atoms strongly coupled to optical cavities\cite{chang2014quantum, birnbaum2005photon, shomroni2014all}, including demonstrations of single photon blockade, optical-switching, Rydberg blockade, and conditional phase shifts\cite{birnbaum2005photon, shomroni2014all, baur2014single, turchette1995measurement}. As explained earlier, to achieve qNLO we need to increase the $Q/V_m$ ratio of the cavities while employing quantum materials with large dipole moments. Solid-state cavities provide an excellent platform as using nano-scale optical cavities can realize extremely small mode-volumes, while many of the quantum emitters exhibit much larger dipole moments compared to cold atoms. However, two limitations of the solid-state platform are moderate Q-factor as well as spectral and spatial inhomogeneity of the emitters. 

\begin{figure}[p]
\centering
\includegraphics[scale=0.5]{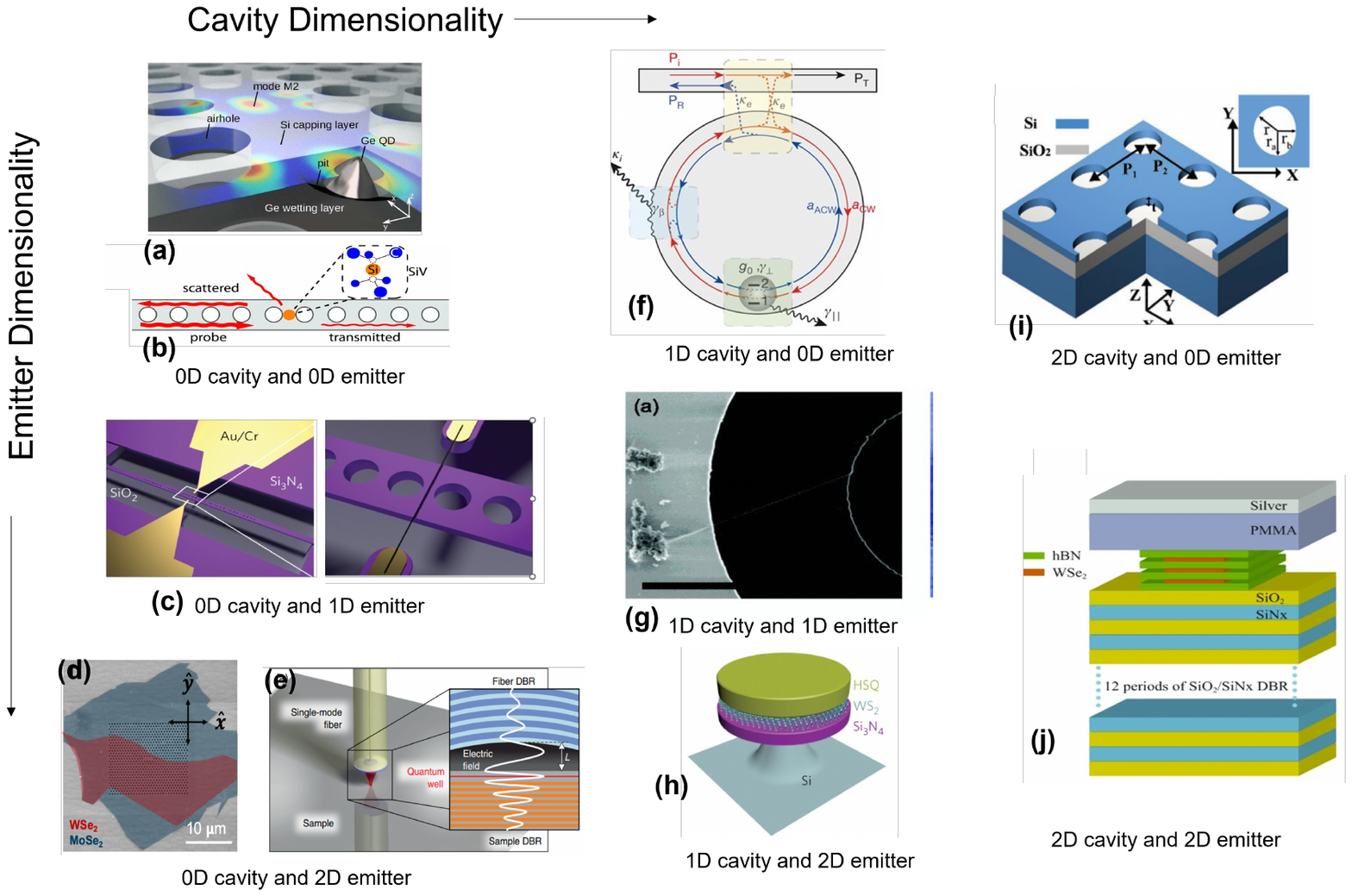}
\caption{Examples of cavity-emitter integration for various cavity and emitter dimensionality, i.e. - the number of dimensions. 0D emitters and 0D cavities - (a) A Germanium quantum dot coupled to a silicon photonic crystal cavity for enhanced emission\cite{schatzl2017enhanced} and (b) SiV centers in diamond inside a nanobeam cavity used as an integrated platform for quantum optical networks including demonstration of single-photon switching\cite{sipahigil2016integrated}. (c) 1D emitters in 0D cavity - A carbon nanotube coupled to a nanobeam cavity for Purcell-enhanced emission.\cite{pyatkov2016cavity} 2D emitters in 0D cavity - (d) Coupling the emission from interlayer excitons in MoSe$_2$-WSe$_2$ heterobilayer to a gallium phosphide photonic crystal cavity\cite{rivera2019coupling} and (e) (In, Ga)As quantum well strongly coupled to a fiber-DBR cavity for polariton-polariton interactions at low input powers\cite{besga2015polariton} (f) 0D emitter in 1D cavity - An epitaxial InAs quantum dot strongly coupled to a GsAs microdisk cavity.\cite{srinivasan2007linear}. A tapered optical fiber is used to perform spectroscopic measurements. (g) 1D emitter in 1D cavity - Emission properties of a carbon nanotube are controlled by coupled it to a silicon microdisk resonator.\cite{imamura2013optical} (h) 2D emitter in 1D cavity - WS$_2$ monolayer integrated inside HSQ/Si$_3$N$_4$ microdisk cavity for excitonic lasing emission due to coupling with whispering gallery modes.\cite{ye2015monolayer} (i) 0D emitter in 2D cavity - Inherent light-emitting G-center defects in crystalline silicon coupled to quasi-bound state in continuum (BIC) resonance supported by the silicon metasurface/metaoptical cavity. \cite{zhu2020manipulating} (j) 2D emitter in 2D cavity - Rydberg excitons in Monolayer $WSe_2$ strongly coupled to a Fabry-Perot cavity formed between a DBR and a silver reflector \cite{gu2021enhanced}}
\label{Current}
\end{figure}

The number of dimensions of the cavity and the emitter has an important role to play in the strength of light-matter coupling. In general, a smaller number of cavity dimensions implies smaller mode volumes. Cavity-emitter coupled systems have been explored for different dimensions of cavity and emitter (Fig. \ref{Current}(a)-(i)). We present some of the prominent results for various cavity dimensions in what follows.

\textbf{Emitters inside zero-dimensional (0D) cavity}

Zero-dimensional photonic cavities provide the strongest photon confinement with mode volumes close to the diffraction limit ($(\lambda/2n)^3$) and typical Q-factors higher than 10$^4$\cite{lalanne2008photon, asano2017photonic}. The most widely used cavities include two-dimensional photonic crystal cavities, nanobeam cavities, nanopillar cavities, and fiber-distributed Bragg reflector (DBR) cavities.\cite{fryett2018encapsulated,bajoni2008polariton, besga2015polariton} A large portion of the cavity QED research in solid-state systems is based on epitaxially grown low-dimensional semiconductors coupled with 0D photonic crystal cavities. These include zero-dimensional (quantum dots), one-dimensional (nanowires), and two-dimensional (quantum wells) semiconductors. These materials are usually grown by molecular beam epitaxy or metalorganic chemical vapor deposition.

Semiconductor quantum dots (QDs) are nanocrystals that possess discrete electronic states due to the three-dimensional confinement of charge carriers. Epitaxial/self-assembled quantum dots (EQDs) form when a semiconductor is epitaxially grown over a substrate with a slight lattice mismatch\cite{garcia2021semiconductor}. The strain due to lattice mismatch leads to the formation of isolated nanoscale islands of the semiconductor.. Among different material systems, III-V-based In(Ga)As EQDs inside (Al)GaAs have been explored most widely. The primary usage for these EQDs had been as single photon sources where the emitters are excited by an above-band laser, and the EQDs emit single photons at the resonant wavelength\cite{buckley2012engineered}. While these EQDs can emit single photons on their own when pumped with an above-band laser, the probability of collecting a single photon is less than unity. Additionally, the relaxation of the carriers from the above-band excitation is probabilistic, giving rise to dephasing and subsequently distinguishable single photons. By coupling the QD transition to a nano-cavity, the decay rate of the exciton can be Purcell enhanced. By enhancing the decay rate significantly faster than the dephasing rate, pure, indistinguishable, and deterministic single-photon emission in the near-infrared can be generated \cite{somaschi2016near,michler2000quantum, senellart2017high}. These single photon sources generally operate in the weak coupling regime. We emphasize that single photon sources do not guarantee qNLO or photon blockade. While in both cases we measure the purity of the light from the cavity-emitter system, in a single photon source the pump laser and the emitted photon are generally at different wavelengths, whereas for qNLO, we essentially filter out a single photon from the pump laser.

Single QDs can be strongly coupled to 0D photonic crystal cavities and Rabi splitting larger than 300 $\mu$eV have been achieved at cryogenic temperatures\cite{ota2018large}. The coupling strength, $g$ in the aforementioned work was estimated to be around 165 $\mu$eV which was significantly larger than the cavity decay rate (28 $\mu$eV) and the exciton linewidth smaller than 18 $\mu$eV (measurement limited by spectrometer resolution). Together these parameters would imply a cooperativity larger than 27. In a photonic crystal cavity, exploiting the small mode-volume, researchers demonstrated photon blockade \cite{faraon2008coherent,reinhard2012strongly}, although the resulting $g^2(0)$ remains high ($>0.9$). 
Strong coupling of EQDs excitons and 0D photonic crystal cavities was experimentally demonstrated in multiple works.\cite{khitrova2006vacuum,yoshie2004vacuum}. The first demonstration of qNLO in solid-state systems was with InGaAs QDs inside two-dimensional photonic crystal cavity in GaAs\cite{faraon2008coherent}. Single photon blockade from a laser adjusted to polariton level and photon-induced tunneling from a laser adjusted to bare cavity resonance were observed, respectively. A similar result was obtained in another work.\cite{reinhard2012strongly} And in one work, photon-induced tunneling was shown for photons slowed down due to high group index in a photonic crystal waveguide with an EQD.\cite{javadi2015single}\par

Micropillar cavities are formed inside micropillars between two DBRs. Micropillars provide the confinement in two lateral directions by virtue of having a higher refractive index than the surrounding medium (usually air/vacuum) while the DBRs provide the confinement in the third dimension \cite{vahala2003optical}.  EQDs coupled with micropillar cavities can be monolithically grown and easily integrated compared to other 0D cavities. While EQD-Nanopillar systems can reach strong coupling as well\cite{schneider2016quantum,reithmaier2004strong}, the QD emission from these cavities is highly directional along the cavity axis based on two distributed Bragg reflectors. The out-of-plane emission makes this platform more compatible with free-space optics but, unsuitable for on-chip integrated devices. Several other emitters including solution-processed materials such as colloidal quantum dots (CQDs) and nanoplatelets (NPLs), 2D materials have been integrated in photonic crystal cavities, nanobeam cavities, and micropillar cavities.\cite{chen2018deterministic,yang2017room,ganesh2007enhanced, wu2007enhanced,fryett2018encapsulated,yu2020optical}However, qNLO has not been shown in these cavities as of now.

Fiber-DBR cavities are another type of 0D cavities in which large cooperativities have been observed. One end of these cavities is a flat DBR mirror, while the other end is a curved fiber end with DBR \cite{besga2015polariton,pfeifer2022achievements}. These cavity resonances can be tuned by controlling the distance between two cavity mirrors and the emitted light is directly coupled to an optical fiber \cite{pfeifer2022achievements}. Recently, near-perfect blockade ($g^2(0) \approx 0.09$) has been demonstrated using a single EQD coupled to a fiber-DBR cavity with a cooperativity close to 150\cite{najer2019gated}. The EQD used in this demonstration showed a transform-limited exciton linewidth which was made possible by gating the EQD in a p-i-n structure along with surface passivation of the p-GaAs layer to reduce the surface-related absorption.Some promising results for qNLO include the use of a single aromatic hydrocarbon molecule inside a host crystal. Strong light-matter interaction has been achieved from a single DBT molecule inside anthracene\cite{pscherer2021single}. In this system, qNLO was shown in the form of saturable absorption and single-photon switching. 

\noindent Other emitters integrated in these cavities for qNLO include epitaxial quantum wells (QWs) which can be grown directly on the DBRs. Quantum wells are a few nm thick semiconductors inside a material of higher bandgap which restricts the charge carriers in two dimensions. The most widely used materials for epitaxial QWs are III-V semiconductors such as GaAs inside AlAs. Other than these, II-VI oxides such as ZnO inside ZnMgO can also be grown and have significantly higher exciton binding energies, which enables room temperature operation \cite{makino2005optical}. Strong exciton-exciton interactions in QWs make them promising candidates to realize polariton blockade. One prominent result in in that direction has shown modest polariton blockade with $g^{(2)}(0) = 0.95$\cite{delteil2019towards}. \par 

\noindent A monolithic approach in 0D cavities involves utilizing the single-photon emitting defects of the cavity material itself has been demonstrated by Sipahigil et. al. for diamond\cite{sipahigil2016integrated}. Silicon vacancy defects were created inside a high-Q freestanding diamond nanobeam cavity with 40 nm precision using ion beam implantation. Using a single cavity-coupled defect, single-photon switching of a weak probe pulse was demonstrated for probe intensities in the single-photon regime. Another work utilized a low-loss 4H-SiC-on-insulator platform\cite{lukin20204h}. The Purcell-enhanced single-photon emission from a single silicon vacancy coupled to high-Q ($7.8 \times 10^5$) nanobeam cavity was shown along with nonlinear frequency conversion. Still, none of those defects in SiC have shown strong coupling yet.

\noindent To summarize, EQDs and epitaxial quantum wells inside 0-dimensional cavities have shown great promise for qNLO. The demonstration of perfect photon blockade shows that it is fundamentally possible to create a solid-state qNLO system\cite{najer2019gated}. However, the random spatial and spectral distribution (due to inhomogeneous size of the QDs) are big hurdles for the scalability of EQD-based qNLO platform. Also, it is difficult to use EQDs for hybrid integration since these emitters are present inside a host semiconductor. Many recent works have tried to achieve deterministic growth and deterministic coupling for monolithic and hybrid integration\cite{pregnolato2020deterministic, choi2023single, sunner2008scalable, han2021ordered, zhang2022chip}. 

\textbf{Emitters inside one-dimensional (1D) cavity}

1D cavities include dielectric micro-resonators, such as microrings and microdisks, that support whispering gallery modes (WGMs). \cite{foreman2015whispering}. Among these, microdisks have primarily been used to enhance light-matter coupling. WGMs are often collected with evanescently-coupled tapered optical fibers. These resonators can have extremely high-Q factors ranging from values of $~10^6$  for GaAs to $~10^8$ in certain silica microresonators\cite{pollinger2009ultrahigh,guha2017surface}. WGM resonators can also be reconfigured by thermal, mechanical, and electro-optical means\cite{strekalov2016nonlinear,von2001tunable}. Such tunability is of huge experimental importance to compensate for the fabrication errors.
Carbon nanotubes coupled with microdisks showed brighter single-photon emission\cite{imamura2013optical}. Multiple works on TMDCs coupled with microdisk cavities have demonstrated low-threshold lasing\cite{ye2015monolayer,salehzadeh2015optically}. Other than these, a lot of other emitters such as CQDs, EQDs, QWs and perovskite nanocrystals inside WGM resonators showed lasing\cite{kryzhanovskaya2014whispering,yan2020stable,li2020stable,duan2022ultralow,wang2017robust}. EQDs can be monolithically integrated inside GaAs microdisks for deterministic single photon emission\cite{michler2000quantum}.
There are very few works exploring coherent light-matter coupling inside WGM resonators. Strong coupling of excitons in a single EQD inside a microdisk was demonstrated by Srinivasan et al\cite{srinivasan2007linear}. Exciton-polariton formation was also observed in ZnO microrods with hexagonal cross-sections that support ‘quasi-WGM’ modes\cite{sun2010quasi}. More recently, Brooks et al. used an on-chip integrated photonic circuit with a single EQD inside GaAs microdisk\cite{brooks2021integrated}. It was shown that the microdisk-EQD structure coherently switches the photon scattering from one output port to another at the pump flux of a single photon per lifetime. This is an encouraging result for the implementation of qNLO on a chip.\\

\textbf{Emitters inside two-dimensional (2D) cavity}

DBR-based microcavities confine light in two dimensions. Planar DBRs act as the cavity mirrors and in some cases one of the mirrors is formed out of a metal. These microcavities have been at the forefront of exciton-polariton studies in QWs, 2D materials, and perovskites. Fabrication of DBR cavities is lithography-free and hence scalable to large areas. EQWs could be integrated in DBR cavities in a monolithic fashion and support exciton polaritons\cite{savona1995quantum,skolnick1998strong,weisbuch1992observation}. Unlike GaAs-based QWs, UV emitting GaN/AlGaN QWs can achieve strong coupling even at room temperature due to larger exciton binding energy compared to the thermal energy\cite{feltin2006room}. 
2D TMDCs in DBR cavities have emerged as an attractive platform for room temperature polaritons\cite{liu2015strong,zhao2020realization}. Furthermore, strongly interacting excited-state/Rydberg polaritons were observed in WSe2 monolayers, which may lead to polariton blockade for higher quantum numbers\cite{gu2021enhanced}.  Perovskite crystals can be integrated inside these 2D cavities by growing them over DBR or dry-tape transfer\cite{wu2021nonlinear,fieramosca2019two,su2021perovskite}. Perovskites can reach the strong coupling regime at room temperature inside DBR microcavities\cite{su2021perovskite}. A few recent studies have demonstrated polariton-polariton interactions in these structures\cite{wu2021nonlinear,fieramosca2019two}. Rydberg exciton-polaritons were observed in a 2D cavity with  CsPbBr$_3$ sandwiched between one DBR and one metal mirror at temperatures lower than 150 K\cite{bao2019observation}. These observations demonstrate perovskites are a promising platform for observing qNLO processes.
Another kind of 2D cavity is metasurfaces that support guided mode resonances \cite{li2018metasurfaces}. Metasurfaces also provide control of far-field emission directionality\cite{chen2020metasurface,zhang2017unidirectional}. Strong coupling was observed for monolayers of WS$_2$ and WSe$_2$ coupled with guided mode resonances of 1D and 2D dielectric gratings in SiN at room temperature and 110 K respectively\cite{zhang2018photonic}. Similarly, for a 2D grating/metasurface with integrated monolayer WSe$_2$, strong coupling was demonstrated at 22 K with ~18 meV Rabi splitting\cite{chen2020metasurface}. Many metasurfaces also support quasi-bound states in the continuum (q-BIC) resonances, which can have high Q-factors\cite{koshelev2019meta}. The research on cavity-emitter interactions in these structures is still at a nascent stage. However, strong interactions with BIC are theoretically predicted for various cavity-emitter systems such as 2D TMDCs in metasurfaces\cite{al2021enhanced} and all perovskite platforms\cite{al2022strong}.

\textbf{Emitters inside Plasmonic cavities}

Photonic cavities are restricted by the diffraction limit to mode volumes $V_m > (\lambda/(2 n))^3$, where $\lambda$ is the free-space wavelength of the confined light and $n$ is the refractive index of the medium inside the cavity, which ultimately sets a limit on the coupling strength $g$ that can be achieved in these cavities. In order to obtain higher coupling strengths, light can be coupled to the coherent oscillations of free electrons in metals or plasmons \cite{Pelton2013}. Plasmonic cavities can be 2D by employing propagating surface plasmon polaritons (SPPs) on metal films, 1D by employing SPPs on metal nanowires, or 0D by employing localized surface plasmon resonances (LSPRs) on metal nanoparticles. Single-photon switching has been proposed based on both SPPs \cite{chang2007single} and LSPRs \cite{gullans2013single}. 

Strong coupling to SPPs was first observed for layers of dye molecules on silver films \cite{Pockrand1982SPP}. Larger splitting was obtained using layers of J-aggregates \cite{Bellessa2004Jagg}, taking advantage of their larger transition dipole moments and narrower excitonic linewidths. Ensembles of J-aggregates were also used to demonstrate strong coupling to ensembles of LSPRs in patterned metal films \cite{dintinger2005strong,sugawara2006Jagg} and in chemically synthesized metal nanoparticles \cite{Wiederrecht:2004kx,fofang2008plexcitonic}. These demonstrations were extended to demonstrations of strong coupling between single metal nanoparticles and ensembles of J-aggregates \cite{delacy2015coherent,das2017exploring,kirschner2017phonon,liu2017strong}. Further experiments were also able to demonstrate strong coupling between individual metal nanoparticles and excitons in 2D transition metal dichalcogenides \cite{zheng2017manipulating,cuadra2018observation,lawless2020influence} and lead halide perovskites \cite{muckel2021tuning}.

\begin{figure}[p]
\centering
\includegraphics[scale=0.35]{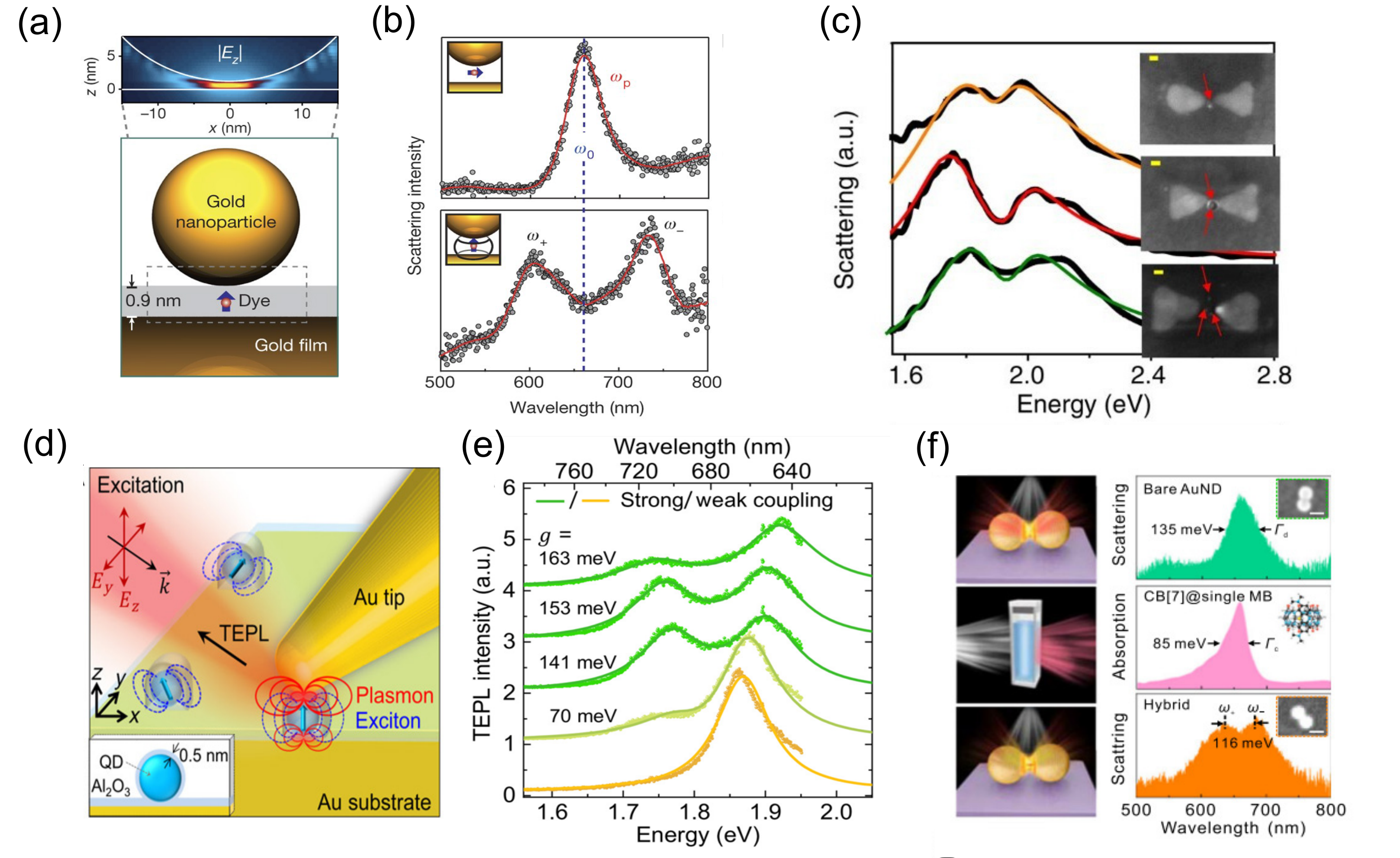}
\caption{Solid-state Cavity QED with Plasmonic Cavities. (a,b) Single dye molecule in a nanoparticle on mirror gap plasmon cavity. (a) Single dye molecule coupled to a gold nanoparticle-on-mirror cavity. The inset shows the simulated value of the vertical polarization component of the electric field in the gap between the gold nanoparticle and gold film, with a maximum field enhancement of 400. (b) Observation of strong coupling for several dye molecules in the nanoparticle-on-mirror cavity. Splitting in the scattering spectrum is seen when the transition dipole moments of the dye molecules are oriented vertically (lower panel) but not when they are oriented horizontally. \cite{chikkaraddy2016single} (c) Scattering spectra showing vacuum Rabi splitting for one, two and three colloidal quantum dots in a silver bowtie cavity. Black lines represent the experimental data and  the colored lines are fits to a coupled oscillator model. The scale bar is 20 nm (yellow). \cite{santhosh2016vacuum} (d,e) Tip-enhanced strong coupling geometry with metal scanning-probe tips. (d) A single quantum dot with a 0.5 nm alumina capping layer located in the strongly confined field between a tilted gold scanning-probe tip and and a gold film. (e) Tip-enhanced photoluminescence (TEPL) spectra of multiple quantum dots with coupling strength varying from 70 meV (at the onset of the strong coupling regime) to 163 meV (well into the strong-coupling regime) \cite{park2019tip} (f) Rabi splitting observed by deterministically placing a single methylene blue molecule in a gold dimer using a cucurbit[7]uril (CB[7])-mediated self-assembly technique. The scale bar in the inset SEM images is 50 nm.\cite{Liu2024dimer}} 
 \label{Plasmonics}
\end{figure}  

The main disadvantage of plasmonic nanocavities is the ohmic losses associated with electron oscillation in the metals as well as the strong radiation of the plasmons to the far field; together, these limit quality factors of plasmonic cavities to the range of 5 to 10. Ensembles of emitters were thus required to demonstrate strong coupling to single metal nanoparticles despite the sub-diffraction-limited mode volumes. In order to overcome these losses and achieve strong coupling at the single-emitter level, it is necessary to confine fields even further by coupling metal nanoparticles to one another or to metal films. For nanoparticle-nanoparticle or nanoparticle-film separations on the order of a few nanometers, mode volumes can be as low as $10^{-6}$ of the free-space diffraction limit, enabling a single emitter to provide high cooperativity and strong coupling despite the low $Q$s. Coupling strengths can be on the order of 100 meV or greater so that strong coupling can be observed at room temperature, compared to the cryogenic temperatures generally required for 0D photonic cavities, opening up the possibility of room-temperature qNLO.

Strong coupling and polariton formation at the single emitter level has been observed at room temperature for single molecules \cite{chikkaraddy2016single,chikkaraddy2018mapping,ojambati2019quantum} (Fig. \ref{Plasmonics}(a,b)) and single QDs \cite{leng2018strong,li2022room} between metal nanoparticles and metal films, for single QDs between metal scanning-probe tips and metal films \cite{gross2018near,park2019tip} (Fig. \ref{Plasmonics}(d,e)), for single QDs between pairs of nanofabricated gold nanoparticles \cite{santhosh2016vacuum,gupta2021complex} (Fig. \ref{Plasmonics}c), and for single dye molecules chemically assembled in the gap between pairs of gold nanoparticles \cite{Liu2024dimer}(Fig. \ref{Plasmonics}f).

Even though several reports of single-emitter strong coupling involving plasmonic nanocavities have been made, there have not yet been reports of qNLO phenomena in these systems. The lossy nature of the plasmonic cavities remains a significant obstacle, motivating the consideration of hybrid plasmonic-photonic approaches, described in the "Emerging Cavities" section.

\section{Emerging Materials}
In recent times, a whole range of solid-state emitters have been pursued as potential candidates for qNLO with a focus on ease of integration, scalability and desired emission properties. We discuss some of the most prominent ones in this section.\par

\textbf{Solution-Processed Materials}\par 

\begin{figure}[p]
\centering
\includegraphics[scale=0.48]{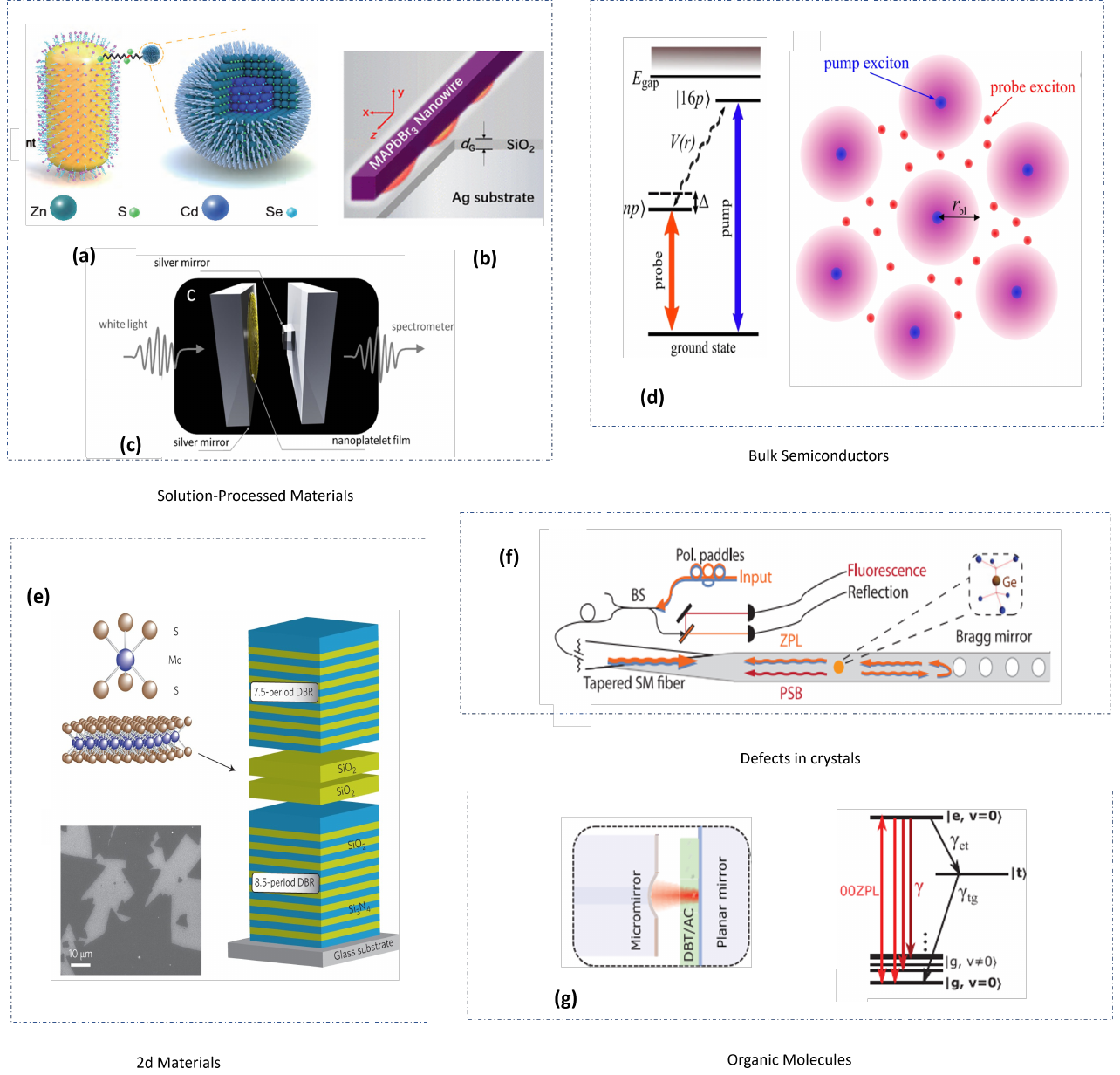}
\caption{Emerging Materials for quantum nonlinear optics and their integration in cavities - Various types of solution processed materials - (a) Single CdSe quantum dot strongly coupled to a gold nanorod at room temperature.\cite{li2022room} (b) A MAPbBr\textsubscript{3} nanowire over a silver film separated by a thin SiO$_2$ layer. The nanowire with flat facets itself acts as a photonic cavity while, plasmonic silver substrate enhances the exciton-photon coupling and a Rabi splitting of 564 meV was observed.\cite{shang2018surface} (c) Colloidal CdSe nanoplatelets integrated in a microcavity formed by two silver mirrors. The structure can exhibit strong-coupling at room temperature \cite{flatten2016strong}. (d) Bulk Materials supporting Rydberg excitons - Example shown for bulk cuprous oxide (Cu$_2$O). The asymmetric Rydberg blockade was observed in a pump-probe measurement in which tw different exciton Rydberg states were coupled to pump and probe beams\cite{heckotter2021asymmetric} (e) 2d Materials - Example shown of a monolayer MoS$_2$ coupled strongly with a DBR cavity\cite{liu2015strong}. (f) Defects/color centers in crystals - A monolithic platform for quantum nonlinear optics including Germanium vacancy diamond waveguide with Bragg mirror at one end coupled to a tapered single mode optical fiber \cite{bhaskar2017quantum} (g) Organic molecules - Single organic molecule dibenzoterrylene (DBT) strongly coupled to a fiber-DBR cavity. The system exhibits optical switching for incident intensity of single-photon\cite{pscherer2021single}. The level structure for the molecule is also shown.}
\label{Materials}
\end{figure}

Excitonic solution-processed materials such as CQDs, NPLs, and perovskite nanomaterials host excitons with large binding energies and dipole moments (Fig. \ref{Materials}(a), (b), and (c)). Unlike EQDs, solution-processed materials can be readily integrated into a wide range of cavities since they do not require lattice matching with the substrate. Techniques such as spin-coating and drop-casting could be used for integration. Additionally, multiple methods for deterministic placement have been developed for these materials\cite{eich2022single, nguyen2022deterministic, barelli2023single, chen2018deterministic}. The synthesis of colloidal nanoparticles is relatively simple and due to their size dependence the emission properties are tunable \cite{litvin2017colloidal}. CQDs can be fabricated from a large variety of materials with single or multi-component compositions\cite{kagan2020colloidal}. CQDs have been extensively used as single-photon sources integrated with a variety of cavities and waveguides on-chip \cite{eich2022single,kagan2020colloidal,chen2018deterministic,kahl2007colloidal, abudayyeh2021single}. Strong coupling with single CQDs in plasmonic cavities has been observed at RT\cite{santhosh2016vacuum}. Colloidal NPLs are quasi-2D semiconductors that provide quantum confinement of charges in one dimension similar to QWs. The most widely explored NPLs are composed of II-VI semiconductors such as CdSe and CdS\cite{diroll20232d}. NPLs have large oscillator strengths in addition to exciton binding energies in 100-300 meV range, making them attractive candidates for enabling strong light-matter interaction\cite{naeem2015giant}. Furthermore, the linewidths are narrower compared to QDs due to reduced inhomogeneous broadening\cite{chen2022integrated}. Polariton formation at room temperature has been demonstrated for an ensemble of CdSe NPLs inside a microcavity\cite{flatten2016strong}.  \par

Perovskite nanomaterials are another well-studied class of light-emitting quantum materials. Solution-processed perovskite materials can be grown with various dimensionalities in the form of QDs/nanocrystals, nanowires, NPLs, and thin films. Lead halide perovskites are the most widely studied platform, which has a chemical formula \(APbX_3\), in which the A is a monovalent cation, e.g. \(Cs^+\), \(CH_3NH_3^+\), and X is a halide anion, e.g. \(Cl^-\), \(Br^-\), \(I^-\). Lead halide perovskite QDs show high photoluminescence (PL) quantum yield and room temperature single-photon emission \cite{chen2019luminescent,li2019purcell,zhu2022room}. These QDs integrated in nanocavities have been used for on-chip lasers\cite{he2020cmos}.  Perovskite materials can themselves function as cavity material for monolithic integration. These cavities can have Q-factors of the order of a few thousand and have been used for lasers with low thresholds\cite{zhu2015lead, kumar2022perovskite}. Such cavities also facilitate strong exciton-photon couplings resulting in the formation of polaritons\cite{zhang2018strong}. Excitons in perovskite NPLs possess large oscillator strength and binding energies. Due to this, these materials are very promising for room-temperature polariton formation\cite{su2021perovskite}. Compared to other two-dimensional emitters such as QWs and II-VI NPLs, the oscillator strength was found to be higher while an exciton binding energy of 120 meV was observed  for CsPbBr$_3$ NPLs\cite{li20172d}. PL emission is tunable as it depends strongly on the number of monolayers in the NPL\cite{otero2022colloidal}. \par      

While solution-processed materials possess excellent emission properties and flexibility of integration, qNLO has not yet been achieved with these materials. Emission instability and large linewidths are major hurdles towards the realization of qNLO, along with difficulty in deterministic placement\cite{chen2022integrated}. Deterministic positioning of ensembles of CQDs on photonic cavities using lithographic techniques and electrohydrodynamic inkjet printing have been demonstrated, but reliable positioning of single CQDs is still under investigation \cite{chen2018deterministic, cohen2022direct}. One route is to fabricate photonic cavities with pockets designed to trap single CQDs with high efficiency \cite{froch2020photonic}. Another is to use lift-off techniques to position single CQDs in lithographically patterned windows, where high yield is either achieved by iterative positioning and integration steps or using “giant” CQDs integrated in windows with capillary-assisted deposition techniques\cite{eich2022single, nguyen2022deterministic, barelli2023single}. Deterministic integration of CQDs with a technique agnostic to the underlying photon structures still remains to be demonstrated.\par

\textbf{Color Centers in Solids}\par 

Another emerging class of materials is color centers in solids. There has been tremendous progress in these color centers for single photon sources or quantum memories. However, their applicability to the qNLO remains limited.

The presence of defects in various insulators and semiconductors can lead to the creation of new electronic energy levels inside the bandgap. The electron transitions due to these new levels can be fluorescent hence these defects are referred to as ``color centers." The most prominent color centers include defects in diamond, silicon carbide (SiC), and hexagonal boron nitride (hBN)\cite{ruf2021quantum,castelletto2020silicon,caldwell2019photonics}. With bandgaps in UV, these color centers can emit light in a large spectral region from UV to the near-IR.  The single photon emission from color centers tends to be robust and potentially very bright. These defects can also have unpaired electrons which have potential applications as spin qubits\cite{ruf2021quantum}. NV centers in diamond are perhaps the most studied color center. Its single-photon emission properties are well-studied, but the emission fraction in the ZPL is only 4$\%$ at room temperature. Therefore, a lot of work on diamond-based single photon emitters instead focuses on group IV defects such as silicon vacancies (SiV) and germanium vacancies (GeV)\cite{iwasaki2015germanium,sipahigil2016integrated} In addition to bright emission in the ZPL, the ZPL linewidth is nearly lifetime-limited at low temperatures\cite{iwasaki2015germanium,sipahigil2016integrated}.\par 
SiC can also host a variety of emitters in NIR\cite{castelletto2021silicon}. SiC can be found in a number of different polytypes such as 4H-SiC, 3C-SiC, and 6H-SiC. Some of the defects in SiC show bright emission at RT. One such emitter of unconfirmed origin was observed to emit in the telecom with a rate of 1 MHz\cite{wang2018bright}. Carbon antisite-vacancy pairs in 4H-SiC were shown to emit at the rate of 2 MHz in the visible band\cite{castelletto2014silicon}. Negative SiVs in 4H-SiC were shown to emit highly indistinguishable photons in the ZPL at low temperature due to lifetime-limited linewidths\cite{morioka2020spin}, but the photon emission rate for these defects was low\cite{castelletto2021silicon}. Additionally, SiC is compatible with wafer-scale fabrication and the fabrication techniques are well-researched \cite{castelletto2020silicon}. Defects in hBN have been studied as single photon emitters both in bulk and single/few layers. These emitters cover a large spectral region due to the large bandgap (6 eV) of hBN\cite{caldwell2019photonics}. Further, these emitters can be deterministically generated using strain in few-layered hBN\cite{proscia2018near}. Stable and bright single photon emission with more than 7 MHz emission rate was observed as an hBN flake\cite{grosso2017tunable}. Single photon emitters in 2D hBN can have transform-limited linewidths even at room temperature due to decoupling of color center from the phonons \cite{dietrich2020solid}.\par
          
Color centers in diamond and SiC have been studied in both bulk materials and nanocrystals. Nanocrystals provide the opportunity for straightforward hybrid integration with photonic structures. Defects in bulk are ideal for monolithic integration which include fabrication of cavities in the material hosting defects itself (Fig. \ref{Materials}(f))\cite{sipahigil2016integrated,lukin20204h,kim2018photonic, bhaskar2017quantum}. Many techniques such as ion-implantation and femtosecond laser irradiation can be used to deterministically generate color centers of various types in diamond and SiC with $<$ 20 nm lateral accuracy \cite{zhou2018direct,riedrich2015nanoimplantation,smith2019colour,castelletto2019deterministic,kraus2017three}. Despite promising single-photon emission properties, the defect center emission is hardly tunable as compared to SPMs, and the dipole moments are relatively small.\par

\textbf{Two-dimensional Materials}\par 

Two-dimensional materials are monolayer crystals. Beginning with graphene, a variety of materials have been prepared in 2D form\cite{miro2014atlas}. For the purposes of cavity-exciton interactions the most interesting 2D materials are semiconducting TMDCs. TMDCs have the molecular formula MX$_2$ where M is a transition metal such as Mo or W, while X is a chalcogen such as S, Se, or Te\cite{raja2017coulomb}. TMDCs support excitons with large exciton binding energies of the order of 100 meV, making room temperature polaritonic operations feasible\cite{ponraj2016photonics}. TMDCs are indirect bandgap semiconductors in bulk, but have a direct bandgap in the monolayer form showing strong light absorption. Furthermore, the dipole orientation of excitons is always in-plane, which allows deterministic coupling to cavities. TMDCs can be transferred to various types of nanocavities. Strong-coupling in TMDCs has been possible for a variety of micro/nano-cavities (An example shown for DBR cavity in Fig. ~\ref{Materials}(e))\cite{al2022recent, liu2015strong}. Nonlinear exciton-polariton interactions have been observed in 2D TMDCs integrated in cavities. These interactions were found to be 10 to 100 times stronger for charged polaritons compared to those forN neutral polaritons\cite{emmanuele2020highly}. It is also possible to pattern 2D materials using lithographic techniques to form islands which restrict the spatial extent of excitons and increase the exciton-exciton interaction strength\cite{ryou2018strong}.\par

TMDC monolayers also host a number of defects that can function as single photon emitters at cryogenic temperatures\cite{tonndorf2015single, chakraborty2019advances}. Deterministic creation of these defects has been shown using methods such as strain engineering and ion implantation\cite{choi2016engineering,branny2017deterministic,chakraborty2019advances}. 

\textbf{Rydberg excitonic materials}\par 

Various materials are being explored that host excited state/Rydberg excitons which can have orders of magnitude stronger exciton-exciton interactions compared to ground state excitons\cite{walther2018giant}. Rydberg excitons have been observed in TMDC monolayers such as up to 5s in WS$_2$\cite{chernikov2014exciton}, up to 3s in MoTe$_2$\cite{biswas2023rydberg}, and highest 11s for WSe$_2$\cite{wang2020giant}. Unfortunately, Rydberg states with a principal quantum number ($n$) larger than 2 have very weak oscillator strengths and short lifetimes for most materials\cite{liu2019magnetophotoluminescence}. Therefore, their observation requires large external magnetic field (typically of the order of 10-100 Tesla) fields which create $n$-dependent diamagnetic shifts\cite{liu2019magnetophotoluminescence}. Polariton formation for 2s excitons in WSe$_2$ was demonstrated in two-dimensional metal-DBR microcavity\cite{gu2021enhanced}. The polariton-polariton interaction for 2s polaritons was estimated to be 4.6 times stronger than the 1s polaritons. Higher exciton states in cavities with smaller mode volumes and higher Q-factors could facilitate single photon Rydberg blockade in TMDC monolayers.
In lead halide perovskites, excitonic Rydberg states have been observed without applying any external fields\cite{bao2019observation,luo2017ultrafast}.
A very promising emerging material for Rydberg blockade is bulk cuprous oxide (Cu$_2$O). Cu$_2$O single crystals have been shown to host giant Rydberg excitons with $n$ as large as 25 at 1.2 K\cite{kazimierczuk2014giant}.
Rydberg blockade has been observed in bulk Cu$_2$O, although demonstrations were far from the single-photon regime (Fig. \ref{Materials}(d)\cite{khazali2017single,heckotter2021asymmetric}. Rydberg excitons in bulk Cu$_2$O were shown to strongly couple to modes of a DBR microcavity.\cite{orfanakis2022rydberg} The polaritons were observed up to quantum number $n=6$. It was estimated that the blockade radius for $n=6$ in this system is larger than 100 nm while the same would be 1 $\mu$m for $n=15$. By reducing the excitation spot size and cavity mode volume, it should be possible to achieve Rydberg blockade at the few/single photon level using Cu$_2$O. Additionally, Rydberg excitons up to $n=10$ have been observed in synthetic Cu$_2$O crystals .\cite{lynch2021rydberg} Synthetic Cu$_2$O could also be grown as thin films by oxidizing a sputtered/evaporated copper thin film. Rydberg excitons up to $n=7$ were observed in 700 nm thick Cu$_2$O films.\cite{delange2023highly} Cu$_2$O thin films are very promising for integration in nanocavities.  \par

\textbf{Organic molecules}\par 

Aromatic organic molecules inside solid-state hosts (such as anthracene) have generated considerable interest for realization of strong light-matter interactions (Fig. \ref{Materials}(g). The zero-phonon transition between ground and first electronic states, also called 00ZPL, can have a Fourier-limited linewidth at cryogenic temperatures. Integrating these molecules inside optical cavities can lead to large emission in the ZPL and minimal in the red-shifted Stokes lines\cite{wang2019turning}. The coherent coupling regime and single-photon nonlinearity in the form of single-photon switching and saturation can then be achieved\cite{pscherer2021single}. Furthermore, thin films and nanocrystals with organic molecules have been developed for flexible integration inside different structures\cite{polisseni2016stable,hail2019nanoprinting}.\par

\section{Emerging Cavities}

Cavities for qNLO need to have a higher Q-factor and smaller mode volumes as we previously discussed. In this section, we will discuss emerging cavities to achieve these properties.

State-of the-art zero-dimensional cavities have mode volumes of the order of $(\lambda/n)^3$, where $\lambda$ is the wavelength in free space and $n$ is the effective refractive index of the cavity mode. But, there have been multiple experimental and theoretical works on dielectric cavities where much smaller mode volumes have been achieved. 

\begin{figure}[p]
\centering
\includegraphics[scale=0.5]{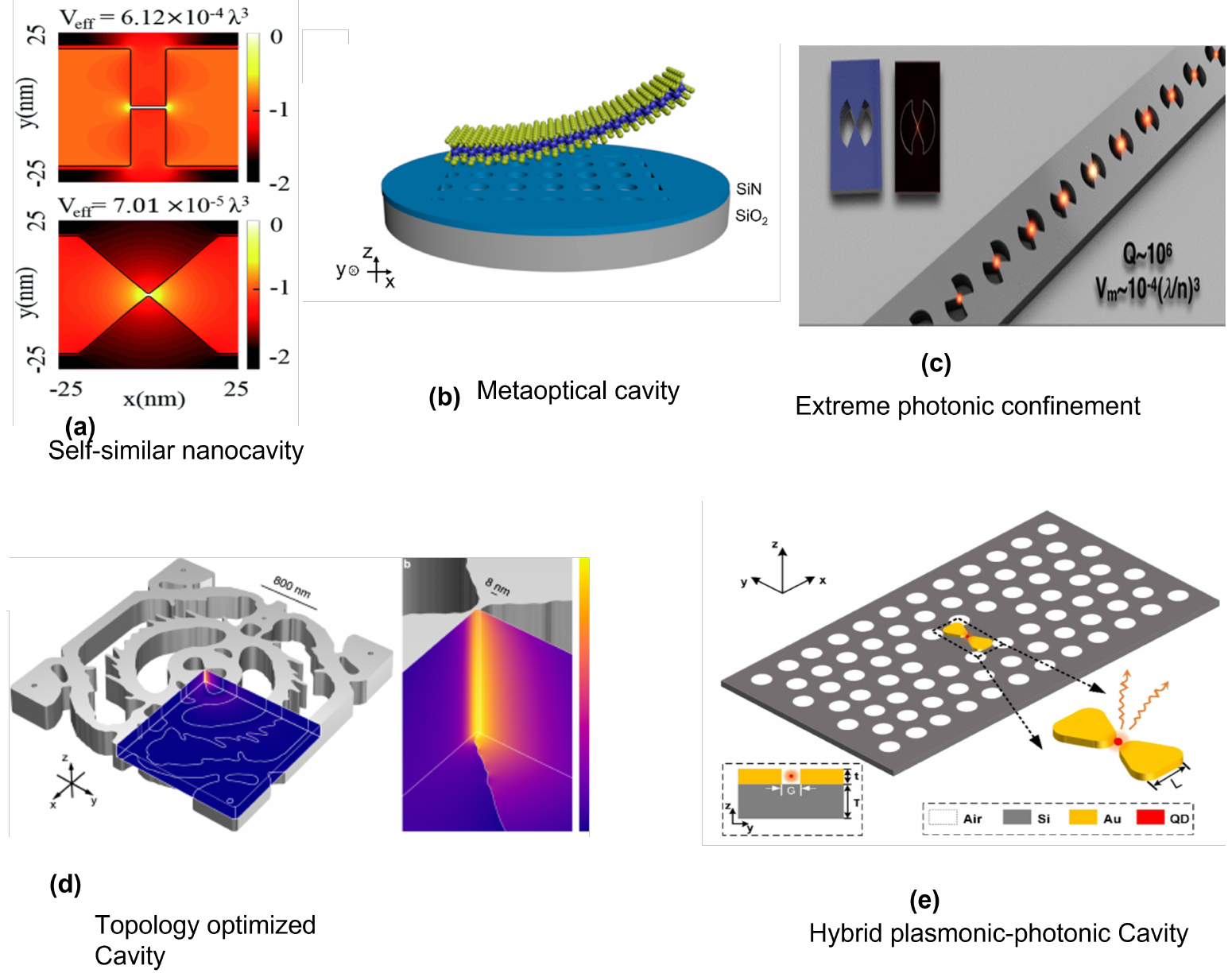}
\caption{Emerging Cavity Designs to achieve Single-Photon Nonlinearity - (a) Self-similar nanocavity.\cite{choi2017self} (b) A silicon nitride metaoptical cavity strongly coupled to monolayer WSe$_2$ excitons.\cite{chen2020metasurface} (c) Photonic bandgap realization gradual rotation of a bowtie unit cell in a nanobeam.\cite{hu2016design} The designed structure has a quality factor of around $10^6$ and a mode volume 4 orders of magnitude smaller than the diffraction limit. (d) Topology-optimized silicon cavity for ultralow mode volume\cite{albrechtsen2022nanometer}. (e) Hybrid metal-dielectric cavity consisting of a gold bowtie inside a silicon photonic crystal cavity.\cite{zhang2020hybrid}}
\label{Cavities}
\end{figure}

A wavelength-independent confinement mechanism was proposed based on the boundary condition for the normal component of the electric displacement field, $\epsilon_1E_1 = \epsilon_2E_2$. In a dielectric mode cavity, this leads to field enhancement inside a lower permittivity slot\cite{robinson2005ultrasmall}. This effect is termed as the `slot effect'. Smaller slots lead to stronger fields. The Purcell enhancement in the infinitesimally wide slot then increases by a factor of ${(\epsilon_2/\epsilon_1)}^{5/2} $. The same concept was used to design a nanobeam photonic crystal cavity with both high-Q ($8.2 \times 10^5$) and low mode volume ($0.0096 (\lambda/n)^3$) such that $Q/V_m$ exceeded $10^7(n/\lambda)^3$\cite{seidler2013slotted}. This method of field confinement only works for transverse magnetic (TM) polarized modes. For transverse electric (TE) polarized modes, the boundary conditions for the tangential component of the electric field result in the enhancement of the electromagnetic energy density inside higher permittivity `anti-slots'. These anti-slots must be oriented perpendicular to the wave-propagation direction. The repetitive application of interlocked slots and anti-slots leads to further localization of the energy. Infinite repetitions will give a bowtie-shaped unit cell. The bowtie field enhancement of 1500 times larger than the zero-order unit, i.e. unit cell without any slots/anti-slots was demonstrated by\cite{hu2016design}. Furthermore, the authors realized a photonic bandgap using bowtie unit cells with the axis gradually rotating from 0 to 180$^{\circ}$ along the length of the cavity. This designed cavity has a Q-factor greater than $10^6$ while having a mode volume of four orders of magnitude smaller than the diffraction limit. 
In a similar theoretical work, it was shown that the repetition of slot and anti-slot effects results in exponentially increasing the energy density\cite{choi2017self}. They used infinite repetitions to produce a bowtie cavity with $45^{\circ}$ taper angle (Fig. \ref{Cavities}(a)). For sub-nanometer gaps between bowties, computational results showed that it can become possible to observe room-temperature Kerr nonlinearity at extremely low power, even at few photon levels for materials such as organic molecules and indium tin oxide.

Recently introduced ``inverse design" of cavities involves finding the design that optimizes certain parameters such as mode volume. There are mainly two routes that are followed in inverse design - geometry optimization and topology optimization. In geometry optimization, the features of design have a fixed geometric shape and the optimization is performed on the size and/or placement of these geometric objects. The topology optimization process on the other hand explores a much larger domain of possible designs\cite{jensen2011topology}. The design is composed of pixels which are assigned values in space to optimize a particular function. 

Topological optimization to minimize the mode volume of cavities has led to different kinds of bowtie-shaped cavities. These designs often involve features that are beyond the reach of current nanofabrication capabilities. In a recent work, topology optimization was performed while taking fabrication constraints in consideration\cite{albrechtsen2022nanometer}. These cavities were fabricated in freestanding silicon films for telecom wavelengths and their design is shown in Fig. \ref{Cavities}(d). From measurements in near and far-field, the cavity mode volume was found to be 12 times smaller than the diffraction limit.

Sub-nanometer field confinement like in the designs above is not suitable for the integration of solution-processed emitters such as CQDs because of their finite sizes, but color centers or dopants such as rare earth ions could effectively couple to the cavity mode. 

There have been a few interesting works towards improving the Q-factors of guided mode resonances (GMRs) in dielectric metasurfaces. Typical Q-factors for these modes have been of the order of $10^2$. These low values have been a limiting factor for applications in qNLO \cite{zhang2018photonic, chen2020metasurface}. One way to improve the Q-factor is by using the phase gradient metasurfaces. In a recent work, a one-dimensional phase gradient metasurface was designed with each period having three bars of different widths. Along with it, one of the bars was made to have subwavelength perturbations. These gratings support GMRs with Q-factors of more than 2500 and were used for steering and splitting light beams \cite{lawrence2020high}. In another work, ultrahigh Q-factor ($2.39 \times 10^5$) was observed for a GMR in a silicon on insulator metasurface at telecom wavelengths\cite{huang2023ultrahigh}. The structure consisted of a patterend low-index photoresist on top of the silicon on insulator substrate. The ultrahigh-Q of this structure has been attributed to (i) reduction in surface roughness due to no dry etching and (ii) large refractive index contrast between the silicon  (3.46 refractive index) and the superstrate and substrate (refractive indices 1.4 and 1.46, respectively). This results in stronger field confinement in the silicon layer.

Another interesting direction in cavity design has been hybrid plasmonic-photonic cavities. The idea behind these is to utilize the extreme field confinement of plasmonic structures while minimizing the losses by using dielectrics for the majority of the structure \cite{palstra2019hybrid}. The earliest hybrid systems consisted of metal films integrated in a dielectric resonator\cite{min2009high}. It was shown that compared to pure plasmonic structures, a few orders of magnitude higher Q-factors were possible in hybrid cavities. However, the presence of metals still significantly reduced the Q-factor of the dielectric cavity mode. More recently, there have been several theoretical proposals for hybrid cavities that provide both strong field concentration and high quality factors by combining gold nanoparticles with whispering-gallery-mode cavities \cite{xiao2012strongly} or photonic-crystal cavities \cite{liu2017nanoantenna,barreda2022hybrid}. For example, it was proposed that that a metallic bowtie with a few nanometer gap could enhance $Q/V_m$ by a factor of 60 compared to a photonic crystal cavity while having a Q-factor of nearly $10^5$ \cite{zhang2020hybrid} (Fig. \ref{Cavities}(e)). 

There has been only a small amount of experimental work towards demonstrating the performance of such hybrid cavities for strong light-matter coupling. For example, a gold nanocylinder on top of a photonic-crystal cavity was shown to reduce the quality factor of the cavity to 900, which is still much larger than the typical plasmonic resonances \cite{barth2010nanoassembled}. Furthermore, the cavity field maxima moved from the dielectric to the surface of the nanocylinder which would lead to enhanced coupling with emitters on the surface. Using a pair of nanospheres with few nanometer gaps enhanced the cavity field by a factor of 34 while retaining a Q-factor of 720. A metal bowtie on top of a photonic-crystal cavity was shown to provide a $Q$ of 800 while giving $Q/V_m \sim 10^6 (n/\lambda)^3$ \cite{conteduca2017ultra}. Recently, gold-silver core-shell nanorods were coupled to leaky modes of a Fabry-Perot cavity, producing hybrid modes with increased quality factors that facilitated strong coupling to molecular J-aggregates \cite{Li2023exceptional}. Further development of these platforms may bridge the gap between plasmonic and photonic cavities, enabling qNLO at room temperature.

\section{Applications of Quantum Nonlinear Optics}

\begin{figure}[p]
\centering
\includegraphics[scale=0.5]{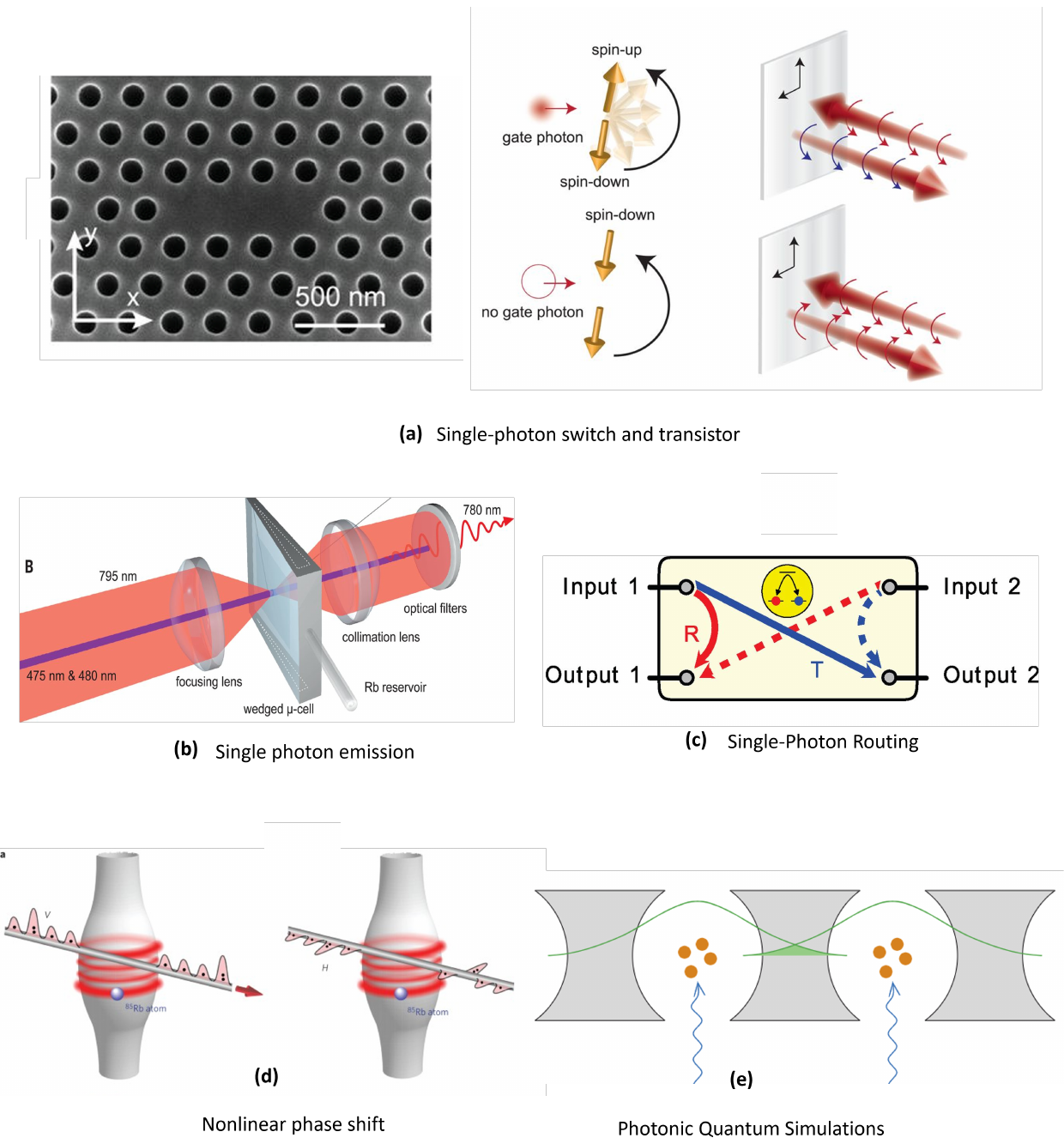}
\caption{Applications of Quantum Nonlinear Optics - (a) A solid-state single-photon switch and transistor with a charged quantum dot in a photonic crystal cavity \cite{sun2018single}. Presence/absence of a gate photon determines the spin-state of QD which in turn determines the spin-state of a reflected signal photon. (b)Experimental demonstration of single-photon generation due to Rydberg blockade in rubidium atoms at room temperature.\cite{ripka2018room} (c) A schematic of a single-photon router - A photon from two input ports could either be reflected or transmitted to two different output ports depending upon the state of the emitter. Image from Shomroni et al.\cite{shomroni2014all} (d) nonlinear $\pi$ phase-shift in a rubidium atom coupled to a bottle microresonator.\cite{volz2014nonlinear} (e) Coupled cavity arrays with strongly interacting polaritons could potentially be used for simulation of quantum many-body phenomena.\cite{hartmann2008quantum, hartmann2006strongly}}
\label{Applications}
\end{figure}

Many quantum technologies require nonlinearity at the single photon level such as quantum information processing and quantum networks. As we saw earlier, qNLO could also be used to generate single photons.
On-demand single-photon sources have applications in quantum key distribution, quantum metrology, and quantum computing \cite{shi2022high}. In photonic quantum networks, it is envisaged that photons act as the information carrier between individual nodes. At the nodes, the interaction between different photons would be used to implement quantum information processing \cite{kimble2008quantum}. Single-photon blockade processes can convert a light source to an antibunched stream of single photons. Single photons could also be generated via Rydberg blockade (Fig. \ref{Applications}b)\cite{ornelas2020demand, ripka2018room}. This process does not require isolated single emitters but instead strongly interacting collection of atoms/emitters.  
A theoretical study was performed to test the feasibility of Rydberg exciton blockade in Cu2O. It was shown that under cw excitation very strong antibunching can occur for n = 24 Rydberg levels inside a cavity in weak coupling regime\cite{khazali2017single}. 
There have been a few photon blockade demonstations in the solid-state with self-assembled QDs and $g^{(2)}(0)$ value up to $0.09$ had been shown.\cite{najer2019gated}

Nonlinear phase shifts at the quantum/single photon level could enable the development of a two qubit quantum phase gate, which is a universal gate for quantum computation\cite{rauschenbeutel1999coherent}. In one work, controlled phase-shifts between two modes of light was demonstrated with single self-assembled QDs inside a photonic crystal cavity. For a single control photon, the phase change was found to be around 12$^{\circ}$ \cite{fushman2008controlled}. For quantum logic gates, this shift needs to be $\pi$ which could be achieved by multiple interactions. More recently large phase shifts have been observed in several experiments with trapped atoms. Using a trapped rubidium atom coupled to a bottle microresonator, a two photon $\pi$ phase shift was observed (Fig. \ref{Applications}d)\cite{volz2014nonlinear}. The other experiment utilized Rydberg blockade with electromagnetically induced transparency in an atomic ensemble with stored excitation\cite{tiarks2016optical}. The control beam with less than a photon per pulse could be used to control the phase of a target beam.

Another application is switching of single photons. Single photon switching is a fundamental operation required for quantum networks. It involves using a gate photon to switch transmission of a different photon through some channel between two states (on and off). Switches based on interacting photons have been demonstrated using trapped atoms. Switching at a single photon level was realized using electromagnetic induced transparency combined with Rydberg blockade in ultracold rubidium atoms \cite{baur2014single}
Similarly, a single-photon switch and transistor using a 4-level atomic system was demonstrated by Chen et al \cite{chen2013all}. 
A single gate photon was stored in the atomic ensemble inside an optical cavity which could switch the transmission of few hundred photons. The gate photon was also recovered with high probability, a fact that could possibly be used for non-destructive detection of single photons. Such detection has applications in quantum computing and communication \cite{reiserer2013nondestructive}.
A promising solid-state system for single photon switching is a single spin qubit in a nanophotonic cavity. Sun et al. showed that using the spin as memory of a single photon could control the transmission of more than 27 photons (Fig. \ref{Applications}a)\cite{sun2018single}.

A related application for photonic quantum networks is routing of single photons in which a photon is sent to one of the many channels by interaction with another photon. A notable example was shown with a rubidium atom coupled to whispering gallery mode of silica microsphere (Fig. \ref{Applications}c) \cite{shomroni2014all}. Two input photons interacting with the microresonator via a tapered fiber could be deterministically routed to one of the two output ports.                     

\section{Discussion}

The advancement of qNLO in solid-state necessitates the progress on primarily three different fronts - cavity design, materials, and integration. Essentially, one needs to have high-Q cavities with small mode volumes and emitters with narrow linewidths and integrate the emitter with cavity so that they interact strongly.

\noindent In the cavity design, extreme dielectric confinement and inversely designed topology-optimized dielectric cavities with extremely large $Q/V_m$ values appear as a promising route toward qNLO if one could integrate emitters in them. While fiber-DBR cavities have somewhat larger mode volumes, they provide the advantage of tunability along with fiber coupling. Hybrid plasmonic-photonic cavities represent a more exploratory direction that may provide especially high $Q/V_m$ values and potentially even enable qNLO at room temperature.

\noindent EQDs in zero-dimensional photonic crystal cavities were used for the earliest demonstrations of solid-state qNLO but lack of deterministic growth with good spectral control limits the scalability of this approach. Emitters that provide more flexibility in integration such as monolayer TMDCs and solution-processed materials have attracted a lot of interest in recent times but larger linewidths have thus far prevented any observation of qNLO phenomena using these. Among the emerging materials, single organic molecules inside a host seem very promising for single-photon blockade. Epitaxial quantum wells are the only materials thus far used for polariton blockade. However, accessing the Rydberg excitons in cuprous oxide, 2d materials, and perovskites could potentially lead to much stronger single-photon nonlinearity. Cuprous oxide, in particular, is a rare material that could host excitons with very large quantum numbers ($>20$) in the solid-state and hold the potential to replicate the qNLO of Rydberg atoms.    

\noindent Zero-dimensional cavities with extreme field confinement would typically only be compatible with zero-dimensional emitters having sizes smaller than the length scale of field confinement. Organic molecules, defect centers, and CQDs could potentially be integrated with them, with solution-processed materials (molecules and CQDs) presenting particular advantages for integration. Other small mode volume cavities such as nanobeam cavities are much more forgiving in that regard and could be used for most of the emitters. Fiber-DBR cavity could also be used with a variety of emitters with zero, one or two dimensions. Initial results in qNLO with fiber-DBR integrated organic molecules and epitaxial quantum wells are promising and the same could be extended for cuprous oxide. 

\noindent Overall, the solid-state qNLO is still at a nascent stage. A number of promising materials and cavity designs have emerged over the past decade or so, but continued concerted effort will be needed for practical realizations to be successful. Progress may be enabled in particular by co-design of materials and cavities, taking into account processing and integration challenges.

\section{Acknowledgment}
This material is based upon work supported by the National Science Foundation under Grant No. DMR-2019444.

\bibliography{bibliography}

\providecommand{\latin}[1]{#1}
\makeatletter
\providecommand{\doi}
  {\begingroup\let\do\@makeother\dospecials
  \catcode`\{=1 \catcode`\}=2 \doi@aux}
\providecommand{\doi@aux}[1]{\endgroup\texttt{#1}}
\makeatother
\providecommand*\mcitethebibliography{\thebibliography}
\csname @ifundefined\endcsname{endmcitethebibliography}  {\let\endmcitethebibliography\endthebibliography}{}
\begin{mcitethebibliography}{229}
\providecommand*\natexlab[1]{#1}
\providecommand*\mciteSetBstSublistMode[1]{}
\providecommand*\mciteSetBstMaxWidthForm[2]{}
\providecommand*\mciteBstWouldAddEndPuncttrue
  {\def\EndOfBibitem{\unskip.}}
\providecommand*\mciteBstWouldAddEndPunctfalse
  {\let\EndOfBibitem\relax}
\providecommand*\mciteSetBstMidEndSepPunct[3]{}
\providecommand*\mciteSetBstSublistLabelBeginEnd[3]{}
\providecommand*\EndOfBibitem{}
\mciteSetBstSublistMode{f}
\mciteSetBstMaxWidthForm{subitem}{(\alph{mcitesubitemcount})}
\mciteSetBstSublistLabelBeginEnd
  {\mcitemaxwidthsubitemform\space}
  {\relax}
  {\relax}

\bibitem[Sch{\"a}fer \latin{et~al.}(2020)Sch{\"a}fer, Fukuhara, Sugawa, Takasu, and Takahashi]{schafer2020tools}
Sch{\"a}fer,~F.; Fukuhara,~T.; Sugawa,~S.; Takasu,~Y.; Takahashi,~Y. Tools for quantum simulation with ultracold atoms in optical lattices. \emph{Nature Reviews Physics} \textbf{2020}, \emph{2}, 411--425\relax
\mciteBstWouldAddEndPuncttrue
\mciteSetBstMidEndSepPunct{\mcitedefaultmidpunct}
{\mcitedefaultendpunct}{\mcitedefaultseppunct}\relax
\EndOfBibitem
\bibitem[Daley \latin{et~al.}(2022)Daley, Bloch, Kokail, Flannigan, Pearson, Troyer, and Zoller]{daley2022practical}
Daley,~A.~J.; Bloch,~I.; Kokail,~C.; Flannigan,~S.; Pearson,~N.; Troyer,~M.; Zoller,~P. Practical quantum advantage in quantum simulation. \emph{Nature} \textbf{2022}, \emph{607}, 667--676\relax
\mciteBstWouldAddEndPuncttrue
\mciteSetBstMidEndSepPunct{\mcitedefaultmidpunct}
{\mcitedefaultendpunct}{\mcitedefaultseppunct}\relax
\EndOfBibitem
\bibitem[Trivedi \latin{et~al.}(2022)Trivedi, Rubio, and Cirac]{trivedi2022quantum}
Trivedi,~R.; Rubio,~A.~F.; Cirac,~J.~I. Quantum advantage and stability to errors in analogue quantum simulators. \emph{arXiv preprint arXiv:2212.04924} \textbf{2022}, \relax
\mciteBstWouldAddEndPunctfalse
\mciteSetBstMidEndSepPunct{\mcitedefaultmidpunct}
{}{\mcitedefaultseppunct}\relax
\EndOfBibitem
\bibitem[Arute \latin{et~al.}(2019)Arute, Arya, Babbush, Bacon, Bardin, Barends, Biswas, Boixo, Brandao, Buell, \latin{et~al.} others]{arute2019quantum}
Arute,~F.; Arya,~K.; Babbush,~R.; Bacon,~D.; Bardin,~J.~C.; Barends,~R.; Biswas,~R.; Boixo,~S.; Brandao,~F.~G.; Buell,~D.~A.; others Quantum supremacy using a programmable superconducting processor. \emph{Nature} \textbf{2019}, \emph{574}, 505--510\relax
\mciteBstWouldAddEndPuncttrue
\mciteSetBstMidEndSepPunct{\mcitedefaultmidpunct}
{\mcitedefaultendpunct}{\mcitedefaultseppunct}\relax
\EndOfBibitem
\bibitem[Choi(2023)]{choi2023ibm}
Choi,~C.~Q. IBM's Quantum Leap: The Company Will Take Quantum Tech Past the 1,000-Qubit Mark in 2023. \emph{IEEE Spectrum} \textbf{2023}, \emph{60}, 46--47\relax
\mciteBstWouldAddEndPuncttrue
\mciteSetBstMidEndSepPunct{\mcitedefaultmidpunct}
{\mcitedefaultendpunct}{\mcitedefaultseppunct}\relax
\EndOfBibitem
\bibitem[Bluvstein \latin{et~al.}(2023)Bluvstein, Evered, Geim, Li, Zhou, Manovitz, Ebadi, Cain, Kalinowski, Hangleiter, \latin{et~al.} others]{bluvstein2023logical}
Bluvstein,~D.; Evered,~S.~J.; Geim,~A.~A.; Li,~S.~H.; Zhou,~H.; Manovitz,~T.; Ebadi,~S.; Cain,~M.; Kalinowski,~M.; Hangleiter,~D.; others Logical quantum processor based on reconfigurable atom arrays. \emph{Nature} \textbf{2023}, 1--3\relax
\mciteBstWouldAddEndPuncttrue
\mciteSetBstMidEndSepPunct{\mcitedefaultmidpunct}
{\mcitedefaultendpunct}{\mcitedefaultseppunct}\relax
\EndOfBibitem
\bibitem[O'brien \latin{et~al.}(2009)O'brien, Furusawa, and Vu{\v{c}}kovi{\'c}]{o2009photonic}
O'brien,~J.~L.; Furusawa,~A.; Vu{\v{c}}kovi{\'c},~J. Photonic quantum technologies. \emph{Nature Photonics} \textbf{2009}, \emph{3}, 687--695\relax
\mciteBstWouldAddEndPuncttrue
\mciteSetBstMidEndSepPunct{\mcitedefaultmidpunct}
{\mcitedefaultendpunct}{\mcitedefaultseppunct}\relax
\EndOfBibitem
\bibitem[Beverland \latin{et~al.}(2022)Beverland, Murali, Troyer, Svore, Hoefler, Kliuchnikov, Low, Soeken, Sundaram, and Vaschillo]{beverland2022assessing}
Beverland,~M.~E.; Murali,~P.; Troyer,~M.; Svore,~K.~M.; Hoefler,~T.; Kliuchnikov,~V.; Low,~G.~H.; Soeken,~M.; Sundaram,~A.; Vaschillo,~A. Assessing requirements to scale to practical quantum advantage. \emph{arXiv preprint arXiv:2211.07629} \textbf{2022}, \relax
\mciteBstWouldAddEndPunctfalse
\mciteSetBstMidEndSepPunct{\mcitedefaultmidpunct}
{}{\mcitedefaultseppunct}\relax
\EndOfBibitem
\bibitem[Vahala(2003)]{vahala2003optical}
Vahala,~K.~J. Optical microcavities. \emph{nature} \textbf{2003}, \emph{424}, 839--846\relax
\mciteBstWouldAddEndPuncttrue
\mciteSetBstMidEndSepPunct{\mcitedefaultmidpunct}
{\mcitedefaultendpunct}{\mcitedefaultseppunct}\relax
\EndOfBibitem
\bibitem[Birnbaum \latin{et~al.}(2005)Birnbaum, Boca, Miller, Boozer, Northup, and Kimble]{birnbaum2005photon}
Birnbaum,~K.~M.; Boca,~A.; Miller,~R.; Boozer,~A.~D.; Northup,~T.~E.; Kimble,~H.~J. Photon blockade in an optical cavity with one trapped atom. \emph{Nature} \textbf{2005}, \emph{436}, 87--90\relax
\mciteBstWouldAddEndPuncttrue
\mciteSetBstMidEndSepPunct{\mcitedefaultmidpunct}
{\mcitedefaultendpunct}{\mcitedefaultseppunct}\relax
\EndOfBibitem
\bibitem[Byrnes \latin{et~al.}(2014)Byrnes, Kim, and Yamamoto]{byrnes2014exciton}
Byrnes,~T.; Kim,~N.~Y.; Yamamoto,~Y. Exciton--polariton condensates. \emph{Nature Physics} \textbf{2014}, \emph{10}, 803--813\relax
\mciteBstWouldAddEndPuncttrue
\mciteSetBstMidEndSepPunct{\mcitedefaultmidpunct}
{\mcitedefaultendpunct}{\mcitedefaultseppunct}\relax
\EndOfBibitem
\bibitem[Fushman \latin{et~al.}(2008)Fushman, Englund, Faraon, Stoltz, Petroff, and Vuckovic]{fushman2008controlled}
Fushman,~I.; Englund,~D.; Faraon,~A.; Stoltz,~N.; Petroff,~P.; Vuckovic,~J. Controlled phase shifts with a single quantum dot. \emph{science} \textbf{2008}, \emph{320}, 769--772\relax
\mciteBstWouldAddEndPuncttrue
\mciteSetBstMidEndSepPunct{\mcitedefaultmidpunct}
{\mcitedefaultendpunct}{\mcitedefaultseppunct}\relax
\EndOfBibitem
\bibitem[Sipahigil \latin{et~al.}(2016)Sipahigil, Evans, Sukachev, Burek, Borregaard, Bhaskar, Nguyen, Pacheco, Atikian, Meuwly, \latin{et~al.} others]{sipahigil2016integrated}
Sipahigil,~A.; Evans,~R.~E.; Sukachev,~D.~D.; Burek,~M.~J.; Borregaard,~J.; Bhaskar,~M.~K.; Nguyen,~C.~T.; Pacheco,~J.~L.; Atikian,~H.~A.; Meuwly,~C.; others An integrated diamond nanophotonics platform for quantum-optical networks. \emph{Science} \textbf{2016}, \emph{354}, 847--850\relax
\mciteBstWouldAddEndPuncttrue
\mciteSetBstMidEndSepPunct{\mcitedefaultmidpunct}
{\mcitedefaultendpunct}{\mcitedefaultseppunct}\relax
\EndOfBibitem
\bibitem[Carusotto \latin{et~al.}(2010)Carusotto, Volz, and Imamo{\u{g}}lu]{carusotto2010feshbach}
Carusotto,~I.; Volz,~T.; Imamo{\u{g}}lu,~A. Feshbach blockade: Single-photon nonlinear optics using resonantly enhanced cavity polariton scattering from biexciton states. \emph{Europhysics Letters} \textbf{2010}, \emph{90}, 37001\relax
\mciteBstWouldAddEndPuncttrue
\mciteSetBstMidEndSepPunct{\mcitedefaultmidpunct}
{\mcitedefaultendpunct}{\mcitedefaultseppunct}\relax
\EndOfBibitem
\bibitem[Flayac and Savona(2017)Flayac, and Savona]{flayac2017unconventional}
Flayac,~H.; Savona,~V. Unconventional photon blockade. \emph{Physical Review A} \textbf{2017}, \emph{96}, 053810\relax
\mciteBstWouldAddEndPuncttrue
\mciteSetBstMidEndSepPunct{\mcitedefaultmidpunct}
{\mcitedefaultendpunct}{\mcitedefaultseppunct}\relax
\EndOfBibitem
\bibitem[Hunger \latin{et~al.}(2010)Hunger, Steinmetz, Colombe, Deutsch, H{\"a}nsch, and Reichel]{hunger2010fiber}
Hunger,~D.; Steinmetz,~T.; Colombe,~Y.; Deutsch,~C.; H{\"a}nsch,~T.~W.; Reichel,~J. A fiber Fabry--Perot cavity with high finesse. \emph{New Journal of Physics} \textbf{2010}, \emph{12}, 065038\relax
\mciteBstWouldAddEndPuncttrue
\mciteSetBstMidEndSepPunct{\mcitedefaultmidpunct}
{\mcitedefaultendpunct}{\mcitedefaultseppunct}\relax
\EndOfBibitem
\bibitem[Reiserer and Rempe(2015)Reiserer, and Rempe]{reiserer2015cavity}
Reiserer,~A.; Rempe,~G. Cavity-based quantum networks with single atoms and optical photons. \emph{Reviews of Modern Physics} \textbf{2015}, \emph{87}, 1379\relax
\mciteBstWouldAddEndPuncttrue
\mciteSetBstMidEndSepPunct{\mcitedefaultmidpunct}
{\mcitedefaultendpunct}{\mcitedefaultseppunct}\relax
\EndOfBibitem
\bibitem[Majumdar \latin{et~al.}(2012)Majumdar, Englund, Bajcsy, and Vu{\v{c}}kovi{\'c}]{majumdar2012nonlinear}
Majumdar,~A.; Englund,~D.; Bajcsy,~M.; Vu{\v{c}}kovi{\'c},~J. Nonlinear temporal dynamics of a strongly coupled quantum-dot--cavity system. \emph{Physical Review A} \textbf{2012}, \emph{85}, 033802\relax
\mciteBstWouldAddEndPuncttrue
\mciteSetBstMidEndSepPunct{\mcitedefaultmidpunct}
{\mcitedefaultendpunct}{\mcitedefaultseppunct}\relax
\EndOfBibitem
\bibitem[Dovzhenko \latin{et~al.}(2018)Dovzhenko, Ryabchuk, Rakovich, and Nabiev]{dovzhenko2018light}
Dovzhenko,~D.; Ryabchuk,~S.; Rakovich,~Y.~P.; Nabiev,~I. Light--matter interaction in the strong coupling regime: configurations, conditions, and applications. \emph{Nanoscale} \textbf{2018}, \emph{10}, 3589--3605\relax
\mciteBstWouldAddEndPuncttrue
\mciteSetBstMidEndSepPunct{\mcitedefaultmidpunct}
{\mcitedefaultendpunct}{\mcitedefaultseppunct}\relax
\EndOfBibitem
\bibitem[Faraon \latin{et~al.}(2008)Faraon, Fushman, Englund, Stoltz, Petroff, and Vu{\v{c}}kovi{\'c}]{faraon2008coherent}
Faraon,~A.; Fushman,~I.; Englund,~D.; Stoltz,~N.; Petroff,~P.; Vu{\v{c}}kovi{\'c},~J. Coherent generation of non-classical light on a chip via photon-induced tunnelling and blockade. \emph{Nature Physics} \textbf{2008}, \emph{4}, 859--863\relax
\mciteBstWouldAddEndPuncttrue
\mciteSetBstMidEndSepPunct{\mcitedefaultmidpunct}
{\mcitedefaultendpunct}{\mcitedefaultseppunct}\relax
\EndOfBibitem
\bibitem[Majumdar \latin{et~al.}(2012)Majumdar, Bajcsy, and Vu{\v{c}}kovi{\'c}]{majumdar2012probing}
Majumdar,~A.; Bajcsy,~M.; Vu{\v{c}}kovi{\'c},~J. Probing the ladder of dressed states and nonclassical light generation in quantum-dot--cavity QED. \emph{Physical Review A} \textbf{2012}, \emph{85}, 041801\relax
\mciteBstWouldAddEndPuncttrue
\mciteSetBstMidEndSepPunct{\mcitedefaultmidpunct}
{\mcitedefaultendpunct}{\mcitedefaultseppunct}\relax
\EndOfBibitem
\bibitem[Delteil \latin{et~al.}(2019)Delteil, Fink, Schade, H{\"o}fling, Schneider, and {\.I}mamo{\u{g}}lu]{delteil2019towards}
Delteil,~A.; Fink,~T.; Schade,~A.; H{\"o}fling,~S.; Schneider,~C.; {\.I}mamo{\u{g}}lu,~A. Towards polariton blockade of confined exciton--polaritons. \emph{Nature materials} \textbf{2019}, \emph{18}, 219--222\relax
\mciteBstWouldAddEndPuncttrue
\mciteSetBstMidEndSepPunct{\mcitedefaultmidpunct}
{\mcitedefaultendpunct}{\mcitedefaultseppunct}\relax
\EndOfBibitem
\bibitem[Verger \latin{et~al.}(2006)Verger, Ciuti, and Carusotto]{verger2006polariton}
Verger,~A.; Ciuti,~C.; Carusotto,~I. Polariton quantum blockade in a photonic dot. \emph{Physical Review B} \textbf{2006}, \emph{73}, 193306\relax
\mciteBstWouldAddEndPuncttrue
\mciteSetBstMidEndSepPunct{\mcitedefaultmidpunct}
{\mcitedefaultendpunct}{\mcitedefaultseppunct}\relax
\EndOfBibitem
\bibitem[Tassone and Yamamoto(1999)Tassone, and Yamamoto]{tassone1999exciton}
Tassone,~F.; Yamamoto,~Y. Exciton-exciton scattering dynamics in a semiconductor microcavity and stimulated scattering into polaritons. \emph{Physical Review B} \textbf{1999}, \emph{59}, 10830\relax
\mciteBstWouldAddEndPuncttrue
\mciteSetBstMidEndSepPunct{\mcitedefaultmidpunct}
{\mcitedefaultendpunct}{\mcitedefaultseppunct}\relax
\EndOfBibitem
\bibitem[Ferretti and Gerace(2012)Ferretti, and Gerace]{ferretti2012single}
Ferretti,~S.; Gerace,~D. Single-photon nonlinear optics with Kerr-type nanostructured materials. \emph{Physical Review B} \textbf{2012}, \emph{85}, 033303\relax
\mciteBstWouldAddEndPuncttrue
\mciteSetBstMidEndSepPunct{\mcitedefaultmidpunct}
{\mcitedefaultendpunct}{\mcitedefaultseppunct}\relax
\EndOfBibitem
\bibitem[Majumdar and Gerace(2013)Majumdar, and Gerace]{majumdar2013single}
Majumdar,~A.; Gerace,~D. Single-photon blockade in doubly resonant nanocavities with second-order nonlinearity. \emph{Physical Review B} \textbf{2013}, \emph{87}, 235319\relax
\mciteBstWouldAddEndPuncttrue
\mciteSetBstMidEndSepPunct{\mcitedefaultmidpunct}
{\mcitedefaultendpunct}{\mcitedefaultseppunct}\relax
\EndOfBibitem
\bibitem[Kazimierczuk \latin{et~al.}(2014)Kazimierczuk, Fr{\"o}hlich, Scheel, Stolz, and Bayer]{kazimierczuk2014giant}
Kazimierczuk,~T.; Fr{\"o}hlich,~D.; Scheel,~S.; Stolz,~H.; Bayer,~M. Giant Rydberg excitons in the copper oxide Cu2O. \emph{Nature} \textbf{2014}, \emph{514}, 343--347\relax
\mciteBstWouldAddEndPuncttrue
\mciteSetBstMidEndSepPunct{\mcitedefaultmidpunct}
{\mcitedefaultendpunct}{\mcitedefaultseppunct}\relax
\EndOfBibitem
\bibitem[Hill \latin{et~al.}(2015)Hill, Rigosi, Roquelet, Chernikov, Berkelbach, Reichman, Hybertsen, Brus, and Heinz]{hill2015observation}
Hill,~H.~M.; Rigosi,~A.~F.; Roquelet,~C.; Chernikov,~A.; Berkelbach,~T.~C.; Reichman,~D.~R.; Hybertsen,~M.~S.; Brus,~L.~E.; Heinz,~T.~F. Observation of excitonic Rydberg states in monolayer MoS2 and WS2 by photoluminescence excitation spectroscopy. \emph{Nano letters} \textbf{2015}, \emph{15}, 2992--2997\relax
\mciteBstWouldAddEndPuncttrue
\mciteSetBstMidEndSepPunct{\mcitedefaultmidpunct}
{\mcitedefaultendpunct}{\mcitedefaultseppunct}\relax
\EndOfBibitem
\bibitem[Bao \latin{et~al.}(2019)Bao, Liu, Xue, Zheng, Tao, Wang, Xia, Zhao, Kim, Yang, \latin{et~al.} others]{bao2019observation}
Bao,~W.; Liu,~X.; Xue,~F.; Zheng,~F.; Tao,~R.; Wang,~S.; Xia,~Y.; Zhao,~M.; Kim,~J.; Yang,~S.; others Observation of Rydberg exciton polaritons and their condensate in a perovskite cavity. \emph{Proceedings of the National Academy of Sciences} \textbf{2019}, \emph{116}, 20274--20279\relax
\mciteBstWouldAddEndPuncttrue
\mciteSetBstMidEndSepPunct{\mcitedefaultmidpunct}
{\mcitedefaultendpunct}{\mcitedefaultseppunct}\relax
\EndOfBibitem
\bibitem[A{\ss}mann and Bayer(2020)A{\ss}mann, and Bayer]{assmann2020semiconductor}
A{\ss}mann,~M.; Bayer,~M. Semiconductor rydberg physics. \emph{Advanced Quantum Technologies} \textbf{2020}, \emph{3}, 1900134\relax
\mciteBstWouldAddEndPuncttrue
\mciteSetBstMidEndSepPunct{\mcitedefaultmidpunct}
{\mcitedefaultendpunct}{\mcitedefaultseppunct}\relax
\EndOfBibitem
\bibitem[Comparat and Pillet(2010)Comparat, and Pillet]{comparat2010dipole}
Comparat,~D.; Pillet,~P. Dipole blockade in a cold Rydberg atomic sample. \emph{JOSA B} \textbf{2010}, \emph{27}, A208--A232\relax
\mciteBstWouldAddEndPuncttrue
\mciteSetBstMidEndSepPunct{\mcitedefaultmidpunct}
{\mcitedefaultendpunct}{\mcitedefaultseppunct}\relax
\EndOfBibitem
\bibitem[Chin \latin{et~al.}(2010)Chin, Grimm, Julienne, and Tiesinga]{chin2010feshbach}
Chin,~C.; Grimm,~R.; Julienne,~P.; Tiesinga,~E. Feshbach resonances in ultracold gases. \emph{Reviews of Modern Physics} \textbf{2010}, \emph{82}, 1225\relax
\mciteBstWouldAddEndPuncttrue
\mciteSetBstMidEndSepPunct{\mcitedefaultmidpunct}
{\mcitedefaultendpunct}{\mcitedefaultseppunct}\relax
\EndOfBibitem
\bibitem[Takemura \latin{et~al.}(2014)Takemura, Trebaol, Wouters, Portella-Oberli, and Deveaud]{takemura2014polaritonic}
Takemura,~N.; Trebaol,~S.; Wouters,~M.; Portella-Oberli,~M.~T.; Deveaud,~B. Polaritonic feshbach resonance. \emph{Nature Physics} \textbf{2014}, \emph{10}, 500--504\relax
\mciteBstWouldAddEndPuncttrue
\mciteSetBstMidEndSepPunct{\mcitedefaultmidpunct}
{\mcitedefaultendpunct}{\mcitedefaultseppunct}\relax
\EndOfBibitem
\bibitem[Chang \latin{et~al.}(2014)Chang, Vuleti{\'c}, and Lukin]{chang2014quantum}
Chang,~D.~E.; Vuleti{\'c},~V.; Lukin,~M.~D. Quantum nonlinear optics—photon by photon. \emph{Nature Photonics} \textbf{2014}, \emph{8}, 685--694\relax
\mciteBstWouldAddEndPuncttrue
\mciteSetBstMidEndSepPunct{\mcitedefaultmidpunct}
{\mcitedefaultendpunct}{\mcitedefaultseppunct}\relax
\EndOfBibitem
\bibitem[Fleischhauer \latin{et~al.}(2005)Fleischhauer, Imamoglu, and Marangos]{fleischhauer2005electromagnetically}
Fleischhauer,~M.; Imamoglu,~A.; Marangos,~J.~P. Electromagnetically induced transparency: Optics in coherent media. \emph{Reviews of modern physics} \textbf{2005}, \emph{77}, 633--673\relax
\mciteBstWouldAddEndPuncttrue
\mciteSetBstMidEndSepPunct{\mcitedefaultmidpunct}
{\mcitedefaultendpunct}{\mcitedefaultseppunct}\relax
\EndOfBibitem
\bibitem[Shomroni \latin{et~al.}(2014)Shomroni, Rosenblum, Lovsky, Bechler, Guendelman, and Dayan]{shomroni2014all}
Shomroni,~I.; Rosenblum,~S.; Lovsky,~Y.; Bechler,~O.; Guendelman,~G.; Dayan,~B. All-optical routing of single photons by a one-atom switch controlled by a single photon. \emph{Science} \textbf{2014}, \emph{345}, 903--906\relax
\mciteBstWouldAddEndPuncttrue
\mciteSetBstMidEndSepPunct{\mcitedefaultmidpunct}
{\mcitedefaultendpunct}{\mcitedefaultseppunct}\relax
\EndOfBibitem
\bibitem[Baur \latin{et~al.}(2014)Baur, Tiarks, Rempe, and D{\"u}rr]{baur2014single}
Baur,~S.; Tiarks,~D.; Rempe,~G.; D{\"u}rr,~S. Single-photon switch based on Rydberg blockade. \emph{Physical review letters} \textbf{2014}, \emph{112}, 073901\relax
\mciteBstWouldAddEndPuncttrue
\mciteSetBstMidEndSepPunct{\mcitedefaultmidpunct}
{\mcitedefaultendpunct}{\mcitedefaultseppunct}\relax
\EndOfBibitem
\bibitem[Turchette \latin{et~al.}(1995)Turchette, Hood, Lange, Mabuchi, and Kimble]{turchette1995measurement}
Turchette,~Q.~A.; Hood,~C.~J.; Lange,~W.; Mabuchi,~H.; Kimble,~H.~J. Measurement of conditional phase shifts for quantum logic. \emph{Physical Review Letters} \textbf{1995}, \emph{75}, 4710\relax
\mciteBstWouldAddEndPuncttrue
\mciteSetBstMidEndSepPunct{\mcitedefaultmidpunct}
{\mcitedefaultendpunct}{\mcitedefaultseppunct}\relax
\EndOfBibitem
\bibitem[Lalanne \latin{et~al.}(2008)Lalanne, Sauvan, and Hugonin]{lalanne2008photon}
Lalanne,~P.; Sauvan,~C.; Hugonin,~J.~P. Photon confinement in photonic crystal nanocavities. \emph{Laser \& Photonics Reviews} \textbf{2008}, \emph{2}, 514--526\relax
\mciteBstWouldAddEndPuncttrue
\mciteSetBstMidEndSepPunct{\mcitedefaultmidpunct}
{\mcitedefaultendpunct}{\mcitedefaultseppunct}\relax
\EndOfBibitem
\bibitem[Asano \latin{et~al.}(2017)Asano, Ochi, Takahashi, Kishimoto, and Noda]{asano2017photonic}
Asano,~T.; Ochi,~Y.; Takahashi,~Y.; Kishimoto,~K.; Noda,~S. Photonic crystal nanocavity with a Q factor exceeding eleven million. \emph{Optics express} \textbf{2017}, \emph{25}, 1769--1777\relax
\mciteBstWouldAddEndPuncttrue
\mciteSetBstMidEndSepPunct{\mcitedefaultmidpunct}
{\mcitedefaultendpunct}{\mcitedefaultseppunct}\relax
\EndOfBibitem
\bibitem[Fryett \latin{et~al.}(2018)Fryett, Chen, Whitehead, Peycke, Xu, and Majumdar]{fryett2018encapsulated}
Fryett,~T.~K.; Chen,~Y.; Whitehead,~J.; Peycke,~Z.~M.; Xu,~X.; Majumdar,~A. Encapsulated silicon nitride nanobeam cavity for hybrid nanophotonics. \emph{ACS Photonics} \textbf{2018}, \emph{5}, 2176--2181\relax
\mciteBstWouldAddEndPuncttrue
\mciteSetBstMidEndSepPunct{\mcitedefaultmidpunct}
{\mcitedefaultendpunct}{\mcitedefaultseppunct}\relax
\EndOfBibitem
\bibitem[Bajoni \latin{et~al.}(2008)Bajoni, Senellart, Wertz, Sagnes, Miard, Lema{\^\i}tre, and Bloch]{bajoni2008polariton}
Bajoni,~D.; Senellart,~P.; Wertz,~E.; Sagnes,~I.; Miard,~A.; Lema{\^\i}tre,~A.; Bloch,~J. Polariton laser using single micropillar GaAs- GaAlAs semiconductor cavities. \emph{Physical review letters} \textbf{2008}, \emph{100}, 047401\relax
\mciteBstWouldAddEndPuncttrue
\mciteSetBstMidEndSepPunct{\mcitedefaultmidpunct}
{\mcitedefaultendpunct}{\mcitedefaultseppunct}\relax
\EndOfBibitem
\bibitem[Besga \latin{et~al.}(2015)Besga, Vaneph, Reichel, Est{\`e}ve, Reinhard, Miguel-S{\'a}nchez, Imamo{\u{g}}lu, and Volz]{besga2015polariton}
Besga,~B.; Vaneph,~C.; Reichel,~J.; Est{\`e}ve,~J.; Reinhard,~A.; Miguel-S{\'a}nchez,~J.; Imamo{\u{g}}lu,~A.; Volz,~T. Polariton boxes in a tunable fiber cavity. \emph{Physical Review Applied} \textbf{2015}, \emph{3}, 014008\relax
\mciteBstWouldAddEndPuncttrue
\mciteSetBstMidEndSepPunct{\mcitedefaultmidpunct}
{\mcitedefaultendpunct}{\mcitedefaultseppunct}\relax
\EndOfBibitem
\bibitem[Garc{\'\i}a~de Arquer \latin{et~al.}(2021)Garc{\'\i}a~de Arquer, Talapin, Klimov, Arakawa, Bayer, and Sargent]{garcia2021semiconductor}
Garc{\'\i}a~de Arquer,~F.~P.; Talapin,~D.~V.; Klimov,~V.~I.; Arakawa,~Y.; Bayer,~M.; Sargent,~E.~H. Semiconductor quantum dots: Technological progress and future challenges. \emph{Science} \textbf{2021}, \emph{373}, eaaz8541\relax
\mciteBstWouldAddEndPuncttrue
\mciteSetBstMidEndSepPunct{\mcitedefaultmidpunct}
{\mcitedefaultendpunct}{\mcitedefaultseppunct}\relax
\EndOfBibitem
\bibitem[Buckley \latin{et~al.}(2012)Buckley, Rivoire, and Vu{\v{c}}kovi{\'c}]{buckley2012engineered}
Buckley,~S.; Rivoire,~K.; Vu{\v{c}}kovi{\'c},~J. Engineered quantum dot single-photon sources. \emph{Reports on Progress in Physics} \textbf{2012}, \emph{75}, 126503\relax
\mciteBstWouldAddEndPuncttrue
\mciteSetBstMidEndSepPunct{\mcitedefaultmidpunct}
{\mcitedefaultendpunct}{\mcitedefaultseppunct}\relax
\EndOfBibitem
\bibitem[Somaschi \latin{et~al.}(2016)Somaschi, Giesz, De~Santis, Loredo, Almeida, Hornecker, Portalupi, Grange, Anton, Demory, \latin{et~al.} others]{somaschi2016near}
Somaschi,~N.; Giesz,~V.; De~Santis,~L.; Loredo,~J.; Almeida,~M.~P.; Hornecker,~G.; Portalupi,~S.~L.; Grange,~T.; Anton,~C.; Demory,~J.; others Near-optimal single-photon sources in the solid state. \emph{Nature Photonics} \textbf{2016}, \emph{10}, 340--345\relax
\mciteBstWouldAddEndPuncttrue
\mciteSetBstMidEndSepPunct{\mcitedefaultmidpunct}
{\mcitedefaultendpunct}{\mcitedefaultseppunct}\relax
\EndOfBibitem
\bibitem[Michler \latin{et~al.}(2000)Michler, Kiraz, Becher, Schoenfeld, Petroff, Zhang, Hu, and Imamoglu]{michler2000quantum}
Michler,~P.; Kiraz,~A.; Becher,~C.; Schoenfeld,~W.; Petroff,~P.; Zhang,~L.; Hu,~E.; Imamoglu,~A. A quantum dot single-photon turnstile device. \emph{science} \textbf{2000}, \emph{290}, 2282--2285\relax
\mciteBstWouldAddEndPuncttrue
\mciteSetBstMidEndSepPunct{\mcitedefaultmidpunct}
{\mcitedefaultendpunct}{\mcitedefaultseppunct}\relax
\EndOfBibitem
\bibitem[Senellart \latin{et~al.}(2017)Senellart, Solomon, and White]{senellart2017high}
Senellart,~P.; Solomon,~G.; White,~A. High-performance semiconductor quantum-dot single-photon sources. \emph{Nature nanotechnology} \textbf{2017}, \emph{12}, 1026--1039\relax
\mciteBstWouldAddEndPuncttrue
\mciteSetBstMidEndSepPunct{\mcitedefaultmidpunct}
{\mcitedefaultendpunct}{\mcitedefaultseppunct}\relax
\EndOfBibitem
\bibitem[Ota \latin{et~al.}(2018)Ota, Takamiya, Ohta, Takagi, Kumagai, Iwamoto, and Arakawa]{ota2018large}
Ota,~Y.; Takamiya,~D.; Ohta,~R.; Takagi,~H.; Kumagai,~N.; Iwamoto,~S.; Arakawa,~Y. Large vacuum Rabi splitting between a single quantum dot and an H0 photonic crystal nanocavity. \emph{Applied Physics Letters} \textbf{2018}, \emph{112}, 093101\relax
\mciteBstWouldAddEndPuncttrue
\mciteSetBstMidEndSepPunct{\mcitedefaultmidpunct}
{\mcitedefaultendpunct}{\mcitedefaultseppunct}\relax
\EndOfBibitem
\bibitem[Reinhard \latin{et~al.}(2012)Reinhard, Volz, Winger, Badolato, Hennessy, Hu, and Imamo{\u{g}}lu]{reinhard2012strongly}
Reinhard,~A.; Volz,~T.; Winger,~M.; Badolato,~A.; Hennessy,~K.~J.; Hu,~E.~L.; Imamo{\u{g}}lu,~A. Strongly correlated photons on a chip. \emph{Nature Photonics} \textbf{2012}, \emph{6}, 93--96\relax
\mciteBstWouldAddEndPuncttrue
\mciteSetBstMidEndSepPunct{\mcitedefaultmidpunct}
{\mcitedefaultendpunct}{\mcitedefaultseppunct}\relax
\EndOfBibitem
\bibitem[Khitrova \latin{et~al.}(2006)Khitrova, Gibbs, Kira, Koch, and Scherer]{khitrova2006vacuum}
Khitrova,~G.; Gibbs,~H.; Kira,~M.; Koch,~S.~W.; Scherer,~A. Vacuum Rabi splitting in semiconductors. \emph{Nature physics} \textbf{2006}, \emph{2}, 81--90\relax
\mciteBstWouldAddEndPuncttrue
\mciteSetBstMidEndSepPunct{\mcitedefaultmidpunct}
{\mcitedefaultendpunct}{\mcitedefaultseppunct}\relax
\EndOfBibitem
\bibitem[Yoshie \latin{et~al.}(2004)Yoshie, Scherer, Hendrickson, Khitrova, Gibbs, Rupper, Ell, Shchekin, and Deppe]{yoshie2004vacuum}
Yoshie,~T.; Scherer,~A.; Hendrickson,~J.; Khitrova,~G.; Gibbs,~H.; Rupper,~G.; Ell,~C.; Shchekin,~O.; Deppe,~D. Vacuum Rabi splitting with a single quantum dot in a photonic crystal nanocavity. \emph{Nature} \textbf{2004}, \emph{432}, 200--203\relax
\mciteBstWouldAddEndPuncttrue
\mciteSetBstMidEndSepPunct{\mcitedefaultmidpunct}
{\mcitedefaultendpunct}{\mcitedefaultseppunct}\relax
\EndOfBibitem
\bibitem[Javadi \latin{et~al.}(2015)Javadi, S{\"o}llner, Arcari, Hansen, Midolo, Mahmoodian, Kir{\v{s}}ansk{\.e}, Pregnolato, Lee, Song, \latin{et~al.} others]{javadi2015single}
Javadi,~A.; S{\"o}llner,~I.; Arcari,~M.; Hansen,~S.~L.; Midolo,~L.; Mahmoodian,~S.; Kir{\v{s}}ansk{\.e},~G.; Pregnolato,~T.; Lee,~E.; Song,~J.; others Single-photon non-linear optics with a quantum dot in a waveguide. \emph{Nature communications} \textbf{2015}, \emph{6}, 8655\relax
\mciteBstWouldAddEndPuncttrue
\mciteSetBstMidEndSepPunct{\mcitedefaultmidpunct}
{\mcitedefaultendpunct}{\mcitedefaultseppunct}\relax
\EndOfBibitem
\bibitem[Schneider \latin{et~al.}(2016)Schneider, Gold, Reitzenstein, Hoefling, and Kamp]{schneider2016quantum}
Schneider,~C.; Gold,~P.; Reitzenstein,~S.; Hoefling,~S.; Kamp,~M. Quantum dot micropillar cavities with quality factors exceeding 250,000. \emph{Applied Physics B} \textbf{2016}, \emph{122}, 1--6\relax
\mciteBstWouldAddEndPuncttrue
\mciteSetBstMidEndSepPunct{\mcitedefaultmidpunct}
{\mcitedefaultendpunct}{\mcitedefaultseppunct}\relax
\EndOfBibitem
\bibitem[Reithmaier \latin{et~al.}(2004)Reithmaier, S{\k{e}}k, L{\"o}ffler, Hofmann, Kuhn, Reitzenstein, Keldysh, Kulakovskii, Reinecke, and Forchel]{reithmaier2004strong}
Reithmaier,~J.~P.; S{\k{e}}k,~G.; L{\"o}ffler,~A.; Hofmann,~C.; Kuhn,~S.; Reitzenstein,~S.; Keldysh,~L.; Kulakovskii,~V.; Reinecke,~T.; Forchel,~A. Strong coupling in a single quantum dot--semiconductor microcavity system. \emph{Nature} \textbf{2004}, \emph{432}, 197--200\relax
\mciteBstWouldAddEndPuncttrue
\mciteSetBstMidEndSepPunct{\mcitedefaultmidpunct}
{\mcitedefaultendpunct}{\mcitedefaultseppunct}\relax
\EndOfBibitem
\bibitem[Chen \latin{et~al.}(2018)Chen, Ryou, Friedfeld, Fryett, Whitehead, Cossairt, and Majumdar]{chen2018deterministic}
Chen,~Y.; Ryou,~A.; Friedfeld,~M.~R.; Fryett,~T.; Whitehead,~J.; Cossairt,~B.~M.; Majumdar,~A. Deterministic positioning of colloidal quantum dots on silicon nitride nanobeam cavities. \emph{Nano letters} \textbf{2018}, \emph{18}, 6404--6410\relax
\mciteBstWouldAddEndPuncttrue
\mciteSetBstMidEndSepPunct{\mcitedefaultmidpunct}
{\mcitedefaultendpunct}{\mcitedefaultseppunct}\relax
\EndOfBibitem
\bibitem[Yang \latin{et~al.}(2017)Yang, Pelton, Fedin, Talapin, and Waks]{yang2017room}
Yang,~Z.; Pelton,~M.; Fedin,~I.; Talapin,~D.~V.; Waks,~E. A room temperature continuous-wave nanolaser using colloidal quantum wells. \emph{Nature Communications} \textbf{2017}, \emph{8}, 143\relax
\mciteBstWouldAddEndPuncttrue
\mciteSetBstMidEndSepPunct{\mcitedefaultmidpunct}
{\mcitedefaultendpunct}{\mcitedefaultseppunct}\relax
\EndOfBibitem
\bibitem[Ganesh \latin{et~al.}(2007)Ganesh, Zhang, Mathias, Chow, Soares, Malyarchuk, Smith, and Cunningham]{ganesh2007enhanced}
Ganesh,~N.; Zhang,~W.; Mathias,~P.~C.; Chow,~E.; Soares,~J.; Malyarchuk,~V.; Smith,~A.~D.; Cunningham,~B.~T. Enhanced fluorescence emission from quantum dots on a photonic crystal surface. \emph{Nature Nanotechnology} \textbf{2007}, \emph{2}, 515--520\relax
\mciteBstWouldAddEndPuncttrue
\mciteSetBstMidEndSepPunct{\mcitedefaultmidpunct}
{\mcitedefaultendpunct}{\mcitedefaultseppunct}\relax
\EndOfBibitem
\bibitem[Wu \latin{et~al.}(2007)Wu, Mi, Bhattacharya, Zhu, and Xu]{wu2007enhanced}
Wu,~Z.; Mi,~Z.; Bhattacharya,~P.; Zhu,~T.; Xu,~J. Enhanced spontaneous emission at 1.55 $\mu$ m from colloidal PbSe quantum dots in a Si photonic crystal microcavity. \emph{Applied Physics Letters} \textbf{2007}, \emph{90}, 171105\relax
\mciteBstWouldAddEndPuncttrue
\mciteSetBstMidEndSepPunct{\mcitedefaultmidpunct}
{\mcitedefaultendpunct}{\mcitedefaultseppunct}\relax
\EndOfBibitem
\bibitem[Yu and Chen(2020)Yu, and Chen]{yu2020optical}
Yu,~J.; Chen,~R. Optical properties and applications of two-dimensional CdSe nanoplatelets. \emph{InfoMat} \textbf{2020}, \emph{2}, 905--927\relax
\mciteBstWouldAddEndPuncttrue
\mciteSetBstMidEndSepPunct{\mcitedefaultmidpunct}
{\mcitedefaultendpunct}{\mcitedefaultseppunct}\relax
\EndOfBibitem
\bibitem[Schatzl \latin{et~al.}(2017)Schatzl, Hackl, Glaser, Rauter, Brehm, Spindlberger, Simbula, Galli, Fromherz, and Schäffler]{schatzl2017enhanced}
Schatzl,~M.; Hackl,~F.; Glaser,~M.; Rauter,~P.; Brehm,~M.; Spindlberger,~L.; Simbula,~A.; Galli,~M.; Fromherz,~T.; Schäffler,~F. Enhanced telecom emission from single group-IV quantum dots by precise CMOS-compatible positioning in photonic crystal cavities. \emph{ACS photonics} \textbf{2017}, \emph{4}, 665--673\relax
\mciteBstWouldAddEndPuncttrue
\mciteSetBstMidEndSepPunct{\mcitedefaultmidpunct}
{\mcitedefaultendpunct}{\mcitedefaultseppunct}\relax
\EndOfBibitem
\bibitem[Pyatkov \latin{et~al.}(2016)Pyatkov, F{\"u}tterling, Khasminskaya, Flavel, Hennrich, Kappes, Krupke, and Pernice]{pyatkov2016cavity}
Pyatkov,~F.; F{\"u}tterling,~V.; Khasminskaya,~S.; Flavel,~B.~S.; Hennrich,~F.; Kappes,~M.~M.; Krupke,~R.; Pernice,~W.~H. Cavity-enhanced light emission from electrically driven carbon nanotubes. \emph{Nature Photonics} \textbf{2016}, \emph{10}, 420--427\relax
\mciteBstWouldAddEndPuncttrue
\mciteSetBstMidEndSepPunct{\mcitedefaultmidpunct}
{\mcitedefaultendpunct}{\mcitedefaultseppunct}\relax
\EndOfBibitem
\bibitem[Rivera \latin{et~al.}(2019)Rivera, Fryett, Chen, Liu, Ray, Hatami, Yan, Mandrus, Yao, Majumdar, \latin{et~al.} others]{rivera2019coupling}
Rivera,~P.; Fryett,~T.~K.; Chen,~Y.; Liu,~C.-H.; Ray,~E.; Hatami,~F.; Yan,~J.; Mandrus,~D.; Yao,~W.; Majumdar,~A.; others Coupling of photonic crystal cavity and interlayer exciton in heterobilayer of transition metal dichalcogenides. \emph{2D Materials} \textbf{2019}, \emph{7}, 015027\relax
\mciteBstWouldAddEndPuncttrue
\mciteSetBstMidEndSepPunct{\mcitedefaultmidpunct}
{\mcitedefaultendpunct}{\mcitedefaultseppunct}\relax
\EndOfBibitem
\bibitem[Srinivasan and Painter(2007)Srinivasan, and Painter]{srinivasan2007linear}
Srinivasan,~K.; Painter,~O. Linear and nonlinear optical spectroscopy of a strongly coupled microdisk--quantum dot system. \emph{Nature} \textbf{2007}, \emph{450}, 862--865\relax
\mciteBstWouldAddEndPuncttrue
\mciteSetBstMidEndSepPunct{\mcitedefaultmidpunct}
{\mcitedefaultendpunct}{\mcitedefaultseppunct}\relax
\EndOfBibitem
\bibitem[Imamura \latin{et~al.}(2013)Imamura, Watahiki, Miura, Shimada, and Kato]{imamura2013optical}
Imamura,~S.; Watahiki,~R.; Miura,~R.; Shimada,~T.; Kato,~Y. Optical control of individual carbon nanotube light emitters by spectral double resonance in silicon microdisk resonators. \emph{Applied Physics Letters} \textbf{2013}, \emph{102}, 161102\relax
\mciteBstWouldAddEndPuncttrue
\mciteSetBstMidEndSepPunct{\mcitedefaultmidpunct}
{\mcitedefaultendpunct}{\mcitedefaultseppunct}\relax
\EndOfBibitem
\bibitem[Ye \latin{et~al.}(2015)Ye, Wong, Lu, Ni, Zhu, Chen, Wang, and Zhang]{ye2015monolayer}
Ye,~Y.; Wong,~Z.~J.; Lu,~X.; Ni,~X.; Zhu,~H.; Chen,~X.; Wang,~Y.; Zhang,~X. Monolayer excitonic laser. \emph{Nature Photonics} \textbf{2015}, \emph{9}, 733--737\relax
\mciteBstWouldAddEndPuncttrue
\mciteSetBstMidEndSepPunct{\mcitedefaultmidpunct}
{\mcitedefaultendpunct}{\mcitedefaultseppunct}\relax
\EndOfBibitem
\bibitem[Zhu \latin{et~al.}(2020)Zhu, Yuan, Zeng, and Xia]{zhu2020manipulating}
Zhu,~L.; Yuan,~S.; Zeng,~C.; Xia,~J. Manipulating Photoluminescence of Carbon G-center in Silicon Metasurface with Optical Bound States in the Continuum. \emph{Advanced Optical Materials} \textbf{2020}, \emph{8}, 1901830\relax
\mciteBstWouldAddEndPuncttrue
\mciteSetBstMidEndSepPunct{\mcitedefaultmidpunct}
{\mcitedefaultendpunct}{\mcitedefaultseppunct}\relax
\EndOfBibitem
\bibitem[Gu \latin{et~al.}(2021)Gu, Walther, Waldecker, Rhodes, Raja, Hone, Heinz, K{\'e}na-Cohen, Pohl, and Menon]{gu2021enhanced}
Gu,~J.; Walther,~V.; Waldecker,~L.; Rhodes,~D.; Raja,~A.; Hone,~J.~C.; Heinz,~T.~F.; K{\'e}na-Cohen,~S.; Pohl,~T.; Menon,~V.~M. Enhanced nonlinear interaction of polaritons via excitonic Rydberg states in monolayer WSe2. \emph{Nature communications} \textbf{2021}, \emph{12}, 2269\relax
\mciteBstWouldAddEndPuncttrue
\mciteSetBstMidEndSepPunct{\mcitedefaultmidpunct}
{\mcitedefaultendpunct}{\mcitedefaultseppunct}\relax
\EndOfBibitem
\bibitem[Pfeifer \latin{et~al.}(2022)Pfeifer, Ratschbacher, Gallego, Saavedra, Fa{\ss}bender, von Haaren, Alt, Hofferberth, K{\"o}hl, Linden, \latin{et~al.} others]{pfeifer2022achievements}
Pfeifer,~H.; Ratschbacher,~L.; Gallego,~J.; Saavedra,~C.; Fa{\ss}bender,~A.; von Haaren,~A.; Alt,~W.; Hofferberth,~S.; K{\"o}hl,~M.; Linden,~S.; others Achievements and perspectives of optical fiber Fabry--Perot cavities. \emph{Applied Physics B} \textbf{2022}, \emph{128}, 29\relax
\mciteBstWouldAddEndPuncttrue
\mciteSetBstMidEndSepPunct{\mcitedefaultmidpunct}
{\mcitedefaultendpunct}{\mcitedefaultseppunct}\relax
\EndOfBibitem
\bibitem[Najer \latin{et~al.}(2019)Najer, S{\"o}llner, Sekatski, Dolique, L{\"o}bl, Riedel, Schott, Starosielec, Valentin, Wieck, \latin{et~al.} others]{najer2019gated}
Najer,~D.; S{\"o}llner,~I.; Sekatski,~P.; Dolique,~V.; L{\"o}bl,~M.~C.; Riedel,~D.; Schott,~R.; Starosielec,~S.; Valentin,~S.~R.; Wieck,~A.~D.; others A gated quantum dot strongly coupled to an optical microcavity. \emph{Nature} \textbf{2019}, \emph{575}, 622--627\relax
\mciteBstWouldAddEndPuncttrue
\mciteSetBstMidEndSepPunct{\mcitedefaultmidpunct}
{\mcitedefaultendpunct}{\mcitedefaultseppunct}\relax
\EndOfBibitem
\bibitem[Pscherer \latin{et~al.}(2021)Pscherer, Meierhofer, Wang, Kelkar, Mart{\'\i}n-Cano, Utikal, G{\"o}tzinger, and Sandoghdar]{pscherer2021single}
Pscherer,~A.; Meierhofer,~M.; Wang,~D.; Kelkar,~H.; Mart{\'\i}n-Cano,~D.; Utikal,~T.; G{\"o}tzinger,~S.; Sandoghdar,~V. Single-molecule vacuum Rabi splitting: Four-wave mixing and optical switching at the single-photon level. \emph{Physical Review Letters} \textbf{2021}, \emph{127}, 133603\relax
\mciteBstWouldAddEndPuncttrue
\mciteSetBstMidEndSepPunct{\mcitedefaultmidpunct}
{\mcitedefaultendpunct}{\mcitedefaultseppunct}\relax
\EndOfBibitem
\bibitem[Makino \latin{et~al.}(2005)Makino, Segawa, Kawasaki, and Koinuma]{makino2005optical}
Makino,~T.; Segawa,~Y.; Kawasaki,~M.; Koinuma,~H. Optical properties of excitons in ZnO-based quantum well heterostructures. \emph{Semiconductor science and technology} \textbf{2005}, \emph{20}, S78\relax
\mciteBstWouldAddEndPuncttrue
\mciteSetBstMidEndSepPunct{\mcitedefaultmidpunct}
{\mcitedefaultendpunct}{\mcitedefaultseppunct}\relax
\EndOfBibitem
\bibitem[Lukin \latin{et~al.}(2020)Lukin, Dory, Guidry, Yang, Mishra, Trivedi, Radulaski, Sun, Vercruysse, Ahn, \latin{et~al.} others]{lukin20204h}
Lukin,~D.~M.; Dory,~C.; Guidry,~M.~A.; Yang,~K.~Y.; Mishra,~S.~D.; Trivedi,~R.; Radulaski,~M.; Sun,~S.; Vercruysse,~D.; Ahn,~G.~H.; others 4H-silicon-carbide-on-insulator for integrated quantum and nonlinear photonics. \emph{Nature Photonics} \textbf{2020}, \emph{14}, 330--334\relax
\mciteBstWouldAddEndPuncttrue
\mciteSetBstMidEndSepPunct{\mcitedefaultmidpunct}
{\mcitedefaultendpunct}{\mcitedefaultseppunct}\relax
\EndOfBibitem
\bibitem[Pregnolato \latin{et~al.}(2020)Pregnolato, Chu, Schr{\"o}der, Schott, Wieck, Ludwig, Lodahl, and Rotenberg]{pregnolato2020deterministic}
Pregnolato,~T.; Chu,~X.-L.; Schr{\"o}der,~T.; Schott,~R.; Wieck,~A.~D.; Ludwig,~A.; Lodahl,~P.; Rotenberg,~N. Deterministic positioning of nanophotonic waveguides around single self-assembled quantum dots. \emph{APL Photonics} \textbf{2020}, \emph{5}, 086101\relax
\mciteBstWouldAddEndPuncttrue
\mciteSetBstMidEndSepPunct{\mcitedefaultmidpunct}
{\mcitedefaultendpunct}{\mcitedefaultseppunct}\relax
\EndOfBibitem
\bibitem[Choi \latin{et~al.}()Choi, Lee, Park, Kim, Jun, Park, Song, Ko, and Cho]{choi2023single}
Choi,~M.; Lee,~M.; Park,~S.-Y.~L.; Kim,~B.~S.; Jun,~S.; Park,~S.~I.; Song,~J.~D.; Ko,~Y.-H.; Cho,~Y.-H. Single Quantum Dot Selection and Tailor-Made Photonic Device Integration using a Nanoscale-Focus Pinspot. \emph{Advanced Materials} 2210667\relax
\mciteBstWouldAddEndPuncttrue
\mciteSetBstMidEndSepPunct{\mcitedefaultmidpunct}
{\mcitedefaultendpunct}{\mcitedefaultseppunct}\relax
\EndOfBibitem
\bibitem[S{\"u}nner \latin{et~al.}(2008)S{\"u}nner, Schneider, Strau{\ss}, Huggenberger, Wiener, H{\"o}fling, Kamp, and Forchel]{sunner2008scalable}
S{\"u}nner,~T.; Schneider,~C.; Strau{\ss},~M.; Huggenberger,~A.; Wiener,~D.; H{\"o}fling,~S.; Kamp,~M.; Forchel,~A. Scalable fabrication of optical resonators with embedded site-controlled quantum dots. \emph{Optics letters} \textbf{2008}, \emph{33}, 1759--1761\relax
\mciteBstWouldAddEndPuncttrue
\mciteSetBstMidEndSepPunct{\mcitedefaultmidpunct}
{\mcitedefaultendpunct}{\mcitedefaultseppunct}\relax
\EndOfBibitem
\bibitem[Han \latin{et~al.}(2021)Han, Wang, and Hopkinson]{han2021ordered}
Han,~I.~S.; Wang,~Y.-R.; Hopkinson,~M. Ordered GaAs quantum dots by droplet epitaxy using in situ direct laser interference patterning. \emph{Applied Physics Letters} \textbf{2021}, \emph{118}, 142101\relax
\mciteBstWouldAddEndPuncttrue
\mciteSetBstMidEndSepPunct{\mcitedefaultmidpunct}
{\mcitedefaultendpunct}{\mcitedefaultseppunct}\relax
\EndOfBibitem
\bibitem[Zhang \latin{et~al.}(2022)Zhang, Chattaraj, Huang, Jordao, Lu, and Madhukar]{zhang2022chip}
Zhang,~J.; Chattaraj,~S.; Huang,~Q.; Jordao,~L.; Lu,~S.; Madhukar,~A. On-chip scalable highly pure and indistinguishable single-photon sources in ordered arrays: Path to quantum optical circuits. \emph{Science Advances} \textbf{2022}, \emph{8}, eabn9252\relax
\mciteBstWouldAddEndPuncttrue
\mciteSetBstMidEndSepPunct{\mcitedefaultmidpunct}
{\mcitedefaultendpunct}{\mcitedefaultseppunct}\relax
\EndOfBibitem
\bibitem[Foreman \latin{et~al.}(2015)Foreman, Swaim, and Vollmer]{foreman2015whispering}
Foreman,~M.~R.; Swaim,~J.~D.; Vollmer,~F. Whispering gallery mode sensors. \emph{Advances in optics and photonics} \textbf{2015}, \emph{7}, 168--240\relax
\mciteBstWouldAddEndPuncttrue
\mciteSetBstMidEndSepPunct{\mcitedefaultmidpunct}
{\mcitedefaultendpunct}{\mcitedefaultseppunct}\relax
\EndOfBibitem
\bibitem[P{\"o}llinger \latin{et~al.}(2009)P{\"o}llinger, O’Shea, Warken, and Rauschenbeutel]{pollinger2009ultrahigh}
P{\"o}llinger,~M.; O’Shea,~D.; Warken,~F.; Rauschenbeutel,~A. Ultrahigh-Q tunable whispering-gallery-mode microresonator. \emph{Physical review letters} \textbf{2009}, \emph{103}, 053901\relax
\mciteBstWouldAddEndPuncttrue
\mciteSetBstMidEndSepPunct{\mcitedefaultmidpunct}
{\mcitedefaultendpunct}{\mcitedefaultseppunct}\relax
\EndOfBibitem
\bibitem[Guha \latin{et~al.}(2017)Guha, Marsault, Cadiz, Morgenroth, Ulin, Berkovitz, Lema{\^\i}tre, Gomez, Amo, Combri{\'e}, \latin{et~al.} others]{guha2017surface}
Guha,~B.; Marsault,~F.; Cadiz,~F.; Morgenroth,~L.; Ulin,~V.; Berkovitz,~V.; Lema{\^\i}tre,~A.; Gomez,~C.; Amo,~A.; Combri{\'e},~S.; others Surface-enhanced gallium arsenide photonic resonator with quality factor of 6$\times$ 10 6. \emph{Optica} \textbf{2017}, \emph{4}, 218--221\relax
\mciteBstWouldAddEndPuncttrue
\mciteSetBstMidEndSepPunct{\mcitedefaultmidpunct}
{\mcitedefaultendpunct}{\mcitedefaultseppunct}\relax
\EndOfBibitem
\bibitem[Strekalov \latin{et~al.}(2016)Strekalov, Marquardt, Matsko, Schwefel, and Leuchs]{strekalov2016nonlinear}
Strekalov,~D.~V.; Marquardt,~C.; Matsko,~A.~B.; Schwefel,~H.~G.; Leuchs,~G. Nonlinear and quantum optics with whispering gallery resonators. \emph{Journal of Optics} \textbf{2016}, \emph{18}, 123002\relax
\mciteBstWouldAddEndPuncttrue
\mciteSetBstMidEndSepPunct{\mcitedefaultmidpunct}
{\mcitedefaultendpunct}{\mcitedefaultseppunct}\relax
\EndOfBibitem
\bibitem[von Klitzing \latin{et~al.}(2001)von Klitzing, Long, Ilchenko, Hare, and Lefevre-Seguin]{von2001tunable}
von Klitzing,~W.; Long,~R.; Ilchenko,~V.~S.; Hare,~J.; Lefevre-Seguin,~V. Tunable whispering gallery modes for spectroscopy and CQED experiments. \emph{New journal of physics} \textbf{2001}, \emph{3}, 14\relax
\mciteBstWouldAddEndPuncttrue
\mciteSetBstMidEndSepPunct{\mcitedefaultmidpunct}
{\mcitedefaultendpunct}{\mcitedefaultseppunct}\relax
\EndOfBibitem
\bibitem[Salehzadeh \latin{et~al.}(2015)Salehzadeh, Djavid, Tran, Shih, and Mi]{salehzadeh2015optically}
Salehzadeh,~O.; Djavid,~M.; Tran,~N.~H.; Shih,~I.; Mi,~Z. Optically pumped two-dimensional MoS2 lasers operating at room-temperature. \emph{Nano letters} \textbf{2015}, \emph{15}, 5302--5306\relax
\mciteBstWouldAddEndPuncttrue
\mciteSetBstMidEndSepPunct{\mcitedefaultmidpunct}
{\mcitedefaultendpunct}{\mcitedefaultseppunct}\relax
\EndOfBibitem
\bibitem[Kryzhanovskaya \latin{et~al.}(2014)Kryzhanovskaya, Maximov, and Zhukov]{kryzhanovskaya2014whispering}
Kryzhanovskaya,~N.~V.; Maximov,~M.; Zhukov,~A.~E. Whispering-gallery mode microcavity quantum-dot lasers. \emph{Quantum Electronics} \textbf{2014}, \emph{44}, 189\relax
\mciteBstWouldAddEndPuncttrue
\mciteSetBstMidEndSepPunct{\mcitedefaultmidpunct}
{\mcitedefaultendpunct}{\mcitedefaultseppunct}\relax
\EndOfBibitem
\bibitem[Yan \latin{et~al.}(2020)Yan, Shi, Zang, Zhao, Du, and Leng]{yan2020stable}
Yan,~D.; Shi,~T.; Zang,~Z.; Zhao,~S.; Du,~J.; Leng,~Y. Stable and low-threshold whispering-gallery-mode lasing from modified CsPbBr3 perovskite quantum dots@ SiO2 sphere. \emph{Chemical Engineering Journal} \textbf{2020}, \emph{401}, 126066\relax
\mciteBstWouldAddEndPuncttrue
\mciteSetBstMidEndSepPunct{\mcitedefaultmidpunct}
{\mcitedefaultendpunct}{\mcitedefaultseppunct}\relax
\EndOfBibitem
\bibitem[Li \latin{et~al.}(2020)Li, Wang, Chen, Wang, Fan, Liang, Song, Xing, and Tang]{li2020stable}
Li,~X.; Wang,~K.; Chen,~M.; Wang,~S.; Fan,~Y.; Liang,~T.; Song,~Q.; Xing,~G.; Tang,~Z. Stable whispering gallery mode lasing from solution-processed formamidinium lead bromide perovskite microdisks. \emph{Advanced Optical Materials} \textbf{2020}, \emph{8}, 2000030\relax
\mciteBstWouldAddEndPuncttrue
\mciteSetBstMidEndSepPunct{\mcitedefaultmidpunct}
{\mcitedefaultendpunct}{\mcitedefaultseppunct}\relax
\EndOfBibitem
\bibitem[Duan \latin{et~al.}(2022)Duan, Zhang, Xiao, Zhao, Thung, Ding, Liu, Yang, Ta, and Sun]{duan2022ultralow}
Duan,~R.; Zhang,~Z.; Xiao,~L.; Zhao,~X.; Thung,~Y.~T.; Ding,~L.; Liu,~Z.; Yang,~J.; Ta,~V.~D.; Sun,~H. Ultralow-Threshold and High-Quality Whispering-Gallery-Mode Lasing from Colloidal Core/Hybrid-Shell Quantum Wells. \emph{Advanced Materials} \textbf{2022}, \emph{34}, 2108884\relax
\mciteBstWouldAddEndPuncttrue
\mciteSetBstMidEndSepPunct{\mcitedefaultmidpunct}
{\mcitedefaultendpunct}{\mcitedefaultseppunct}\relax
\EndOfBibitem
\bibitem[Wang \latin{et~al.}(2017)Wang, Ta, Leck, Tan, Wang, He, Ohl, Demir, and Sun]{wang2017robust}
Wang,~Y.; Ta,~V.~D.; Leck,~K.~S.; Tan,~B. H.~I.; Wang,~Z.; He,~T.; Ohl,~C.-D.; Demir,~H.~V.; Sun,~H. Robust whispering-gallery-mode microbubble lasers from colloidal quantum dots. \emph{Nano letters} \textbf{2017}, \emph{17}, 2640--2646\relax
\mciteBstWouldAddEndPuncttrue
\mciteSetBstMidEndSepPunct{\mcitedefaultmidpunct}
{\mcitedefaultendpunct}{\mcitedefaultseppunct}\relax
\EndOfBibitem
\bibitem[Sun \latin{et~al.}(2010)Sun, Dong, Xie, An, Shen, and Chen]{sun2010quasi}
Sun,~L.; Dong,~H.; Xie,~W.; An,~Z.; Shen,~X.; Chen,~Z. Quasi-whispering gallery modes of exciton-polaritons in a ZnO microrod. \emph{Optics express} \textbf{2010}, \emph{18}, 15371--15376\relax
\mciteBstWouldAddEndPuncttrue
\mciteSetBstMidEndSepPunct{\mcitedefaultmidpunct}
{\mcitedefaultendpunct}{\mcitedefaultseppunct}\relax
\EndOfBibitem
\bibitem[Brooks \latin{et~al.}(2021)Brooks, Chu, Liu, Schott, Ludwig, Wieck, Midolo, Lodahl, and Rotenberg]{brooks2021integrated}
Brooks,~A.; Chu,~X.-L.; Liu,~Z.; Schott,~R.; Ludwig,~A.; Wieck,~A.~D.; Midolo,~L.; Lodahl,~P.; Rotenberg,~N. Integrated Whispering-Gallery-Mode Resonator for Solid-State Coherent Quantum Photonics. \emph{Nano Letters} \textbf{2021}, \emph{21}, 8707--8714\relax
\mciteBstWouldAddEndPuncttrue
\mciteSetBstMidEndSepPunct{\mcitedefaultmidpunct}
{\mcitedefaultendpunct}{\mcitedefaultseppunct}\relax
\EndOfBibitem
\bibitem[Savona \latin{et~al.}(1995)Savona, Andreani, Schwendimann, and Quattropani]{savona1995quantum}
Savona,~V.; Andreani,~L.; Schwendimann,~P.; Quattropani,~A. Quantum well excitons in semiconductor microcavities: Unified treatment of weak and strong coupling regimes. \emph{Solid State Communications} \textbf{1995}, \emph{93}, 733--739\relax
\mciteBstWouldAddEndPuncttrue
\mciteSetBstMidEndSepPunct{\mcitedefaultmidpunct}
{\mcitedefaultendpunct}{\mcitedefaultseppunct}\relax
\EndOfBibitem
\bibitem[Skolnick \latin{et~al.}(1998)Skolnick, Fisher, and Whittaker]{skolnick1998strong}
Skolnick,~M.; Fisher,~T.; Whittaker,~D. Strong coupling phenomena in quantum microcavity structures. \emph{Semiconductor Science and Technology} \textbf{1998}, \emph{13}, 645\relax
\mciteBstWouldAddEndPuncttrue
\mciteSetBstMidEndSepPunct{\mcitedefaultmidpunct}
{\mcitedefaultendpunct}{\mcitedefaultseppunct}\relax
\EndOfBibitem
\bibitem[Weisbuch \latin{et~al.}(1992)Weisbuch, Nishioka, Ishikawa, and Arakawa]{weisbuch1992observation}
Weisbuch,~C.; Nishioka,~M.; Ishikawa,~A.; Arakawa,~Y. Observation of the coupled exciton-photon mode splitting in a semiconductor quantum microcavity. \emph{Physical review letters} \textbf{1992}, \emph{69}, 3314\relax
\mciteBstWouldAddEndPuncttrue
\mciteSetBstMidEndSepPunct{\mcitedefaultmidpunct}
{\mcitedefaultendpunct}{\mcitedefaultseppunct}\relax
\EndOfBibitem
\bibitem[Feltin \latin{et~al.}(2006)Feltin, Christmann, Butt{\'e}, Carlin, Mosca, and Grandjean]{feltin2006room}
Feltin,~E.; Christmann,~G.; Butt{\'e},~R.; Carlin,~J.-F.; Mosca,~M.; Grandjean,~N. Room temperature polariton luminescence from a Ga N/ Al Ga N quantum well microcavity. \emph{Applied physics letters} \textbf{2006}, \emph{89}, 071107\relax
\mciteBstWouldAddEndPuncttrue
\mciteSetBstMidEndSepPunct{\mcitedefaultmidpunct}
{\mcitedefaultendpunct}{\mcitedefaultseppunct}\relax
\EndOfBibitem
\bibitem[Liu \latin{et~al.}(2015)Liu, Galfsky, Sun, Xia, Lin, Lee, K{\'e}na-Cohen, and Menon]{liu2015strong}
Liu,~X.; Galfsky,~T.; Sun,~Z.; Xia,~F.; Lin,~E.-c.; Lee,~Y.-H.; K{\'e}na-Cohen,~S.; Menon,~V.~M. Strong light--matter coupling in two-dimensional atomic crystals. \emph{Nature Photonics} \textbf{2015}, \emph{9}, 30--34\relax
\mciteBstWouldAddEndPuncttrue
\mciteSetBstMidEndSepPunct{\mcitedefaultmidpunct}
{\mcitedefaultendpunct}{\mcitedefaultseppunct}\relax
\EndOfBibitem
\bibitem[Zhao \latin{et~al.}(2020)Zhao, Yan, Cui, Liu, Wang, Sun, Chen, and Lu]{zhao2020realization}
Zhao,~X.; Yan,~Y.; Cui,~Z.; Liu,~F.; Wang,~S.; Sun,~L.; Chen,~Y.; Lu,~W. Realization of strong coupling between 2D excitons and cavity photons at room temperature. \emph{Optics Letters} \textbf{2020}, \emph{45}, 6571--6574\relax
\mciteBstWouldAddEndPuncttrue
\mciteSetBstMidEndSepPunct{\mcitedefaultmidpunct}
{\mcitedefaultendpunct}{\mcitedefaultseppunct}\relax
\EndOfBibitem
\bibitem[Wu \latin{et~al.}(2021)Wu, Ghosh, Su, Fieramosca, Liew, and Xiong]{wu2021nonlinear}
Wu,~J.; Ghosh,~S.; Su,~R.; Fieramosca,~A.; Liew,~T.~C.; Xiong,~Q. Nonlinear parametric scattering of exciton polaritons in perovskite microcavities. \emph{Nano Letters} \textbf{2021}, \emph{21}, 3120--3126\relax
\mciteBstWouldAddEndPuncttrue
\mciteSetBstMidEndSepPunct{\mcitedefaultmidpunct}
{\mcitedefaultendpunct}{\mcitedefaultseppunct}\relax
\EndOfBibitem
\bibitem[Fieramosca \latin{et~al.}(2019)Fieramosca, Polimeno, Ardizzone, De~Marco, Pugliese, Maiorano, De~Giorgi, Dominici, Gigli, Gerace, \latin{et~al.} others]{fieramosca2019two}
Fieramosca,~A.; Polimeno,~L.; Ardizzone,~V.; De~Marco,~L.; Pugliese,~M.; Maiorano,~V.; De~Giorgi,~M.; Dominici,~L.; Gigli,~G.; Gerace,~D.; others Two-dimensional hybrid perovskites sustaining strong polariton interactions at room temperature. \emph{Science advances} \textbf{2019}, \emph{5}, eaav9967\relax
\mciteBstWouldAddEndPuncttrue
\mciteSetBstMidEndSepPunct{\mcitedefaultmidpunct}
{\mcitedefaultendpunct}{\mcitedefaultseppunct}\relax
\EndOfBibitem
\bibitem[Su \latin{et~al.}(2021)Su, Fieramosca, Zhang, Nguyen, Deleporte, Chen, Sanvitto, Liew, and Xiong]{su2021perovskite}
Su,~R.; Fieramosca,~A.; Zhang,~Q.; Nguyen,~H.~S.; Deleporte,~E.; Chen,~Z.; Sanvitto,~D.; Liew,~T.~C.; Xiong,~Q. Perovskite semiconductors for room-temperature exciton-polaritonics. \emph{Nature Materials} \textbf{2021}, \emph{20}, 1315--1324\relax
\mciteBstWouldAddEndPuncttrue
\mciteSetBstMidEndSepPunct{\mcitedefaultmidpunct}
{\mcitedefaultendpunct}{\mcitedefaultseppunct}\relax
\EndOfBibitem
\bibitem[Li \latin{et~al.}(2018)Li, Singh, and Sievenpiper]{li2018metasurfaces}
Li,~A.; Singh,~S.; Sievenpiper,~D. Metasurfaces and their applications. \emph{Nanophotonics} \textbf{2018}, \emph{7}, 989--1011\relax
\mciteBstWouldAddEndPuncttrue
\mciteSetBstMidEndSepPunct{\mcitedefaultmidpunct}
{\mcitedefaultendpunct}{\mcitedefaultseppunct}\relax
\EndOfBibitem
\bibitem[Chen \latin{et~al.}(2020)Chen, Miao, Wang, Zhong, Saxena, Chow, Whitehead, Gerace, Xu, Shi, \latin{et~al.} others]{chen2020metasurface}
Chen,~Y.; Miao,~S.; Wang,~T.; Zhong,~D.; Saxena,~A.; Chow,~C.; Whitehead,~J.; Gerace,~D.; Xu,~X.; Shi,~S.-F.; others Metasurface integrated monolayer exciton polariton. \emph{Nano Letters} \textbf{2020}, \emph{20}, 5292--5300\relax
\mciteBstWouldAddEndPuncttrue
\mciteSetBstMidEndSepPunct{\mcitedefaultmidpunct}
{\mcitedefaultendpunct}{\mcitedefaultseppunct}\relax
\EndOfBibitem
\bibitem[Zhang \latin{et~al.}(2017)Zhang, Choi, Wang, Naylor, Johnson, and Cubukcu]{zhang2017unidirectional}
Zhang,~X.; Choi,~S.; Wang,~D.; Naylor,~C.~H.; Johnson,~A.~C.; Cubukcu,~E. Unidirectional doubly enhanced MoS2 emission via photonic Fano resonances. \emph{Nano Letters} \textbf{2017}, \emph{17}, 6715--6720\relax
\mciteBstWouldAddEndPuncttrue
\mciteSetBstMidEndSepPunct{\mcitedefaultmidpunct}
{\mcitedefaultendpunct}{\mcitedefaultseppunct}\relax
\EndOfBibitem
\bibitem[Zhang \latin{et~al.}(2018)Zhang, Gogna, Burg, Tutuc, and Deng]{zhang2018photonic}
Zhang,~L.; Gogna,~R.; Burg,~W.; Tutuc,~E.; Deng,~H. Photonic-crystal exciton-polaritons in monolayer semiconductors. \emph{Nature communications} \textbf{2018}, \emph{9}, 713\relax
\mciteBstWouldAddEndPuncttrue
\mciteSetBstMidEndSepPunct{\mcitedefaultmidpunct}
{\mcitedefaultendpunct}{\mcitedefaultseppunct}\relax
\EndOfBibitem
\bibitem[Koshelev \latin{et~al.}(2019)Koshelev, Bogdanov, and Kivshar]{koshelev2019meta}
Koshelev,~K.; Bogdanov,~A.; Kivshar,~Y. Meta-optics and bound states in the continuum. \emph{Science Bulletin} \textbf{2019}, \emph{64}, 836--842\relax
\mciteBstWouldAddEndPuncttrue
\mciteSetBstMidEndSepPunct{\mcitedefaultmidpunct}
{\mcitedefaultendpunct}{\mcitedefaultseppunct}\relax
\EndOfBibitem
\bibitem[Al-Ani \latin{et~al.}(2021)Al-Ani, As'~Ham, Huang, Miroshnichenko, and Hattori]{al2021enhanced}
Al-Ani,~I.~A.; As'~Ham,~K.; Huang,~L.; Miroshnichenko,~A.~E.; Hattori,~H.~T. Enhanced strong coupling of TMDC monolayers by bound state in the continuum. \emph{Laser \& Photonics Reviews} \textbf{2021}, \emph{15}, 2100240\relax
\mciteBstWouldAddEndPuncttrue
\mciteSetBstMidEndSepPunct{\mcitedefaultmidpunct}
{\mcitedefaultendpunct}{\mcitedefaultseppunct}\relax
\EndOfBibitem
\bibitem[Al-Ani \latin{et~al.}(2022)Al-Ani, As'~Ham, Huang, Miroshnichenko, Lei, and Hattori]{al2022strong}
Al-Ani,~I.~A.; As'~Ham,~K.; Huang,~L.; Miroshnichenko,~A.~E.; Lei,~W.; Hattori,~H.~T. Strong coupling of exciton and high-Q mode in all-perovskite metasurfaces. \emph{Advanced Optical Materials} \textbf{2022}, \emph{10}, 2101120\relax
\mciteBstWouldAddEndPuncttrue
\mciteSetBstMidEndSepPunct{\mcitedefaultmidpunct}
{\mcitedefaultendpunct}{\mcitedefaultseppunct}\relax
\EndOfBibitem
\bibitem[Pelton and Bryant(2013)Pelton, and Bryant]{Pelton2013}
Pelton,~M.; Bryant,~G.~W. \emph{Introduction to Metal-Nanoparticle Plasmonics}; John Wiley \& Sons.: Hoboken, NJ, U.S.A., 2013\relax
\mciteBstWouldAddEndPuncttrue
\mciteSetBstMidEndSepPunct{\mcitedefaultmidpunct}
{\mcitedefaultendpunct}{\mcitedefaultseppunct}\relax
\EndOfBibitem
\bibitem[Chang \latin{et~al.}(2007)Chang, S{\o}rensen, Demler, and Lukin]{chang2007single}
Chang,~D.~E.; S{\o}rensen,~A.~S.; Demler,~E.~A.; Lukin,~M.~D. A single-photon transistor using nanoscale surface plasmons. \emph{Nature physics} \textbf{2007}, \emph{3}, 807--812\relax
\mciteBstWouldAddEndPuncttrue
\mciteSetBstMidEndSepPunct{\mcitedefaultmidpunct}
{\mcitedefaultendpunct}{\mcitedefaultseppunct}\relax
\EndOfBibitem
\bibitem[Gullans \latin{et~al.}(2013)Gullans, Chang, Koppens, de~Abajo, and Lukin]{gullans2013single}
Gullans,~M.; Chang,~D.; Koppens,~F.; de~Abajo,~F.~G.; Lukin,~M.~D. Single-photon nonlinear optics with graphene plasmons. \emph{Physical review letters} \textbf{2013}, \emph{111}, 247401\relax
\mciteBstWouldAddEndPuncttrue
\mciteSetBstMidEndSepPunct{\mcitedefaultmidpunct}
{\mcitedefaultendpunct}{\mcitedefaultseppunct}\relax
\EndOfBibitem
\bibitem[Pockrand \latin{et~al.}(1982)Pockrand, Brillante, and Möbius]{Pockrand1982SPP}
Pockrand,~I.; Brillante,~A.; Möbius,~D. {Exciton–surface plasmon coupling: An experimental investigation}. \emph{The Journal of Chemical Physics} \textbf{1982}, \emph{77}, 6289--6295\relax
\mciteBstWouldAddEndPuncttrue
\mciteSetBstMidEndSepPunct{\mcitedefaultmidpunct}
{\mcitedefaultendpunct}{\mcitedefaultseppunct}\relax
\EndOfBibitem
\bibitem[Bellessa \latin{et~al.}(2004)Bellessa, Bonnand, Plenet, and Mugnier]{Bellessa2004Jagg}
Bellessa,~J.; Bonnand,~C.; Plenet,~J.~C.; Mugnier,~J. Strong Coupling between Surface Plasmons and Excitons in an Organic Semiconductor. \emph{Phys. Rev. Lett.} \textbf{2004}, \emph{93}, 036404\relax
\mciteBstWouldAddEndPuncttrue
\mciteSetBstMidEndSepPunct{\mcitedefaultmidpunct}
{\mcitedefaultendpunct}{\mcitedefaultseppunct}\relax
\EndOfBibitem
\bibitem[Dintinger \latin{et~al.}(2005)Dintinger, Klein, Bustos, Barnes, and Ebbesen]{dintinger2005strong}
Dintinger,~J.; Klein,~S.; Bustos,~F.; Barnes,~W.~L.; Ebbesen,~T. Strong coupling between surface plasmon-polaritons and organic molecules in subwavelength hole arrays. \emph{Physical Review B} \textbf{2005}, \emph{71}, 035424\relax
\mciteBstWouldAddEndPuncttrue
\mciteSetBstMidEndSepPunct{\mcitedefaultmidpunct}
{\mcitedefaultendpunct}{\mcitedefaultseppunct}\relax
\EndOfBibitem
\bibitem[Sugawara \latin{et~al.}(2006)Sugawara, Kelf, Baumberg, Abdelsalam, and Bartlett]{sugawara2006Jagg}
Sugawara,~Y.; Kelf,~T.~A.; Baumberg,~J.~J.; Abdelsalam,~M.~E.; Bartlett,~P.~N. Strong Coupling between Localized Plasmons and Organic Excitons in Metal Nanovoids. \emph{Phys. Rev. Lett.} \textbf{2006}, \emph{97}, 266808\relax
\mciteBstWouldAddEndPuncttrue
\mciteSetBstMidEndSepPunct{\mcitedefaultmidpunct}
{\mcitedefaultendpunct}{\mcitedefaultseppunct}\relax
\EndOfBibitem
\bibitem[Wiederrecht \latin{et~al.}(2004)Wiederrecht, Wurtz, and Hranisavljevic]{Wiederrecht:2004kx}
Wiederrecht,~G.; Wurtz,~G.; Hranisavljevic,~J. {Coherent coupling of molecular excitons to electronic polarizations of noble metal nanoparticles}. \emph{Nano Letters} \textbf{2004}, \emph{4}, 2121 2125\relax
\mciteBstWouldAddEndPuncttrue
\mciteSetBstMidEndSepPunct{\mcitedefaultmidpunct}
{\mcitedefaultendpunct}{\mcitedefaultseppunct}\relax
\EndOfBibitem
\bibitem[Fofang \latin{et~al.}(2008)Fofang, Park, Neumann, Mirin, Nordlander, and Halas]{fofang2008plexcitonic}
Fofang,~N.~T.; Park,~T.-H.; Neumann,~O.; Mirin,~N.~A.; Nordlander,~P.; Halas,~N.~J. Plexcitonic nanoparticles: plasmon- exciton coupling in nanoshell- J-aggregate complexes. \emph{Nano letters} \textbf{2008}, \emph{8}, 3481--3487\relax
\mciteBstWouldAddEndPuncttrue
\mciteSetBstMidEndSepPunct{\mcitedefaultmidpunct}
{\mcitedefaultendpunct}{\mcitedefaultseppunct}\relax
\EndOfBibitem
\bibitem[DeLacy \latin{et~al.}(2015)DeLacy, Miller, Hsu, Zander, Lacey, Yagloski, Fountain, Valdes, Anquillare, Soljačić, \latin{et~al.} others]{delacy2015coherent}
DeLacy,~B.~G.; Miller,~O.~D.; Hsu,~C.~W.; Zander,~Z.; Lacey,~S.; Yagloski,~R.; Fountain,~A.~W.; Valdes,~E.; Anquillare,~E.; Soljačić,~M.; others Coherent plasmon-exciton coupling in silver platelet-J-aggregate nanocomposites. \emph{Nano letters} \textbf{2015}, \emph{15}, 2588--2593\relax
\mciteBstWouldAddEndPuncttrue
\mciteSetBstMidEndSepPunct{\mcitedefaultmidpunct}
{\mcitedefaultendpunct}{\mcitedefaultseppunct}\relax
\EndOfBibitem
\bibitem[Das \latin{et~al.}(2017)Das, Hazra, and Chandra]{das2017exploring}
Das,~K.; Hazra,~B.; Chandra,~M. Exploring the coherent interaction in a hybrid system of hollow gold nanoprisms and cyanine dye J-aggregates: role of plasmon-hybridization mediated local electric-field enhancement. \emph{Physical Chemistry Chemical Physics} \textbf{2017}, \emph{19}, 27997--28005\relax
\mciteBstWouldAddEndPuncttrue
\mciteSetBstMidEndSepPunct{\mcitedefaultmidpunct}
{\mcitedefaultendpunct}{\mcitedefaultseppunct}\relax
\EndOfBibitem
\bibitem[Kirschner \latin{et~al.}(2017)Kirschner, Ding, Li, Chapman, Lei, Lin, Chen, Schatz, and Schaller]{kirschner2017phonon}
Kirschner,~M.~S.; Ding,~W.; Li,~Y.; Chapman,~C.~T.; Lei,~A.; Lin,~X.-M.; Chen,~L.~X.; Schatz,~G.~C.; Schaller,~R.~D. Phonon-driven oscillatory plasmonic excitonic nanomaterials. \emph{Nano letters} \textbf{2017}, \emph{18}, 442--448\relax
\mciteBstWouldAddEndPuncttrue
\mciteSetBstMidEndSepPunct{\mcitedefaultmidpunct}
{\mcitedefaultendpunct}{\mcitedefaultseppunct}\relax
\EndOfBibitem
\bibitem[Liu \latin{et~al.}(2017)Liu, Zhou, Yu, Zhang, Wang, Liu, Wei, Chen, and Wang]{liu2017strong}
Liu,~R.; Zhou,~Z.-K.; Yu,~Y.-C.; Zhang,~T.; Wang,~H.; Liu,~G.; Wei,~Y.; Chen,~H.; Wang,~X.-H. Strong light-matter interactions in single open plasmonic nanocavities at the quantum optics limit. \emph{Physical review letters} \textbf{2017}, \emph{118}, 237401\relax
\mciteBstWouldAddEndPuncttrue
\mciteSetBstMidEndSepPunct{\mcitedefaultmidpunct}
{\mcitedefaultendpunct}{\mcitedefaultseppunct}\relax
\EndOfBibitem
\bibitem[Zheng \latin{et~al.}(2017)Zheng, Zhang, Deng, Kang, Nordlander, and Xu]{zheng2017manipulating}
Zheng,~D.; Zhang,~S.; Deng,~Q.; Kang,~M.; Nordlander,~P.; Xu,~H. Manipulating coherent plasmon--exciton interaction in a single silver nanorod on monolayer WSe2. \emph{Nano letters} \textbf{2017}, \emph{17}, 3809--3814\relax
\mciteBstWouldAddEndPuncttrue
\mciteSetBstMidEndSepPunct{\mcitedefaultmidpunct}
{\mcitedefaultendpunct}{\mcitedefaultseppunct}\relax
\EndOfBibitem
\bibitem[Cuadra \latin{et~al.}(2018)Cuadra, Baranov, Wersall, Verre, Antosiewicz, and Shegai]{cuadra2018observation}
Cuadra,~J.; Baranov,~D.~G.; Wersall,~M.; Verre,~R.; Antosiewicz,~T.~J.; Shegai,~T. Observation of tunable charged exciton polaritons in hybrid monolayer WS2- plasmonic nanoantenna system. \emph{Nano letters} \textbf{2018}, \emph{18}, 1777--1785\relax
\mciteBstWouldAddEndPuncttrue
\mciteSetBstMidEndSepPunct{\mcitedefaultmidpunct}
{\mcitedefaultendpunct}{\mcitedefaultseppunct}\relax
\EndOfBibitem
\bibitem[Lawless \latin{et~al.}(2020)Lawless, Hrelescu, Elliott, Peters, McEvoy, and Bradley]{lawless2020influence}
Lawless,~J.; Hrelescu,~C.; Elliott,~C.; Peters,~L.; McEvoy,~N.; Bradley,~A.~L. Influence of gold nano-bipyramid dimensions on strong coupling with excitons of monolayer MoS2. \emph{ACS Applied Materials \& Interfaces} \textbf{2020}, \emph{12}, 46406--46415\relax
\mciteBstWouldAddEndPuncttrue
\mciteSetBstMidEndSepPunct{\mcitedefaultmidpunct}
{\mcitedefaultendpunct}{\mcitedefaultseppunct}\relax
\EndOfBibitem
\bibitem[Muckel \latin{et~al.}(2021)Muckel, Guye, Gallagher, Liu, and Ginger]{muckel2021tuning}
Muckel,~F.; Guye,~K.~N.; Gallagher,~S.~M.; Liu,~Y.; Ginger,~D.~S. Tuning Hybrid exciton--Photon Fano Resonances in Two-Dimensional Organic--Inorganic Perovskite Thin Films. \emph{Nano Letters} \textbf{2021}, \emph{21}, 6124--6131\relax
\mciteBstWouldAddEndPuncttrue
\mciteSetBstMidEndSepPunct{\mcitedefaultmidpunct}
{\mcitedefaultendpunct}{\mcitedefaultseppunct}\relax
\EndOfBibitem
\bibitem[Chikkaraddy \latin{et~al.}(2016)Chikkaraddy, De~Nijs, Benz, Barrow, Scherman, Rosta, Demetriadou, Fox, Hess, and Baumberg]{chikkaraddy2016single}
Chikkaraddy,~R.; De~Nijs,~B.; Benz,~F.; Barrow,~S.~J.; Scherman,~O.~A.; Rosta,~E.; Demetriadou,~A.; Fox,~P.; Hess,~O.; Baumberg,~J.~J. Single-molecule strong coupling at room temperature in plasmonic nanocavities. \emph{Nature} \textbf{2016}, \emph{535}, 127--130\relax
\mciteBstWouldAddEndPuncttrue
\mciteSetBstMidEndSepPunct{\mcitedefaultmidpunct}
{\mcitedefaultendpunct}{\mcitedefaultseppunct}\relax
\EndOfBibitem
\bibitem[Santhosh \latin{et~al.}(2016)Santhosh, Bitton, Chuntonov, and Haran]{santhosh2016vacuum}
Santhosh,~K.; Bitton,~O.; Chuntonov,~L.; Haran,~G. Vacuum Rabi splitting in a plasmonic cavity at the single quantum emitter limit. \emph{Nature communications} \textbf{2016}, \emph{7}, ncomms11823\relax
\mciteBstWouldAddEndPuncttrue
\mciteSetBstMidEndSepPunct{\mcitedefaultmidpunct}
{\mcitedefaultendpunct}{\mcitedefaultseppunct}\relax
\EndOfBibitem
\bibitem[Park \latin{et~al.}(2019)Park, May, Leng, Wang, Kropp, Gougousi, Pelton, and Raschke]{park2019tip}
Park,~K.-D.; May,~M.~A.; Leng,~H.; Wang,~J.; Kropp,~J.~A.; Gougousi,~T.; Pelton,~M.; Raschke,~M.~B. Tip-enhanced strong coupling spectroscopy, imaging, and control of a single quantum emitter. \emph{Science advances} \textbf{2019}, \emph{5}, eaav5931\relax
\mciteBstWouldAddEndPuncttrue
\mciteSetBstMidEndSepPunct{\mcitedefaultmidpunct}
{\mcitedefaultendpunct}{\mcitedefaultseppunct}\relax
\EndOfBibitem
\bibitem[Liu \latin{et~al.}(2024)Liu, Geng, Ai, Fan, Liu, Lu, Kuang, Liu, Guo, and Wu]{Liu2024dimer}
Liu,~R.; Geng,~M.; Ai,~J.; Fan,~X.; Liu,~Z.; Lu,~Y.-W.; Kuang,~Y.; Liu,~J.-F.; Guo,~L.; Wu,~L. {Deterministic positioning and alignment of a single-molecule exciton in plasmonic nanodimer for strong coupling}. \emph{Nature Communications} \textbf{2024}, \emph{15}, 4103\relax
\mciteBstWouldAddEndPuncttrue
\mciteSetBstMidEndSepPunct{\mcitedefaultmidpunct}
{\mcitedefaultendpunct}{\mcitedefaultseppunct}\relax
\EndOfBibitem
\bibitem[Chikkaraddy \latin{et~al.}(2018)Chikkaraddy, Turek, Kongsuwan, Benz, Carnegie, Van De~Goor, De~Nijs, Demetriadou, Hess, Keyser, \latin{et~al.} others]{chikkaraddy2018mapping}
Chikkaraddy,~R.; Turek,~V.; Kongsuwan,~N.; Benz,~F.; Carnegie,~C.; Van De~Goor,~T.; De~Nijs,~B.; Demetriadou,~A.; Hess,~O.; Keyser,~U.~F.; others Mapping nanoscale hotspots with single-molecule emitters assembled into plasmonic nanocavities using DNA origami. \emph{Nano letters} \textbf{2018}, \emph{18}, 405--411\relax
\mciteBstWouldAddEndPuncttrue
\mciteSetBstMidEndSepPunct{\mcitedefaultmidpunct}
{\mcitedefaultendpunct}{\mcitedefaultseppunct}\relax
\EndOfBibitem
\bibitem[Ojambati \latin{et~al.}(2019)Ojambati, Chikkaraddy, Deacon, Horton, Kos, Turek, Keyser, and Baumberg]{ojambati2019quantum}
Ojambati,~O.~S.; Chikkaraddy,~R.; Deacon,~W.~D.; Horton,~M.; Kos,~D.; Turek,~V.~A.; Keyser,~U.~F.; Baumberg,~J.~J. Quantum electrodynamics at room temperature coupling a single vibrating molecule with a plasmonic nanocavity. \emph{Nature communications} \textbf{2019}, \emph{10}, 1049\relax
\mciteBstWouldAddEndPuncttrue
\mciteSetBstMidEndSepPunct{\mcitedefaultmidpunct}
{\mcitedefaultendpunct}{\mcitedefaultseppunct}\relax
\EndOfBibitem
\bibitem[Leng \latin{et~al.}(2018)Leng, Szychowski, Daniel, and Pelton]{leng2018strong}
Leng,~H.; Szychowski,~B.; Daniel,~M.-C.; Pelton,~M. Strong coupling and induced transparency at room temperature with single quantum dots and gap plasmons. \emph{Nature communications} \textbf{2018}, \emph{9}, 4012\relax
\mciteBstWouldAddEndPuncttrue
\mciteSetBstMidEndSepPunct{\mcitedefaultmidpunct}
{\mcitedefaultendpunct}{\mcitedefaultseppunct}\relax
\EndOfBibitem
\bibitem[Li \latin{et~al.}(2022)Li, Li, Liu, Zhong, Liu, Chen, and Wang]{li2022room}
Li,~J.-Y.; Li,~W.; Liu,~J.; Zhong,~J.; Liu,~R.; Chen,~H.; Wang,~X.-H. Room-temperature strong coupling between a single quantum dot and a single plasmonic nanoparticle. \emph{Nano Letters} \textbf{2022}, \emph{22}, 4686--4693\relax
\mciteBstWouldAddEndPuncttrue
\mciteSetBstMidEndSepPunct{\mcitedefaultmidpunct}
{\mcitedefaultendpunct}{\mcitedefaultseppunct}\relax
\EndOfBibitem
\bibitem[Gro{\ss} \latin{et~al.}(2018)Gro{\ss}, Hamm, Tufarelli, Hess, and Hecht]{gross2018near}
Gro{\ss},~H.; Hamm,~J.~M.; Tufarelli,~T.; Hess,~O.; Hecht,~B. Near-field strong coupling of single quantum dots. \emph{Science advances} \textbf{2018}, \emph{4}, eaar4906\relax
\mciteBstWouldAddEndPuncttrue
\mciteSetBstMidEndSepPunct{\mcitedefaultmidpunct}
{\mcitedefaultendpunct}{\mcitedefaultseppunct}\relax
\EndOfBibitem
\bibitem[Gupta \latin{et~al.}(2021)Gupta, Bitton, Neuman, Esteban, Chuntonov, Aizpurua, and Haran]{gupta2021complex}
Gupta,~S.~N.; Bitton,~O.; Neuman,~T.; Esteban,~R.; Chuntonov,~L.; Aizpurua,~J.; Haran,~G. Complex plasmon-exciton dynamics revealed through quantum dot light emission in a nanocavity. \emph{Nature Communications} \textbf{2021}, \emph{12}, 1310\relax
\mciteBstWouldAddEndPuncttrue
\mciteSetBstMidEndSepPunct{\mcitedefaultmidpunct}
{\mcitedefaultendpunct}{\mcitedefaultseppunct}\relax
\EndOfBibitem
\bibitem[Eich \latin{et~al.}(2022)Eich, Spiekermann, Gehring, Sommer, Bankwitz, Schrinner, Preu{\ss}, Michaelis~de Vasconcellos, Bratschitsch, Pernice, \latin{et~al.} others]{eich2022single}
Eich,~A.; Spiekermann,~T.~C.; Gehring,~H.; Sommer,~L.; Bankwitz,~J.~R.; Schrinner,~P.~P.; Preu{\ss},~J.~A.; Michaelis~de Vasconcellos,~S.; Bratschitsch,~R.; Pernice,~W.~H.; others Single-photon emission from individual nanophotonic-integrated colloidal quantum dots. \emph{ACS Photonics} \textbf{2022}, \emph{9}, 551--558\relax
\mciteBstWouldAddEndPuncttrue
\mciteSetBstMidEndSepPunct{\mcitedefaultmidpunct}
{\mcitedefaultendpunct}{\mcitedefaultseppunct}\relax
\EndOfBibitem
\bibitem[Nguyen \latin{et~al.}(2022)Nguyen, Sharp, Fröch, Cai, Wu, Monahan, Munley, Manna, Majumdar, Kagan, \latin{et~al.} others]{nguyen2022deterministic}
Nguyen,~H.~A.; Sharp,~D.; Fröch,~J.~E.; Cai,~Y.-Y.; Wu,~S.; Monahan,~M.; Munley,~C.; Manna,~A.; Majumdar,~A.; Kagan,~C.~R.; others Deterministic Quantum Light Arrays from Giant Silica-Shelled Quantum Dots. \emph{ACS Applied Materials \& Interfaces} \textbf{2022}, \emph{15}, 4294--4302\relax
\mciteBstWouldAddEndPuncttrue
\mciteSetBstMidEndSepPunct{\mcitedefaultmidpunct}
{\mcitedefaultendpunct}{\mcitedefaultseppunct}\relax
\EndOfBibitem
\bibitem[Barelli \latin{et~al.}(2023)Barelli, Vidal, Fiorito, Myslovska, Cielecki, Aglieri, Moreels, Sapienza, and Di~Stasio]{barelli2023single}
Barelli,~M.; Vidal,~C.; Fiorito,~S.; Myslovska,~A.; Cielecki,~D.; Aglieri,~V.; Moreels,~I.; Sapienza,~R.; Di~Stasio,~F. Single-Photon Emitting Arrays by Capillary Assembly of Colloidal Semiconductor CdSe/CdS/SiO2 Nanocrystals. \emph{ACS photonics} \textbf{2023}, \emph{10}, 1662--1670\relax
\mciteBstWouldAddEndPuncttrue
\mciteSetBstMidEndSepPunct{\mcitedefaultmidpunct}
{\mcitedefaultendpunct}{\mcitedefaultseppunct}\relax
\EndOfBibitem
\bibitem[Litvin \latin{et~al.}(2017)Litvin, Martynenko, Purcell-Milton, Baranov, Fedorov, and Gun'Ko]{litvin2017colloidal}
Litvin,~A.; Martynenko,~I.; Purcell-Milton,~F.; Baranov,~A.; Fedorov,~A.; Gun'Ko,~Y. Colloidal quantum dots for optoelectronics. \emph{Journal of Materials Chemistry A} \textbf{2017}, \emph{5}, 13252--13275\relax
\mciteBstWouldAddEndPuncttrue
\mciteSetBstMidEndSepPunct{\mcitedefaultmidpunct}
{\mcitedefaultendpunct}{\mcitedefaultseppunct}\relax
\EndOfBibitem
\bibitem[Kagan \latin{et~al.}(2020)Kagan, Bassett, Murray, and Thompson]{kagan2020colloidal}
Kagan,~C.~R.; Bassett,~L.~C.; Murray,~C.~B.; Thompson,~S.~M. Colloidal quantum dots as platforms for quantum information science. \emph{Chemical reviews} \textbf{2020}, \emph{121}, 3186--3233\relax
\mciteBstWouldAddEndPuncttrue
\mciteSetBstMidEndSepPunct{\mcitedefaultmidpunct}
{\mcitedefaultendpunct}{\mcitedefaultseppunct}\relax
\EndOfBibitem
\bibitem[Kahl \latin{et~al.}(2007)Kahl, Thomay, Kohnle, Beha, Merlein, Hagner, Halm, Ziegler, Nann, Fedutik, \latin{et~al.} others]{kahl2007colloidal}
Kahl,~M.; Thomay,~T.; Kohnle,~V.; Beha,~K.; Merlein,~J.; Hagner,~M.; Halm,~A.; Ziegler,~J.; Nann,~T.; Fedutik,~Y.; others Colloidal quantum dots in all-dielectric high-Q pillar microcavities. \emph{Nano letters} \textbf{2007}, \emph{7}, 2897--2900\relax
\mciteBstWouldAddEndPuncttrue
\mciteSetBstMidEndSepPunct{\mcitedefaultmidpunct}
{\mcitedefaultendpunct}{\mcitedefaultseppunct}\relax
\EndOfBibitem
\bibitem[Abudayyeh \latin{et~al.}(2021)Abudayyeh, Lubotzky, Blake, Wang, Majumder, Hu, Kim, Htoon, Bose, Malko, \latin{et~al.} others]{abudayyeh2021single}
Abudayyeh,~H.; Lubotzky,~B.; Blake,~A.; Wang,~J.; Majumder,~S.; Hu,~Z.; Kim,~Y.; Htoon,~H.; Bose,~R.; Malko,~A.~V.; others Single photon sources with near unity collection efficiencies by deterministic placement of quantum dots in nanoantennas. \emph{APL Photonics} \textbf{2021}, \emph{6}, 036109\relax
\mciteBstWouldAddEndPuncttrue
\mciteSetBstMidEndSepPunct{\mcitedefaultmidpunct}
{\mcitedefaultendpunct}{\mcitedefaultseppunct}\relax
\EndOfBibitem
\bibitem[Diroll \latin{et~al.}(2023)Diroll, Guzelturk, Po, Dabard, Fu, Makke, Lhuillier, and Ithurria]{diroll20232d}
Diroll,~B.~T.; Guzelturk,~B.; Po,~H.; Dabard,~C.; Fu,~N.; Makke,~L.; Lhuillier,~E.; Ithurria,~S. 2D II--VI Semiconductor Nanoplatelets: From Material Synthesis to Optoelectronic Integration. \emph{Chemical Reviews} \textbf{2023}, \relax
\mciteBstWouldAddEndPunctfalse
\mciteSetBstMidEndSepPunct{\mcitedefaultmidpunct}
{}{\mcitedefaultseppunct}\relax
\EndOfBibitem
\bibitem[Naeem \latin{et~al.}(2015)Naeem, Masia, Christodoulou, Moreels, Borri, and Langbein]{naeem2015giant}
Naeem,~A.; Masia,~F.; Christodoulou,~S.; Moreels,~I.; Borri,~P.; Langbein,~W. Giant exciton oscillator strength and radiatively limited dephasing in two-dimensional platelets. \emph{Physical Review B} \textbf{2015}, \emph{91}, 121302\relax
\mciteBstWouldAddEndPuncttrue
\mciteSetBstMidEndSepPunct{\mcitedefaultmidpunct}
{\mcitedefaultendpunct}{\mcitedefaultseppunct}\relax
\EndOfBibitem
\bibitem[Chen \latin{et~al.}(2022)Chen, Sharp, Saxena, Nguyen, Cossairt, and Majumdar]{chen2022integrated}
Chen,~Y.; Sharp,~D.; Saxena,~A.; Nguyen,~H.; Cossairt,~B.~M.; Majumdar,~A. Integrated Quantum Nanophotonics with Solution-Processed Materials. \emph{Advanced Quantum Technologies} \textbf{2022}, \emph{5}, 2100078\relax
\mciteBstWouldAddEndPuncttrue
\mciteSetBstMidEndSepPunct{\mcitedefaultmidpunct}
{\mcitedefaultendpunct}{\mcitedefaultseppunct}\relax
\EndOfBibitem
\bibitem[Flatten \latin{et~al.}(2016)Flatten, Christodoulou, Patel, Buccheri, Coles, Reid, Taylor, Moreels, and Smith]{flatten2016strong}
Flatten,~L.~C.; Christodoulou,~S.; Patel,~R.~K.; Buccheri,~A.; Coles,~D.~M.; Reid,~B.~P.; Taylor,~R.~A.; Moreels,~I.; Smith,~J.~M. Strong exciton--photon coupling with colloidal nanoplatelets in an open microcavity. \emph{Nano Letters} \textbf{2016}, \emph{16}, 7137--7141\relax
\mciteBstWouldAddEndPuncttrue
\mciteSetBstMidEndSepPunct{\mcitedefaultmidpunct}
{\mcitedefaultendpunct}{\mcitedefaultseppunct}\relax
\EndOfBibitem
\bibitem[Chen and Chen(2019)Chen, and Chen]{chen2019luminescent}
Chen,~D.; Chen,~X. Luminescent perovskite quantum dots: synthesis, microstructures, optical properties and applications. \emph{Journal of Materials Chemistry C} \textbf{2019}, \emph{7}, 1413--1446\relax
\mciteBstWouldAddEndPuncttrue
\mciteSetBstMidEndSepPunct{\mcitedefaultmidpunct}
{\mcitedefaultendpunct}{\mcitedefaultseppunct}\relax
\EndOfBibitem
\bibitem[Li \latin{et~al.}(2019)Li, He, Ji, Zhu, Xu, Zhang, Meng, Fang, and Ding]{li2019purcell}
Li,~H.; He,~F.; Ji,~C.; Zhu,~W.; Xu,~Y.; Zhang,~W.; Meng,~X.; Fang,~X.; Ding,~T. Purcell-enhanced spontaneous emission from perovskite quantum dots coupled to plasmonic crystal. \emph{The Journal of Physical Chemistry C} \textbf{2019}, \emph{123}, 25359--25365\relax
\mciteBstWouldAddEndPuncttrue
\mciteSetBstMidEndSepPunct{\mcitedefaultmidpunct}
{\mcitedefaultendpunct}{\mcitedefaultseppunct}\relax
\EndOfBibitem
\bibitem[Zhu \latin{et~al.}(2022)Zhu, Marczak, Feld, Boehme, Bernasconi, Moskalenko, Cherniukh, Dirin, Bodnarchuk, Kovalenko, \latin{et~al.} others]{zhu2022room}
Zhu,~C.; Marczak,~M.; Feld,~L.; Boehme,~S.~C.; Bernasconi,~C.; Moskalenko,~A.; Cherniukh,~I.; Dirin,~D.; Bodnarchuk,~M.~I.; Kovalenko,~M.~V.; others Room-Temperature, Highly Pure Single-Photon Sources from All-Inorganic Lead Halide Perovskite Quantum Dots. \emph{Nano Letters} \textbf{2022}, \emph{22}, 3751--3760\relax
\mciteBstWouldAddEndPuncttrue
\mciteSetBstMidEndSepPunct{\mcitedefaultmidpunct}
{\mcitedefaultendpunct}{\mcitedefaultseppunct}\relax
\EndOfBibitem
\bibitem[He \latin{et~al.}(2020)He, Chen, Hua, Liu, Wei, Liu, Hu, Shen, Zhang, Gao, \latin{et~al.} others]{he2020cmos}
He,~Z.; Chen,~B.; Hua,~Y.; Liu,~Z.; Wei,~Y.; Liu,~S.; Hu,~A.; Shen,~X.; Zhang,~Y.; Gao,~Y.; others CMOS Compatible High-Performance Nanolasing Based on Perovskite--SiN Hybrid Integration. \emph{Advanced Optical Materials} \textbf{2020}, \emph{8}, 2000453\relax
\mciteBstWouldAddEndPuncttrue
\mciteSetBstMidEndSepPunct{\mcitedefaultmidpunct}
{\mcitedefaultendpunct}{\mcitedefaultseppunct}\relax
\EndOfBibitem
\bibitem[Zhu \latin{et~al.}(2015)Zhu, Fu, Meng, Wu, Gong, Ding, Gustafsson, Trinh, Jin, and Zhu]{zhu2015lead}
Zhu,~H.; Fu,~Y.; Meng,~F.; Wu,~X.; Gong,~Z.; Ding,~Q.; Gustafsson,~M.~V.; Trinh,~M.~T.; Jin,~S.; Zhu,~X. Lead halide perovskite nanowire lasers with low lasing thresholds and high quality factors. \emph{Nature materials} \textbf{2015}, \emph{14}, 636--642\relax
\mciteBstWouldAddEndPuncttrue
\mciteSetBstMidEndSepPunct{\mcitedefaultmidpunct}
{\mcitedefaultendpunct}{\mcitedefaultseppunct}\relax
\EndOfBibitem
\bibitem[Kumar \latin{et~al.}(2022)Kumar, Sumukam, Rajaboina, Savu, Srinivas, and Banavoth]{kumar2022perovskite}
Kumar,~G.~S.; Sumukam,~R.~R.; Rajaboina,~R.~K.; Savu,~R.~N.; Srinivas,~M.; Banavoth,~M. Perovskite nanowires for next-generation optoelectronic devices: Lab to fab. \emph{ACS Applied Energy Materials} \textbf{2022}, \emph{5}, 1342--1377\relax
\mciteBstWouldAddEndPuncttrue
\mciteSetBstMidEndSepPunct{\mcitedefaultmidpunct}
{\mcitedefaultendpunct}{\mcitedefaultseppunct}\relax
\EndOfBibitem
\bibitem[Zhang \latin{et~al.}(2018)Zhang, Shang, Du, Shi, Wu, Mi, Chen, Liu, Li, Liu, \latin{et~al.} others]{zhang2018strong}
Zhang,~S.; Shang,~Q.; Du,~W.; Shi,~J.; Wu,~Z.; Mi,~Y.; Chen,~J.; Liu,~F.; Li,~Y.; Liu,~M.; others Strong exciton--photon coupling in hybrid inorganic--organic perovskite micro/nanowires. \emph{Advanced Optical Materials} \textbf{2018}, \emph{6}, 1701032\relax
\mciteBstWouldAddEndPuncttrue
\mciteSetBstMidEndSepPunct{\mcitedefaultmidpunct}
{\mcitedefaultendpunct}{\mcitedefaultseppunct}\relax
\EndOfBibitem
\bibitem[Li \latin{et~al.}(2017)Li, Luo, Huang, Ma, Ye, Zeng, and He]{li20172d}
Li,~J.; Luo,~L.; Huang,~H.; Ma,~C.; Ye,~Z.; Zeng,~J.; He,~H. 2D behaviors of excitons in cesium lead halide perovskite nanoplatelets. \emph{The Journal of Physical Chemistry Letters} \textbf{2017}, \emph{8}, 1161--1168\relax
\mciteBstWouldAddEndPuncttrue
\mciteSetBstMidEndSepPunct{\mcitedefaultmidpunct}
{\mcitedefaultendpunct}{\mcitedefaultseppunct}\relax
\EndOfBibitem
\bibitem[Otero-Mart{\'\i}nez \latin{et~al.}(2022)Otero-Mart{\'\i}nez, Ye, Sung, Pastoriza-Santos, P{\'e}rez-Juste, Xia, Rao, Hoye, and Polavarapu]{otero2022colloidal}
Otero-Mart{\'\i}nez,~C.; Ye,~J.; Sung,~J.; Pastoriza-Santos,~I.; P{\'e}rez-Juste,~J.; Xia,~Z.; Rao,~A.; Hoye,~R.~L.; Polavarapu,~L. Colloidal Metal-Halide Perovskite Nanoplatelets: Thickness-Controlled Synthesis, Properties, and Application in Light-Emitting Diodes. \emph{Advanced Materials} \textbf{2022}, \emph{34}, 2107105\relax
\mciteBstWouldAddEndPuncttrue
\mciteSetBstMidEndSepPunct{\mcitedefaultmidpunct}
{\mcitedefaultendpunct}{\mcitedefaultseppunct}\relax
\EndOfBibitem
\bibitem[Cohen \latin{et~al.}(2022)Cohen, Sharp, Kluherz, Chen, Munley, Anderson, Swanson, De~Yoreo, Luscombe, Majumdar, \latin{et~al.} others]{cohen2022direct}
Cohen,~T.~A.; Sharp,~D.; Kluherz,~K.~T.; Chen,~Y.; Munley,~C.; Anderson,~R.~T.; Swanson,~C.~J.; De~Yoreo,~J.~J.; Luscombe,~C.~K.; Majumdar,~A.; others Direct patterning of perovskite nanocrystals on nanophotonic cavities with electrohydrodynamic inkjet printing. \emph{Nano Letters} \textbf{2022}, \emph{22}, 5681--5688\relax
\mciteBstWouldAddEndPuncttrue
\mciteSetBstMidEndSepPunct{\mcitedefaultmidpunct}
{\mcitedefaultendpunct}{\mcitedefaultseppunct}\relax
\EndOfBibitem
\bibitem[Froch \latin{et~al.}(2020)Froch, Kim, Stewart, Xu, Du, Lockrey, Toth, and Aharonovich]{froch2020photonic}
Froch,~J.~E.; Kim,~S.; Stewart,~C.; Xu,~X.; Du,~Z.; Lockrey,~M.; Toth,~M.; Aharonovich,~I. Photonic nanobeam cavities with nanopockets for efficient integration of fluorescent nanoparticles. \emph{Nano letters} \textbf{2020}, \emph{20}, 2784--2790\relax
\mciteBstWouldAddEndPuncttrue
\mciteSetBstMidEndSepPunct{\mcitedefaultmidpunct}
{\mcitedefaultendpunct}{\mcitedefaultseppunct}\relax
\EndOfBibitem
\bibitem[Shang \latin{et~al.}(2018)Shang, Zhang, Liu, Chen, Yang, Li, Li, Zhang, Xiong, Liu, \latin{et~al.} others]{shang2018surface}
Shang,~Q.; Zhang,~S.; Liu,~Z.; Chen,~J.; Yang,~P.; Li,~C.; Li,~W.; Zhang,~Y.; Xiong,~Q.; Liu,~X.; others Surface plasmon enhanced strong exciton--photon coupling in hybrid inorganic--organic perovskite nanowires. \emph{Nano letters} \textbf{2018}, \emph{18}, 3335--3343\relax
\mciteBstWouldAddEndPuncttrue
\mciteSetBstMidEndSepPunct{\mcitedefaultmidpunct}
{\mcitedefaultendpunct}{\mcitedefaultseppunct}\relax
\EndOfBibitem
\bibitem[Heck{\"o}tter \latin{et~al.}(2021)Heck{\"o}tter, Walther, Scheel, Bayer, Pohl, and A{\ss}mann]{heckotter2021asymmetric}
Heck{\"o}tter,~J.; Walther,~V.; Scheel,~S.; Bayer,~M.; Pohl,~T.; A{\ss}mann,~M. Asymmetric Rydberg blockade of giant excitons in Cuprous Oxide. \emph{Nature communications} \textbf{2021}, \emph{12}, 3556\relax
\mciteBstWouldAddEndPuncttrue
\mciteSetBstMidEndSepPunct{\mcitedefaultmidpunct}
{\mcitedefaultendpunct}{\mcitedefaultseppunct}\relax
\EndOfBibitem
\bibitem[Bhaskar \latin{et~al.}(2017)Bhaskar, Sukachev, Sipahigil, Evans, Burek, Nguyen, Rogers, Siyushev, Metsch, Park, \latin{et~al.} others]{bhaskar2017quantum}
Bhaskar,~M.~K.; Sukachev,~D.~D.; Sipahigil,~A.; Evans,~R.~E.; Burek,~M.~J.; Nguyen,~C.~T.; Rogers,~L.~J.; Siyushev,~P.; Metsch,~M.~H.; Park,~H.; others Quantum nonlinear optics with a germanium-vacancy color center in a nanoscale diamond waveguide. \emph{Physical review letters} \textbf{2017}, \emph{118}, 223603\relax
\mciteBstWouldAddEndPuncttrue
\mciteSetBstMidEndSepPunct{\mcitedefaultmidpunct}
{\mcitedefaultendpunct}{\mcitedefaultseppunct}\relax
\EndOfBibitem
\bibitem[Ruf \latin{et~al.}(2021)Ruf, Wan, Choi, Englund, and Hanson]{ruf2021quantum}
Ruf,~M.; Wan,~N.~H.; Choi,~H.; Englund,~D.; Hanson,~R. Quantum networks based on color centers in diamond. \emph{Journal of Applied Physics} \textbf{2021}, \emph{130}, 070901\relax
\mciteBstWouldAddEndPuncttrue
\mciteSetBstMidEndSepPunct{\mcitedefaultmidpunct}
{\mcitedefaultendpunct}{\mcitedefaultseppunct}\relax
\EndOfBibitem
\bibitem[Castelletto and Boretti(2020)Castelletto, and Boretti]{castelletto2020silicon}
Castelletto,~S.; Boretti,~A. Silicon carbide color centers for quantum applications. \emph{Journal of Physics: Photonics} \textbf{2020}, \emph{2}, 022001\relax
\mciteBstWouldAddEndPuncttrue
\mciteSetBstMidEndSepPunct{\mcitedefaultmidpunct}
{\mcitedefaultendpunct}{\mcitedefaultseppunct}\relax
\EndOfBibitem
\bibitem[Caldwell \latin{et~al.}(2019)Caldwell, Aharonovich, Cassabois, Edgar, Gil, and Basov]{caldwell2019photonics}
Caldwell,~J.~D.; Aharonovich,~I.; Cassabois,~G.; Edgar,~J.~H.; Gil,~B.; Basov,~D. Photonics with hexagonal boron nitride. \emph{Nature Reviews Materials} \textbf{2019}, \emph{4}, 552--567\relax
\mciteBstWouldAddEndPuncttrue
\mciteSetBstMidEndSepPunct{\mcitedefaultmidpunct}
{\mcitedefaultendpunct}{\mcitedefaultseppunct}\relax
\EndOfBibitem
\bibitem[Iwasaki \latin{et~al.}(2015)Iwasaki, Ishibashi, Miyamoto, Doi, Kobayashi, Miyazaki, Tahara, Jahnke, Rogers, Naydenov, \latin{et~al.} others]{iwasaki2015germanium}
Iwasaki,~T.; Ishibashi,~F.; Miyamoto,~Y.; Doi,~Y.; Kobayashi,~S.; Miyazaki,~T.; Tahara,~K.; Jahnke,~K.~D.; Rogers,~L.~J.; Naydenov,~B.; others Germanium-vacancy single color centers in diamond. \emph{Scientific reports} \textbf{2015}, \emph{5}, 12882\relax
\mciteBstWouldAddEndPuncttrue
\mciteSetBstMidEndSepPunct{\mcitedefaultmidpunct}
{\mcitedefaultendpunct}{\mcitedefaultseppunct}\relax
\EndOfBibitem
\bibitem[Castelletto(2021)]{castelletto2021silicon}
Castelletto,~S. Silicon carbide single-photon sources: challenges and prospects. \emph{Materials for Quantum Technology} \textbf{2021}, \emph{1}, 023001\relax
\mciteBstWouldAddEndPuncttrue
\mciteSetBstMidEndSepPunct{\mcitedefaultmidpunct}
{\mcitedefaultendpunct}{\mcitedefaultseppunct}\relax
\EndOfBibitem
\bibitem[Wang \latin{et~al.}(2018)Wang, Zhou, Wang, Rasmita, Yang, Li, von Bardeleben, and Gao]{wang2018bright}
Wang,~J.; Zhou,~Y.; Wang,~Z.; Rasmita,~A.; Yang,~J.; Li,~X.; von Bardeleben,~H.~J.; Gao,~W. Bright room temperature single photon source at telecom range in cubic silicon carbide. \emph{Nature communications} \textbf{2018}, \emph{9}, 4106\relax
\mciteBstWouldAddEndPuncttrue
\mciteSetBstMidEndSepPunct{\mcitedefaultmidpunct}
{\mcitedefaultendpunct}{\mcitedefaultseppunct}\relax
\EndOfBibitem
\bibitem[Castelletto \latin{et~al.}(2014)Castelletto, Johnson, Iv{\'a}dy, Stavrias, Umeda, Gali, and Ohshima]{castelletto2014silicon}
Castelletto,~S.; Johnson,~B.; Iv{\'a}dy,~V.; Stavrias,~N.; Umeda,~T.; Gali,~A.; Ohshima,~T. A silicon carbide room-temperature single-photon source. \emph{Nature materials} \textbf{2014}, \emph{13}, 151--156\relax
\mciteBstWouldAddEndPuncttrue
\mciteSetBstMidEndSepPunct{\mcitedefaultmidpunct}
{\mcitedefaultendpunct}{\mcitedefaultseppunct}\relax
\EndOfBibitem
\bibitem[Morioka \latin{et~al.}(2020)Morioka, Babin, Nagy, Gediz, Hesselmeier, Liu, Joliffe, Niethammer, Dasari, Vorobyov, \latin{et~al.} others]{morioka2020spin}
Morioka,~N.; Babin,~C.; Nagy,~R.; Gediz,~I.; Hesselmeier,~E.; Liu,~D.; Joliffe,~M.; Niethammer,~M.; Dasari,~D.; Vorobyov,~V.; others Spin-controlled generation of indistinguishable and distinguishable photons from silicon vacancy centres in silicon carbide. \emph{Nature communications} \textbf{2020}, \emph{11}, 2516\relax
\mciteBstWouldAddEndPuncttrue
\mciteSetBstMidEndSepPunct{\mcitedefaultmidpunct}
{\mcitedefaultendpunct}{\mcitedefaultseppunct}\relax
\EndOfBibitem
\bibitem[Proscia \latin{et~al.}(2018)Proscia, Shotan, Jayakumar, Reddy, Cohen, Dollar, Alkauskas, Doherty, Meriles, and Menon]{proscia2018near}
Proscia,~N.~V.; Shotan,~Z.; Jayakumar,~H.; Reddy,~P.; Cohen,~C.; Dollar,~M.; Alkauskas,~A.; Doherty,~M.; Meriles,~C.~A.; Menon,~V.~M. Near-deterministic activation of room-temperature quantum emitters in hexagonal boron nitride. \emph{Optica} \textbf{2018}, \emph{5}, 1128--1134\relax
\mciteBstWouldAddEndPuncttrue
\mciteSetBstMidEndSepPunct{\mcitedefaultmidpunct}
{\mcitedefaultendpunct}{\mcitedefaultseppunct}\relax
\EndOfBibitem
\bibitem[Grosso \latin{et~al.}(2017)Grosso, Moon, Lienhard, Ali, Efetov, Furchi, Jarillo-Herrero, Ford, Aharonovich, and Englund]{grosso2017tunable}
Grosso,~G.; Moon,~H.; Lienhard,~B.; Ali,~S.; Efetov,~D.~K.; Furchi,~M.~M.; Jarillo-Herrero,~P.; Ford,~M.~J.; Aharonovich,~I.; Englund,~D. Tunable and high-purity room temperature single-photon emission from atomic defects in hexagonal boron nitride. \emph{Nature communications} \textbf{2017}, \emph{8}, 1--8\relax
\mciteBstWouldAddEndPuncttrue
\mciteSetBstMidEndSepPunct{\mcitedefaultmidpunct}
{\mcitedefaultendpunct}{\mcitedefaultseppunct}\relax
\EndOfBibitem
\bibitem[Dietrich \latin{et~al.}(2020)Dietrich, Doherty, Aharonovich, and Kubanek]{dietrich2020solid}
Dietrich,~A.; Doherty,~M.; Aharonovich,~I.; Kubanek,~A. Solid-state single photon source with Fourier transform limited lines at room temperature. \emph{Physical Review B} \textbf{2020}, \emph{101}, 081401\relax
\mciteBstWouldAddEndPuncttrue
\mciteSetBstMidEndSepPunct{\mcitedefaultmidpunct}
{\mcitedefaultendpunct}{\mcitedefaultseppunct}\relax
\EndOfBibitem
\bibitem[Kim \latin{et~al.}(2018)Kim, Fr{\"o}ch, Christian, Straw, Bishop, Totonjian, Watanabe, Taniguchi, Toth, and Aharonovich]{kim2018photonic}
Kim,~S.; Fr{\"o}ch,~J.~E.; Christian,~J.; Straw,~M.; Bishop,~J.; Totonjian,~D.; Watanabe,~K.; Taniguchi,~T.; Toth,~M.; Aharonovich,~I. Photonic crystal cavities from hexagonal boron nitride. \emph{Nature communications} \textbf{2018}, \emph{9}, 2623\relax
\mciteBstWouldAddEndPuncttrue
\mciteSetBstMidEndSepPunct{\mcitedefaultmidpunct}
{\mcitedefaultendpunct}{\mcitedefaultseppunct}\relax
\EndOfBibitem
\bibitem[Zhou \latin{et~al.}(2018)Zhou, Mu, Adamo, Bauerdick, Rudzinski, Aharonovich, and Gao]{zhou2018direct}
Zhou,~Y.; Mu,~Z.; Adamo,~G.; Bauerdick,~S.; Rudzinski,~A.; Aharonovich,~I.; Gao,~W.-b. Direct writing of single germanium vacancy center arrays in diamond. \emph{New Journal of Physics} \textbf{2018}, \emph{20}, 125004\relax
\mciteBstWouldAddEndPuncttrue
\mciteSetBstMidEndSepPunct{\mcitedefaultmidpunct}
{\mcitedefaultendpunct}{\mcitedefaultseppunct}\relax
\EndOfBibitem
\bibitem[Riedrich-M{\"o}ller \latin{et~al.}(2015)Riedrich-M{\"o}ller, Pezzagna, Meijer, Pauly, M{\"u}cklich, Markham, Edmonds, and Becher]{riedrich2015nanoimplantation}
Riedrich-M{\"o}ller,~J.; Pezzagna,~S.; Meijer,~J.; Pauly,~C.; M{\"u}cklich,~F.; Markham,~M.; Edmonds,~A.~M.; Becher,~C. Nanoimplantation and Purcell enhancement of single nitrogen-vacancy centers in photonic crystal cavities in diamond. \emph{Applied Physics Letters} \textbf{2015}, \emph{106}, 221103\relax
\mciteBstWouldAddEndPuncttrue
\mciteSetBstMidEndSepPunct{\mcitedefaultmidpunct}
{\mcitedefaultendpunct}{\mcitedefaultseppunct}\relax
\EndOfBibitem
\bibitem[Smith \latin{et~al.}(2019)Smith, Meynell, Bleszynski~Jayich, and Meijer]{smith2019colour}
Smith,~J.~M.; Meynell,~S.~A.; Bleszynski~Jayich,~A.~C.; Meijer,~J. Colour centre generation in diamond for quantum technologies. \emph{Nanophotonics} \textbf{2019}, \emph{8}, 1889--1906\relax
\mciteBstWouldAddEndPuncttrue
\mciteSetBstMidEndSepPunct{\mcitedefaultmidpunct}
{\mcitedefaultendpunct}{\mcitedefaultseppunct}\relax
\EndOfBibitem
\bibitem[Castelletto \latin{et~al.}(2019)Castelletto, Al~Atem, Inam, von Bardeleben, Hameau, Almutairi, Guillot, Sato, Boretti, and Bluet]{castelletto2019deterministic}
Castelletto,~S.; Al~Atem,~A.~S.; Inam,~F.~A.; von Bardeleben,~H.~J.; Hameau,~S.; Almutairi,~A.~F.; Guillot,~G.; Sato,~S.-i.; Boretti,~A.; Bluet,~J.~M. Deterministic placement of ultra-bright near-infrared color centers in arrays of silicon carbide micropillars. \emph{Beilstein Journal of Nanotechnology} \textbf{2019}, \emph{10}, 2383--2395\relax
\mciteBstWouldAddEndPuncttrue
\mciteSetBstMidEndSepPunct{\mcitedefaultmidpunct}
{\mcitedefaultendpunct}{\mcitedefaultseppunct}\relax
\EndOfBibitem
\bibitem[Kraus \latin{et~al.}(2017)Kraus, Simin, Kasper, Suda, Kawabata, Kada, Honda, Hijikata, Ohshima, Dyakonov, \latin{et~al.} others]{kraus2017three}
Kraus,~H.; Simin,~D.; Kasper,~C.; Suda,~Y.; Kawabata,~S.; Kada,~W.; Honda,~T.; Hijikata,~Y.; Ohshima,~T.; Dyakonov,~V.; others Three-dimensional proton beam writing of optically active coherent vacancy spins in silicon carbide. \emph{Nano letters} \textbf{2017}, \emph{17}, 2865--2870\relax
\mciteBstWouldAddEndPuncttrue
\mciteSetBstMidEndSepPunct{\mcitedefaultmidpunct}
{\mcitedefaultendpunct}{\mcitedefaultseppunct}\relax
\EndOfBibitem
\bibitem[Mir{\'o} \latin{et~al.}(2014)Mir{\'o}, Audiffred, and Heine]{miro2014atlas}
Mir{\'o},~P.; Audiffred,~M.; Heine,~T. An atlas of two-dimensional materials. \emph{Chemical Society Reviews} \textbf{2014}, \emph{43}, 6537--6554\relax
\mciteBstWouldAddEndPuncttrue
\mciteSetBstMidEndSepPunct{\mcitedefaultmidpunct}
{\mcitedefaultendpunct}{\mcitedefaultseppunct}\relax
\EndOfBibitem
\bibitem[Raja \latin{et~al.}(2017)Raja, Chaves, Yu, Arefe, Hill, Rigosi, Berkelbach, Nagler, Sch{\"u}ller, Korn, \latin{et~al.} others]{raja2017coulomb}
Raja,~A.; Chaves,~A.; Yu,~J.; Arefe,~G.; Hill,~H.~M.; Rigosi,~A.~F.; Berkelbach,~T.~C.; Nagler,~P.; Sch{\"u}ller,~C.; Korn,~T.; others Coulomb engineering of the bandgap and excitons in two-dimensional materials. \emph{Nature communications} \textbf{2017}, \emph{8}, 15251\relax
\mciteBstWouldAddEndPuncttrue
\mciteSetBstMidEndSepPunct{\mcitedefaultmidpunct}
{\mcitedefaultendpunct}{\mcitedefaultseppunct}\relax
\EndOfBibitem
\bibitem[Ponraj \latin{et~al.}(2016)Ponraj, Xu, Dhanabalan, Mu, Wang, Yuan, Li, Thakur, Ashrafi, Mccoubrey, \latin{et~al.} others]{ponraj2016photonics}
Ponraj,~J.~S.; Xu,~Z.-Q.; Dhanabalan,~S.~C.; Mu,~H.; Wang,~Y.; Yuan,~J.; Li,~P.; Thakur,~S.; Ashrafi,~M.; Mccoubrey,~K.; others Photonics and optoelectronics of two-dimensional materials beyond graphene. \emph{Nanotechnology} \textbf{2016}, \emph{27}, 462001\relax
\mciteBstWouldAddEndPuncttrue
\mciteSetBstMidEndSepPunct{\mcitedefaultmidpunct}
{\mcitedefaultendpunct}{\mcitedefaultseppunct}\relax
\EndOfBibitem
\bibitem[Al-Ani \latin{et~al.}(2022)Al-Ani, As'~Ham, Klochan, Hattori, Huang, and Miroshnichenko]{al2022recent}
Al-Ani,~I.~A.; As'~Ham,~K.; Klochan,~O.; Hattori,~H.~T.; Huang,~L.; Miroshnichenko,~A. Recent advances on strong light-matter coupling in atomically thin TMDC semiconductor materials. \emph{Journal of Optics} \textbf{2022}, \relax
\mciteBstWouldAddEndPunctfalse
\mciteSetBstMidEndSepPunct{\mcitedefaultmidpunct}
{}{\mcitedefaultseppunct}\relax
\EndOfBibitem
\bibitem[Emmanuele \latin{et~al.}(2020)Emmanuele, Sich, Kyriienko, Shahnazaryan, Withers, Catanzaro, Walker, Benimetskiy, Skolnick, Tartakovskii, \latin{et~al.} others]{emmanuele2020highly}
Emmanuele,~R.; Sich,~M.; Kyriienko,~O.; Shahnazaryan,~V.; Withers,~F.; Catanzaro,~A.; Walker,~P.; Benimetskiy,~F.; Skolnick,~M.; Tartakovskii,~A.; others Highly nonlinear trion-polaritons in a monolayer semiconductor. \emph{Nature communications} \textbf{2020}, \emph{11}, 3589\relax
\mciteBstWouldAddEndPuncttrue
\mciteSetBstMidEndSepPunct{\mcitedefaultmidpunct}
{\mcitedefaultendpunct}{\mcitedefaultseppunct}\relax
\EndOfBibitem
\bibitem[Ryou \latin{et~al.}(2018)Ryou, Rosser, Saxena, Fryett, and Majumdar]{ryou2018strong}
Ryou,~A.; Rosser,~D.; Saxena,~A.; Fryett,~T.; Majumdar,~A. Strong photon antibunching in weakly nonlinear two-dimensional exciton-polaritons. \emph{Physical Review B} \textbf{2018}, \emph{97}, 235307\relax
\mciteBstWouldAddEndPuncttrue
\mciteSetBstMidEndSepPunct{\mcitedefaultmidpunct}
{\mcitedefaultendpunct}{\mcitedefaultseppunct}\relax
\EndOfBibitem
\bibitem[Tonndorf \latin{et~al.}(2015)Tonndorf, Schmidt, Schneider, Kern, Buscema, Steele, Castellanos-Gomez, van~der Zant, de~Vasconcellos, and Bratschitsch]{tonndorf2015single}
Tonndorf,~P.; Schmidt,~R.; Schneider,~R.; Kern,~J.; Buscema,~M.; Steele,~G.~A.; Castellanos-Gomez,~A.; van~der Zant,~H.~S.; de~Vasconcellos,~S.~M.; Bratschitsch,~R. Single-photon emission from localized excitons in an atomically thin semiconductor. \emph{Optica} \textbf{2015}, \emph{2}, 347--352\relax
\mciteBstWouldAddEndPuncttrue
\mciteSetBstMidEndSepPunct{\mcitedefaultmidpunct}
{\mcitedefaultendpunct}{\mcitedefaultseppunct}\relax
\EndOfBibitem
\bibitem[Chakraborty \latin{et~al.}(2019)Chakraborty, Vamivakas, and Englund]{chakraborty2019advances}
Chakraborty,~C.; Vamivakas,~N.; Englund,~D. Advances in quantum light emission from 2D materials. \emph{Nanophotonics} \textbf{2019}, \emph{8}, 2017--2032\relax
\mciteBstWouldAddEndPuncttrue
\mciteSetBstMidEndSepPunct{\mcitedefaultmidpunct}
{\mcitedefaultendpunct}{\mcitedefaultseppunct}\relax
\EndOfBibitem
\bibitem[Choi \latin{et~al.}(2016)Choi, Tran, Elbadawi, Lobo, Wang, Juodkazis, Seniutinas, Toth, and Aharonovich]{choi2016engineering}
Choi,~S.; Tran,~T.~T.; Elbadawi,~C.; Lobo,~C.; Wang,~X.; Juodkazis,~S.; Seniutinas,~G.; Toth,~M.; Aharonovich,~I. Engineering and localization of quantum emitters in large hexagonal boron nitride layers. \emph{ACS applied materials \& interfaces} \textbf{2016}, \emph{8}, 29642--29648\relax
\mciteBstWouldAddEndPuncttrue
\mciteSetBstMidEndSepPunct{\mcitedefaultmidpunct}
{\mcitedefaultendpunct}{\mcitedefaultseppunct}\relax
\EndOfBibitem
\bibitem[Branny \latin{et~al.}(2017)Branny, Kumar, Proux, and Gerardot]{branny2017deterministic}
Branny,~A.; Kumar,~S.; Proux,~R.; Gerardot,~B.~D. Deterministic strain-induced arrays of quantum emitters in a two-dimensional semiconductor. \emph{Nature communications} \textbf{2017}, \emph{8}, 15053\relax
\mciteBstWouldAddEndPuncttrue
\mciteSetBstMidEndSepPunct{\mcitedefaultmidpunct}
{\mcitedefaultendpunct}{\mcitedefaultseppunct}\relax
\EndOfBibitem
\bibitem[Walther \latin{et~al.}(2018)Walther, Johne, and Pohl]{walther2018giant}
Walther,~V.; Johne,~R.; Pohl,~T. Giant optical nonlinearities from Rydberg excitons in semiconductor microcavities. \emph{Nature communications} \textbf{2018}, \emph{9}, 1309\relax
\mciteBstWouldAddEndPuncttrue
\mciteSetBstMidEndSepPunct{\mcitedefaultmidpunct}
{\mcitedefaultendpunct}{\mcitedefaultseppunct}\relax
\EndOfBibitem
\bibitem[Chernikov \latin{et~al.}(2014)Chernikov, Berkelbach, Hill, Rigosi, Li, Aslan, Reichman, Hybertsen, and Heinz]{chernikov2014exciton}
Chernikov,~A.; Berkelbach,~T.~C.; Hill,~H.~M.; Rigosi,~A.; Li,~Y.; Aslan,~B.; Reichman,~D.~R.; Hybertsen,~M.~S.; Heinz,~T.~F. Exciton binding energy and nonhydrogenic Rydberg series in monolayer WS 2. \emph{Physical review letters} \textbf{2014}, \emph{113}, 076802\relax
\mciteBstWouldAddEndPuncttrue
\mciteSetBstMidEndSepPunct{\mcitedefaultmidpunct}
{\mcitedefaultendpunct}{\mcitedefaultseppunct}\relax
\EndOfBibitem
\bibitem[Biswas \latin{et~al.}(2023)Biswas, Champagne, Haber, Pokawanvit, Wong, Akbari, Krylyuk, Watanabe, Taniguchi, Davydov, \latin{et~al.} others]{biswas2023rydberg}
Biswas,~S.; Champagne,~A.; Haber,~J.~B.; Pokawanvit,~S.; Wong,~J.; Akbari,~H.; Krylyuk,~S.; Watanabe,~K.; Taniguchi,~T.; Davydov,~A.~V.; others Rydberg excitons and trions in monolayer MoTe2. \emph{ACS nano} \textbf{2023}, \emph{17}, 7685--7694\relax
\mciteBstWouldAddEndPuncttrue
\mciteSetBstMidEndSepPunct{\mcitedefaultmidpunct}
{\mcitedefaultendpunct}{\mcitedefaultseppunct}\relax
\EndOfBibitem
\bibitem[Wang \latin{et~al.}(2020)Wang, Li, Li, Lu, Miao, Lian, Meng, Blei, Taniguchi, Watanabe, \latin{et~al.} others]{wang2020giant}
Wang,~T.; Li,~Z.; Li,~Y.; Lu,~Z.; Miao,~S.; Lian,~Z.; Meng,~Y.; Blei,~M.; Taniguchi,~T.; Watanabe,~K.; others Giant valley-polarized Rydberg excitons in monolayer WSe2 revealed by magneto-photocurrent spectroscopy. \emph{Nano Letters} \textbf{2020}, \emph{20}, 7635--7641\relax
\mciteBstWouldAddEndPuncttrue
\mciteSetBstMidEndSepPunct{\mcitedefaultmidpunct}
{\mcitedefaultendpunct}{\mcitedefaultseppunct}\relax
\EndOfBibitem
\bibitem[Liu \latin{et~al.}(2019)Liu, van Baren, Taniguchi, Watanabe, Chang, and Lui]{liu2019magnetophotoluminescence}
Liu,~E.; van Baren,~J.; Taniguchi,~T.; Watanabe,~K.; Chang,~Y.-C.; Lui,~C.~H. Magnetophotoluminescence of exciton Rydberg states in monolayer WS e 2. \emph{Physical Review B} \textbf{2019}, \emph{99}, 205420\relax
\mciteBstWouldAddEndPuncttrue
\mciteSetBstMidEndSepPunct{\mcitedefaultmidpunct}
{\mcitedefaultendpunct}{\mcitedefaultseppunct}\relax
\EndOfBibitem
\bibitem[Luo \latin{et~al.}(2017)Luo, Men, Liu, Mudryk, Zhao, Yao, Park, Shinar, Shinar, Ho, \latin{et~al.} others]{luo2017ultrafast}
Luo,~L.; Men,~L.; Liu,~Z.; Mudryk,~Y.; Zhao,~X.; Yao,~Y.; Park,~J.~M.; Shinar,~R.; Shinar,~J.; Ho,~K.-M.; others Ultrafast terahertz snapshots of excitonic Rydberg states and electronic coherence in an organometal halide perovskite. \emph{Nature Communications} \textbf{2017}, \emph{8}, 15565\relax
\mciteBstWouldAddEndPuncttrue
\mciteSetBstMidEndSepPunct{\mcitedefaultmidpunct}
{\mcitedefaultendpunct}{\mcitedefaultseppunct}\relax
\EndOfBibitem
\bibitem[Khazali \latin{et~al.}(2017)Khazali, Heshami, and Simon]{khazali2017single}
Khazali,~M.; Heshami,~K.; Simon,~C. Single-photon source based on Rydberg exciton blockade. \emph{Journal of Physics B: Atomic, Molecular and Optical Physics} \textbf{2017}, \emph{50}, 215301\relax
\mciteBstWouldAddEndPuncttrue
\mciteSetBstMidEndSepPunct{\mcitedefaultmidpunct}
{\mcitedefaultendpunct}{\mcitedefaultseppunct}\relax
\EndOfBibitem
\bibitem[Orfanakis \latin{et~al.}(2022)Orfanakis, Rajendran, Walther, Volz, Pohl, and Ohadi]{orfanakis2022rydberg}
Orfanakis,~K.; Rajendran,~S.~K.; Walther,~V.; Volz,~T.; Pohl,~T.; Ohadi,~H. Rydberg exciton--polaritons in a Cu2O microcavity. \emph{Nature Materials} \textbf{2022}, \emph{21}, 767--772\relax
\mciteBstWouldAddEndPuncttrue
\mciteSetBstMidEndSepPunct{\mcitedefaultmidpunct}
{\mcitedefaultendpunct}{\mcitedefaultseppunct}\relax
\EndOfBibitem
\bibitem[Lynch \latin{et~al.}(2021)Lynch, Hodges, Mandal, Langbein, Singh, Gallagher, Pritchett, Pizzey, Rogers, Adams, \latin{et~al.} others]{lynch2021rydberg}
Lynch,~S.~A.; Hodges,~C.; Mandal,~S.; Langbein,~W.; Singh,~R.~P.; Gallagher,~L.~A.; Pritchett,~J.~D.; Pizzey,~D.; Rogers,~J.~P.; Adams,~C.~S.; others Rydberg excitons in synthetic cuprous oxide Cu 2 O. \emph{Physical Review Materials} \textbf{2021}, \emph{5}, 084602\relax
\mciteBstWouldAddEndPuncttrue
\mciteSetBstMidEndSepPunct{\mcitedefaultmidpunct}
{\mcitedefaultendpunct}{\mcitedefaultseppunct}\relax
\EndOfBibitem
\bibitem[DeLange \latin{et~al.}(2023)DeLange, Barua, Paul, Ohadi, Zwiller, Steinhauer, and Alaeian]{delange2023highly}
DeLange,~J.; Barua,~K.; Paul,~A.~S.; Ohadi,~H.; Zwiller,~V.; Steinhauer,~S.; Alaeian,~H. Highly-excited Rydberg excitons in synthetic thin-film cuprous oxide. \emph{Scientific Reports} \textbf{2023}, \emph{13}, 16881\relax
\mciteBstWouldAddEndPuncttrue
\mciteSetBstMidEndSepPunct{\mcitedefaultmidpunct}
{\mcitedefaultendpunct}{\mcitedefaultseppunct}\relax
\EndOfBibitem
\bibitem[Wang \latin{et~al.}(2019)Wang, Kelkar, Martin-Cano, Rattenbacher, Shkarin, Utikal, G{\"o}tzinger, and Sandoghdar]{wang2019turning}
Wang,~D.; Kelkar,~H.; Martin-Cano,~D.; Rattenbacher,~D.; Shkarin,~A.; Utikal,~T.; G{\"o}tzinger,~S.; Sandoghdar,~V. Turning a molecule into a coherent two-level quantum system. \emph{Nature Physics} \textbf{2019}, \emph{15}, 483--489\relax
\mciteBstWouldAddEndPuncttrue
\mciteSetBstMidEndSepPunct{\mcitedefaultmidpunct}
{\mcitedefaultendpunct}{\mcitedefaultseppunct}\relax
\EndOfBibitem
\bibitem[Polisseni \latin{et~al.}(2016)Polisseni, Major, Boissier, Grandi, Clark, and Hinds]{polisseni2016stable}
Polisseni,~C.; Major,~K.~D.; Boissier,~S.; Grandi,~S.; Clark,~A.~S.; Hinds,~E. Stable, single-photon emitter in a thin organic crystal for application to quantum-photonic devices. \emph{Optics Express} \textbf{2016}, \emph{24}, 5615--5627\relax
\mciteBstWouldAddEndPuncttrue
\mciteSetBstMidEndSepPunct{\mcitedefaultmidpunct}
{\mcitedefaultendpunct}{\mcitedefaultseppunct}\relax
\EndOfBibitem
\bibitem[Hail \latin{et~al.}(2019)Hail, H{\"o}ller, Matsuzaki, Rohner, Renger, Sandoghdar, Poulikakos, and Eghlidi]{hail2019nanoprinting}
Hail,~C.~U.; H{\"o}ller,~C.; Matsuzaki,~K.; Rohner,~P.; Renger,~J.; Sandoghdar,~V.; Poulikakos,~D.; Eghlidi,~H. Nanoprinting organic molecules at the quantum level. \emph{Nature communications} \textbf{2019}, \emph{10}, 1--8\relax
\mciteBstWouldAddEndPuncttrue
\mciteSetBstMidEndSepPunct{\mcitedefaultmidpunct}
{\mcitedefaultendpunct}{\mcitedefaultseppunct}\relax
\EndOfBibitem
\bibitem[Robinson \latin{et~al.}(2005)Robinson, Manolatou, Chen, and Lipson]{robinson2005ultrasmall}
Robinson,~J.~T.; Manolatou,~C.; Chen,~L.; Lipson,~M. Ultrasmall mode volumes in dielectric optical microcavities. \emph{Physical review letters} \textbf{2005}, \emph{95}, 143901\relax
\mciteBstWouldAddEndPuncttrue
\mciteSetBstMidEndSepPunct{\mcitedefaultmidpunct}
{\mcitedefaultendpunct}{\mcitedefaultseppunct}\relax
\EndOfBibitem
\bibitem[Seidler \latin{et~al.}(2013)Seidler, Lister, Drechsler, Hofrichter, and St{\"o}ferle]{seidler2013slotted}
Seidler,~P.; Lister,~K.; Drechsler,~U.; Hofrichter,~J.; St{\"o}ferle,~T. Slotted photonic crystal nanobeam cavity with an ultrahigh quality factor-to-mode volume ratio. \emph{Optics Express} \textbf{2013}, \emph{21}, 32468--32483\relax
\mciteBstWouldAddEndPuncttrue
\mciteSetBstMidEndSepPunct{\mcitedefaultmidpunct}
{\mcitedefaultendpunct}{\mcitedefaultseppunct}\relax
\EndOfBibitem
\bibitem[Hu and Weiss(2016)Hu, and Weiss]{hu2016design}
Hu,~S.; Weiss,~S.~M. Design of photonic crystal cavities for extreme light concentration. \emph{ACS photonics} \textbf{2016}, \emph{3}, 1647--1653\relax
\mciteBstWouldAddEndPuncttrue
\mciteSetBstMidEndSepPunct{\mcitedefaultmidpunct}
{\mcitedefaultendpunct}{\mcitedefaultseppunct}\relax
\EndOfBibitem
\bibitem[Choi \latin{et~al.}(2017)Choi, Heuck, and Englund]{choi2017self}
Choi,~H.; Heuck,~M.; Englund,~D. Self-similar nanocavity design with ultrasmall mode volume for single-photon nonlinearities. \emph{Physical review letters} \textbf{2017}, \emph{118}, 223605\relax
\mciteBstWouldAddEndPuncttrue
\mciteSetBstMidEndSepPunct{\mcitedefaultmidpunct}
{\mcitedefaultendpunct}{\mcitedefaultseppunct}\relax
\EndOfBibitem
\bibitem[Jensen and Sigmund(2011)Jensen, and Sigmund]{jensen2011topology}
Jensen,~J.~S.; Sigmund,~O. Topology optimization for nano-photonics. \emph{Laser \& Photonics Reviews} \textbf{2011}, \emph{5}, 308--321\relax
\mciteBstWouldAddEndPuncttrue
\mciteSetBstMidEndSepPunct{\mcitedefaultmidpunct}
{\mcitedefaultendpunct}{\mcitedefaultseppunct}\relax
\EndOfBibitem
\bibitem[Albrechtsen \latin{et~al.}(2022)Albrechtsen, Vosoughi~Lahijani, Christiansen, Nguyen, Casses, Hansen, Stenger, Sigmund, Jansen, M{\o}rk, \latin{et~al.} others]{albrechtsen2022nanometer}
Albrechtsen,~M.; Vosoughi~Lahijani,~B.; Christiansen,~R.~E.; Nguyen,~V. T.~H.; Casses,~L.~N.; Hansen,~S.~E.; Stenger,~N.; Sigmund,~O.; Jansen,~H.; M{\o}rk,~J.; others Nanometer-scale photon confinement in topology-optimized dielectric cavities. \emph{Nature Communications} \textbf{2022}, \emph{13}, 6281\relax
\mciteBstWouldAddEndPuncttrue
\mciteSetBstMidEndSepPunct{\mcitedefaultmidpunct}
{\mcitedefaultendpunct}{\mcitedefaultseppunct}\relax
\EndOfBibitem
\bibitem[Zhang \latin{et~al.}(2020)Zhang, Liu, Wang, Zhang, and Lu]{zhang2020hybrid}
Zhang,~H.; Liu,~Y.-C.; Wang,~C.; Zhang,~N.; Lu,~C. Hybrid photonic-plasmonic nano-cavity with ultra-high Q/V. \emph{Optics Letters} \textbf{2020}, \emph{45}, 4794--4797\relax
\mciteBstWouldAddEndPuncttrue
\mciteSetBstMidEndSepPunct{\mcitedefaultmidpunct}
{\mcitedefaultendpunct}{\mcitedefaultseppunct}\relax
\EndOfBibitem
\bibitem[Lawrence \latin{et~al.}(2020)Lawrence, Barton~III, Dixon, Song, van~de Groep, Brongersma, and Dionne]{lawrence2020high}
Lawrence,~M.; Barton~III,~D.~R.; Dixon,~J.; Song,~J.-H.; van~de Groep,~J.; Brongersma,~M.~L.; Dionne,~J.~A. High quality factor phase gradient metasurfaces. \emph{Nature Nanotechnology} \textbf{2020}, \emph{15}, 956--961\relax
\mciteBstWouldAddEndPuncttrue
\mciteSetBstMidEndSepPunct{\mcitedefaultmidpunct}
{\mcitedefaultendpunct}{\mcitedefaultseppunct}\relax
\EndOfBibitem
\bibitem[Huang \latin{et~al.}(2023)Huang, Jin, Zhou, Li, Xu, Overvig, Deng, Chen, Lu, Al{\`u}, \latin{et~al.} others]{huang2023ultrahigh}
Huang,~L.; Jin,~R.; Zhou,~C.; Li,~G.; Xu,~L.; Overvig,~A.; Deng,~F.; Chen,~X.; Lu,~W.; Al{\`u},~A.; others Ultrahigh-Q guided mode resonances in an All-dielectric metasurface. \emph{Nature Communications} \textbf{2023}, \emph{14}, 3433\relax
\mciteBstWouldAddEndPuncttrue
\mciteSetBstMidEndSepPunct{\mcitedefaultmidpunct}
{\mcitedefaultendpunct}{\mcitedefaultseppunct}\relax
\EndOfBibitem
\bibitem[Palstra \latin{et~al.}(2019)Palstra, Doeleman, and Koenderink]{palstra2019hybrid}
Palstra,~I.~M.; Doeleman,~H.~M.; Koenderink,~A.~F. Hybrid cavity-antenna systems for quantum optics outside the cryostat? \emph{Nanophotonics} \textbf{2019}, \emph{8}, 1513--1531\relax
\mciteBstWouldAddEndPuncttrue
\mciteSetBstMidEndSepPunct{\mcitedefaultmidpunct}
{\mcitedefaultendpunct}{\mcitedefaultseppunct}\relax
\EndOfBibitem
\bibitem[Min \latin{et~al.}(2009)Min, Ostby, Sorger, Ulin-Avila, Yang, Zhang, and Vahala]{min2009high}
Min,~B.; Ostby,~E.; Sorger,~V.; Ulin-Avila,~E.; Yang,~L.; Zhang,~X.; Vahala,~K. High-Q surface-plasmon-polariton whispering-gallery microcavity. \emph{Nature} \textbf{2009}, \emph{457}, 455--458\relax
\mciteBstWouldAddEndPuncttrue
\mciteSetBstMidEndSepPunct{\mcitedefaultmidpunct}
{\mcitedefaultendpunct}{\mcitedefaultseppunct}\relax
\EndOfBibitem
\bibitem[Xiao \latin{et~al.}(2012)Xiao, Liu, Li, Chen, Li, and Gong]{xiao2012strongly}
Xiao,~Y.-F.; Liu,~Y.-C.; Li,~B.-B.; Chen,~Y.-L.; Li,~Y.; Gong,~Q. Strongly enhanced light-matter interaction in a hybrid photonic-plasmonic resonator. \emph{Physical Review A} \textbf{2012}, \emph{85}, 031805\relax
\mciteBstWouldAddEndPuncttrue
\mciteSetBstMidEndSepPunct{\mcitedefaultmidpunct}
{\mcitedefaultendpunct}{\mcitedefaultseppunct}\relax
\EndOfBibitem
\bibitem[Liu \latin{et~al.}(2017)Liu, Huang, Liu, Singamaneni, and Cunningham]{liu2017nanoantenna}
Liu,~J.-N.; Huang,~Q.; Liu,~K.-K.; Singamaneni,~S.; Cunningham,~B.~T. Nanoantenna--microcavity hybrids with highly cooperative plasmonic--photonic coupling. \emph{Nano letters} \textbf{2017}, \emph{17}, 7569--7577\relax
\mciteBstWouldAddEndPuncttrue
\mciteSetBstMidEndSepPunct{\mcitedefaultmidpunct}
{\mcitedefaultendpunct}{\mcitedefaultseppunct}\relax
\EndOfBibitem
\bibitem[Barreda \latin{et~al.}(2022)Barreda, Mercad{\'e}, Zapata-Herrera, Aizpurua, and Mart{\'\i}nez]{barreda2022hybrid}
Barreda,~A.; Mercad{\'e},~L.; Zapata-Herrera,~M.; Aizpurua,~J.; Mart{\'\i}nez,~A. Hybrid photonic-plasmonic cavity design for very large purcell factors at telecommunication wavelengths. \emph{Physical Review Applied} \textbf{2022}, \emph{18}, 044066\relax
\mciteBstWouldAddEndPuncttrue
\mciteSetBstMidEndSepPunct{\mcitedefaultmidpunct}
{\mcitedefaultendpunct}{\mcitedefaultseppunct}\relax
\EndOfBibitem
\bibitem[Barth \latin{et~al.}(2010)Barth, Schietinger, Fischer, Becker, Nusse, Aichele, Lochel, Sonnichsen, and Benson]{barth2010nanoassembled}
Barth,~M.; Schietinger,~S.; Fischer,~S.; Becker,~J.; Nusse,~N.; Aichele,~T.; Lochel,~B.; Sonnichsen,~C.; Benson,~O. Nanoassembled plasmonic-photonic hybrid cavity for tailored light-matter coupling. \emph{Nano letters} \textbf{2010}, \emph{10}, 891--895\relax
\mciteBstWouldAddEndPuncttrue
\mciteSetBstMidEndSepPunct{\mcitedefaultmidpunct}
{\mcitedefaultendpunct}{\mcitedefaultseppunct}\relax
\EndOfBibitem
\bibitem[Conteduca \latin{et~al.}(2017)Conteduca, Reardon, Scullion, Dell’Olio, Armenise, Krauss, and Ciminelli]{conteduca2017ultra}
Conteduca,~D.; Reardon,~C.; Scullion,~M.~G.; Dell’Olio,~F.; Armenise,~M.~N.; Krauss,~T.~F.; Ciminelli,~C. Ultra-high Q/V hybrid cavity for strong light-matter interaction. \emph{APL Photonics} \textbf{2017}, \emph{2}\relax
\mciteBstWouldAddEndPuncttrue
\mciteSetBstMidEndSepPunct{\mcitedefaultmidpunct}
{\mcitedefaultendpunct}{\mcitedefaultseppunct}\relax
\EndOfBibitem
\bibitem[Li \latin{et~al.}(2023)Li, Liu, Li, Zhong, Lu, Chen, and Wang]{Li2023exceptional}
Li,~W.; Liu,~R.; Li,~J.; Zhong,~J.; Lu,~Y.-W.; Chen,~H.; Wang,~X.-H. {Highly Efficient Single-Exciton Strong Coupling with Plasmons by Lowering Critical Interaction Strength at an Exceptional Point}. \emph{Physical Review Letters} \textbf{2023}, \emph{130}, 143601\relax
\mciteBstWouldAddEndPuncttrue
\mciteSetBstMidEndSepPunct{\mcitedefaultmidpunct}
{\mcitedefaultendpunct}{\mcitedefaultseppunct}\relax
\EndOfBibitem
\bibitem[Shi \latin{et~al.}(2022)Shi, Xu, Zhang, Ye, Xiang, Liu, Wang, Su, and Li]{shi2022high}
Shi,~S.; Xu,~B.; Zhang,~K.; Ye,~G.-S.; Xiang,~D.-S.; Liu,~Y.; Wang,~J.; Su,~D.; Li,~L. High-fidelity photonic quantum logic gate based on near-optimal Rydberg single-photon source. \emph{Nature Communications} \textbf{2022}, \emph{13}, 4454\relax
\mciteBstWouldAddEndPuncttrue
\mciteSetBstMidEndSepPunct{\mcitedefaultmidpunct}
{\mcitedefaultendpunct}{\mcitedefaultseppunct}\relax
\EndOfBibitem
\bibitem[Kimble(2008)]{kimble2008quantum}
Kimble,~H.~J. The quantum internet. \emph{Nature} \textbf{2008}, \emph{453}, 1023--1030\relax
\mciteBstWouldAddEndPuncttrue
\mciteSetBstMidEndSepPunct{\mcitedefaultmidpunct}
{\mcitedefaultendpunct}{\mcitedefaultseppunct}\relax
\EndOfBibitem
\bibitem[Ornelas-Huerta \latin{et~al.}(2020)Ornelas-Huerta, Craddock, Goldschmidt, Hachtel, Wang, Bienias, Gorshkov, Rolston, and Porto]{ornelas2020demand}
Ornelas-Huerta,~D.~P.; Craddock,~A.~N.; Goldschmidt,~E.~A.; Hachtel,~A.~J.; Wang,~Y.; Bienias,~P.; Gorshkov,~A.~V.; Rolston,~S.~L.; Porto,~J.~V. On-demand indistinguishable single photons from an efficient and pure source based on a Rydberg ensemble. \emph{Optica} \textbf{2020}, \emph{7}, 813--819\relax
\mciteBstWouldAddEndPuncttrue
\mciteSetBstMidEndSepPunct{\mcitedefaultmidpunct}
{\mcitedefaultendpunct}{\mcitedefaultseppunct}\relax
\EndOfBibitem
\bibitem[Ripka \latin{et~al.}(2018)Ripka, K{\"u}bler, L{\"o}w, and Pfau]{ripka2018room}
Ripka,~F.; K{\"u}bler,~H.; L{\"o}w,~R.; Pfau,~T. A room-temperature single-photon source based on strongly interacting Rydberg atoms. \emph{Science} \textbf{2018}, \emph{362}, 446--449\relax
\mciteBstWouldAddEndPuncttrue
\mciteSetBstMidEndSepPunct{\mcitedefaultmidpunct}
{\mcitedefaultendpunct}{\mcitedefaultseppunct}\relax
\EndOfBibitem
\bibitem[Rauschenbeutel \latin{et~al.}(1999)Rauschenbeutel, Nogues, Osnaghi, Bertet, Brune, Raimond, and Haroche]{rauschenbeutel1999coherent}
Rauschenbeutel,~A.; Nogues,~G.; Osnaghi,~S.; Bertet,~P.; Brune,~M.; Raimond,~J.-M.; Haroche,~S. Coherent operation of a tunable quantum phase gate in cavity QED. \emph{Physical Review Letters} \textbf{1999}, \emph{83}, 5166\relax
\mciteBstWouldAddEndPuncttrue
\mciteSetBstMidEndSepPunct{\mcitedefaultmidpunct}
{\mcitedefaultendpunct}{\mcitedefaultseppunct}\relax
\EndOfBibitem
\bibitem[Volz \latin{et~al.}(2014)Volz, Scheucher, Junge, and Rauschenbeutel]{volz2014nonlinear}
Volz,~J.; Scheucher,~M.; Junge,~C.; Rauschenbeutel,~A. Nonlinear $\pi$ phase shift for single fibre-guided photons interacting with a single resonator-enhanced atom. \emph{Nature Photonics} \textbf{2014}, \emph{8}, 965--970\relax
\mciteBstWouldAddEndPuncttrue
\mciteSetBstMidEndSepPunct{\mcitedefaultmidpunct}
{\mcitedefaultendpunct}{\mcitedefaultseppunct}\relax
\EndOfBibitem
\bibitem[Tiarks \latin{et~al.}(2016)Tiarks, Schmidt, Rempe, and D{\"u}rr]{tiarks2016optical}
Tiarks,~D.; Schmidt,~S.; Rempe,~G.; D{\"u}rr,~S. Optical $\pi$ phase shift created with a single-photon pulse. \emph{Science Advances} \textbf{2016}, \emph{2}, e1600036\relax
\mciteBstWouldAddEndPuncttrue
\mciteSetBstMidEndSepPunct{\mcitedefaultmidpunct}
{\mcitedefaultendpunct}{\mcitedefaultseppunct}\relax
\EndOfBibitem
\bibitem[Sun \latin{et~al.}(2018)Sun, Kim, Luo, Solomon, and Waks]{sun2018single}
Sun,~S.; Kim,~H.; Luo,~Z.; Solomon,~G.~S.; Waks,~E. A single-photon switch and transistor enabled by a solid-state quantum memory. \emph{Science} \textbf{2018}, \emph{361}, 57--60\relax
\mciteBstWouldAddEndPuncttrue
\mciteSetBstMidEndSepPunct{\mcitedefaultmidpunct}
{\mcitedefaultendpunct}{\mcitedefaultseppunct}\relax
\EndOfBibitem
\bibitem[Hartmann \latin{et~al.}(2008)Hartmann, Brandao, and Plenio]{hartmann2008quantum}
Hartmann,~M.~J.; Brandao,~F.~G.; Plenio,~M.~B. Quantum many-body phenomena in coupled cavity arrays. \emph{Laser \& Photonics Reviews} \textbf{2008}, \emph{2}, 527--556\relax
\mciteBstWouldAddEndPuncttrue
\mciteSetBstMidEndSepPunct{\mcitedefaultmidpunct}
{\mcitedefaultendpunct}{\mcitedefaultseppunct}\relax
\EndOfBibitem
\bibitem[Hartmann \latin{et~al.}(2006)Hartmann, Brandao, and Plenio]{hartmann2006strongly}
Hartmann,~M.~J.; Brandao,~F.~G.; Plenio,~M.~B. Strongly interacting polaritons in coupled arrays of cavities. \emph{Nature Physics} \textbf{2006}, \emph{2}, 849--855\relax
\mciteBstWouldAddEndPuncttrue
\mciteSetBstMidEndSepPunct{\mcitedefaultmidpunct}
{\mcitedefaultendpunct}{\mcitedefaultseppunct}\relax
\EndOfBibitem
\bibitem[Chen \latin{et~al.}(2013)Chen, Beck, B{\"u}cker, Gullans, Lukin, Tanji-Suzuki, and Vuleti{\'c}]{chen2013all}
Chen,~W.; Beck,~K.~M.; B{\"u}cker,~R.; Gullans,~M.; Lukin,~M.~D.; Tanji-Suzuki,~H.; Vuleti{\'c},~V. All-optical switch and transistor gated by one stored photon. \emph{Science} \textbf{2013}, \emph{341}, 768--770\relax
\mciteBstWouldAddEndPuncttrue
\mciteSetBstMidEndSepPunct{\mcitedefaultmidpunct}
{\mcitedefaultendpunct}{\mcitedefaultseppunct}\relax
\EndOfBibitem
\bibitem[Reiserer \latin{et~al.}(2013)Reiserer, Ritter, and Rempe]{reiserer2013nondestructive}
Reiserer,~A.; Ritter,~S.; Rempe,~G. Nondestructive detection of an optical photon. \emph{Science} \textbf{2013}, \emph{342}, 1349--1351\relax
\mciteBstWouldAddEndPuncttrue
\mciteSetBstMidEndSepPunct{\mcitedefaultmidpunct}
{\mcitedefaultendpunct}{\mcitedefaultseppunct}\relax
\EndOfBibitem
\end{mcitethebibliography}

\end{document}